\documentclass[12pt]{article}
\newcommand{\refLem}[2]{ Lemma~\ref{#2}.\ref{#1} }

\usepackage{amssymb}
\usepackage{amsfonts}
\usepackage{latexsym}
\usepackage{amsmath}
\usepackage{amsthm}
\usepackage{graphicx}

\renewcommand{\theequation}{\arabic{section}.\arabic{equation}}
\newcommand{\er}[1]{{\rm(\ref{#1})}}
\def\lb{\label}

\theoremstyle{plain}
\newtheorem{theorem}{\bf Theorem}[section]
\newtheorem{lemma}[theorem]{\bf Lemma}
\newtheorem{corollary}[theorem]{\bf Corollary}
\newtheorem{proposition}[theorem]{\bf Proposition}
\theoremstyle{remark}

\newcommand{\abs}[1]{\lvert#1\rvert}
\newcommand{\norm}[1]{\lVert#1\rVert}

\begin{document}
\def\a{\alpha}  \def\cA{{\cal A}}     \def\bA{{\bf A}}  \def\mA{{\mathscr A}}
\def\b{\beta}   \def\cB{{\cal B}}     \def\bB{{\bf B}}  \def\mB{{\mathscr B}}
\def\g{\gamma}  \def\cC{{\cal C}}     \def\bC{{\bf C}}  \def\mC{{\mathscr C}}
\def\G{\Gamma}  \def\cD{{\cal D}}     \def\bD{{\bf D}}  \def\mD{{\mathscr D}}
\def\d{\delta}  \def\cE{{\cal E}}     \def\bE{{\bf E}}  \def\mE{{\mathscr E}}
\def\D{\Delta}  \def\cF{{\cal F}}     \def\bF{{\bf F}}  \def\mF{{\mathscr F}}
\def\c{\chi}    \def\cG{{\cal G}}     \def\bG{{\bf G}}  \def\mG{{\mathscr G}}
\def\z{\zeta}   \def\cH{{\cal H}}     \def\bH{{\bf H}}  \def\mH{{\mathscr H}}
\def\e{\eta}    \def\cI{{\cal I}}     \def\bI{{\bf I}}  \def\mI{{\mathscr I}}
\def\p{\psi}    \def\cJ{{\cal J}}     \def\bJ{{\bf J}}  \def\mJ{{\mathscr J}}
\def\vT{\Theta} \def\cK{{\cal K}}     \def\bK{{\bf K}}  \def\mK{{\mathscr K}}
\def\k{\kappa}  \def\cL{{\cal L}}     \def\bL{{\bf L}}  \def\mL{{\mathscr L}}
\def\l{\lambda} \def\cM{{\cal M}}     \def\bM{{\bf M}}  \def\mM{{\mathscr M}}
\def\L{\Lambda} \def\cN{{\cal N}}     \def\bN{{\bf N}}  \def\mN{{\mathscr N}}
\def\m{\mu}     \def\cO{{\cal O}}     \def\bO{{\bf O}}  \def\mO{{\mathscr O}}
\def\n{\nu}     \def\cP{{\cal P}}     \def\bP{{\bf P}}  \def\mP{{\mathscr P}}
\def\r{\rho}    \def\cQ{{\cal Q}}     \def\bQ{{\bf Q}}  \def\mQ{{\mathscr Q}}
\def\s{\sigma}  \def\cR{{\cal R}}     \def\bR{{\bf R}}  \def\mR{{\mathscr R}}
\def\S{\Sigma}  \def\cS{{\cal S}}     \def\bS{{\bf S}}  \def\mS{{\mathscr S}}
\def\t{\tau}    \def\cT{{\cal T}}     \def\bT{{\bf T}}  \def\mT{{\mathscr T}}
\def\f{\phi}    \def\cU{{\cal U}}     \def\bU{{\bf U}}  \def\mU{{\mathscr U}}
\def\F{\Phi}    \def\cV{{\cal V}}     \def\bV{{\bf V}}  \def\mV{{\mathscr V}}
\def\P{\Psi}    \def\cW{{\cal W}}     \def\bW{{\bf W}}  \def\mW{{\mathscr W}}
\def\o{\omega}  \def\cX{{\cal X}}     \def\bX{{\bf X}}  \def\mX{{\mathscr X}}
\def\x{\xi}     \def\cY{{\cal Y}}     \def\bY{{\bf Y}}  \def\mY{{\mathscr Y}}
\def\X{\Xi}     \def\cZ{{\cal Z}}     \def\bZ{{\bf Z}}  \def\Z{{\mathscr Z}}
\def\O{\Omega}
\def\ve{\varepsilon}
\def\vt{\vartheta}
\def\vp{\varphi}
\def\vk{\varkappa}

\def\Z{{\Bbb Z}}
\def\R{{\Bbb R}}
\def\C{{\Bbb C}}
\def\T{{\Bbb T}}
\def\N{{\Bbb N}}

\def\Z{{\Bbb Z}}
\def\R{{\Bbb R}}
\def\C{{\Bbb C}}
\def\T{{\Bbb T}}
\def\N{{\Bbb N}}
\def\dD{{\Bbb D}}

\def\ma{\left(\begin{array}{cc}}    \def\am{\end{array}\right)}
\def\iint{\int\!\!\!\int}
\def\lt{\biggl}                    \def\rt{\biggr}
\let\ge\geqslant                   \let\le\leqslant
\def\[{\begin{equation}}            \def\]{\end{equation}}
\def\wt{\widetilde}                 \def\pa{\partial}
\def\sm{\setminus}                  \def\es{\emptyset}
\def\no{\noindent}                  \def\ol{\overline}
\def\iy{\infty}                     \def\ev{\equiv}
\def\/{\over}
\def\we{\wedge}
\def\ts{\times}
\def\os{\oplus}
\def\ss{\subset}
\def\h{\hat}
\def\wh{\widehat}
\def\Ra{\Rightarrow}
\def\ra{\rightarrow}
\def\la{\leftarrow}
\def\da{\downarrow}
\def\ua{\uparrow}
\def\lra{\leftrightarrow}
\def\Lra{\Leftrightarrow}
\def\Re{\mathop{\rm Re}\nolimits}
\def\Im{\mathop{\rm Im}\nolimits}
\def\supp{\mathop{\rm supp}\nolimits}
\def\sign{\mathop{\rm sign}\nolimits}
\def\Ran{\mathop{\rm Ran}\nolimits}
\def\Ker{\mathop{\rm Ker}\nolimits}
\def\Tr{\mathop{\rm Tr}\nolimits}
\def\const{\mathop{\rm const}\nolimits}
\def\Wr{\mathop{\rm Wr}\nolimits}
\def\la{\langle} \def\ra{\rangle}

\def\th{\theta}
\def\dlint{\displaystyle\int\limits}
\def\iintt{\mathop{\int\!\!\int\!\!\dots\!\!\int}\limits}
\def\intt{\mathop{\int\int}\limits}
\def\lim{\mathop{\rm lim}\limits}
\def\mult{\!\cdot\!}
\def\BBox{\blacksquare}
\def\1{1\!\!1}
\newcommand{\bwt}[1]{{\mathop{#1}\limits^{{}_{\,\bf{\sim}}}}\vphantom{#1}}
\newcommand{\bhat}[1]{{\mathop{#1}\limits^{{}_{\,\bf{\wedge}}}}\vphantom{#1}}
\newcommand{\bcheck}[1]{{\mathop{#1}\limits^{{}_{\,\bf{\vee}}}}\vphantom{#1}}
\def\nh{\bhat}
\def\nc{\bcheck}
\newcommand{\oo}[1]{{\mathop{#1}\limits^{\,\circ}}\vphantom{#1}}
\newcommand{\po}[1]{{\mathop{#1}\limits^{\phantom{\circ}}}\vphantom{#1}}
\def\notto{\to\!\!\!\!\!\!\!/\,\,\,}

\def\pgbrk{\pagebreak}


\def\Twelve{
\font\Tenmsa=msam10 scaled 1200 \font\Sevenmsa=msam7 scaled 1200
\font\Fivemsa=msam5 scaled 1200
\textfont\msafam=\Tenmsa \scriptfont\msafam=\Sevenmsa
\scriptscriptfont\msafam=\Fivemsa

\font\Tenmsb=msbm10 scaled 1200 \font\Sevenmsb=msbm7 scaled 1200
\font\Fivemsb=msbm5 scaled 1200
\textfont\msbfam=\Tenmsb \scriptfont\msbfam=\Sevenmsb
\scriptscriptfont\msbfam=\Fivemsb

\font\Teneufm=eufm10 scaled 1200 \font\Seveneufm=eufm7 scaled 1200
\font\Fiveeufm=eufm5 scaled 1200
\textfont\eufmfam=\Teneufm \scriptfont\eufmfam=\Seveneufm
\scriptscriptfont\eufmfam=\Fiveeufm}

\def\Ten{
\textfont\msafam=\tenmsa \scriptfont\msafam=\sevenmsa
\scriptscriptfont\msafam=\fivemsa

\textfont\msbfam=\tenmsb \scriptfont\msbfam=\sevenmsb
\scriptscriptfont\msbfam=\fivemsb

\textfont\eufmfam=\teneufm \scriptfont\eufmfam=\seveneufm
\scriptscriptfont\eufmfam=\fiveeufm}

\renewcommand{\(}{\left( }
\renewcommand{\)}{\right) }
\renewcommand{\t}{\theta}
\newcommand{\co}{C}
\newcommand{\Ai}{\hbox{\rm Ai}}
\newcommand{\Bi}{\hbox{\rm Bi}}
\newcommand{\ai}{ a}
\newcommand{\bi}{\hbox{\rm bi}}
\newcommand{\re}{\Re}

\title {Spectral asymptotics of harmonic oscillator perturbed by
bounded potential}

\author{  Markus Klein
\begin{footnote}
{Institut f\"ur  Mathematik, Universit\"at Potsdam, Am Neuen
Palais 10, 14469 Potsdam, Germany}
\end{footnote}
and Evgeni Korotyaev
\begin{footnote}
{Institut f\"ur  Mathematik,  Humboldt Universit\"at zu Berlin,
Rudower Chaussee 25, 12489, Berlin, Germany}\end{footnote}
  and Alexis Pokrovski \begin{footnote}
{ Institut f\"ur  Mathematik, Universit\"at Potsdam, and
Laboratory of Quantum Networks, Institute for Physics,
St.-Petersburg State University, Ulyanovskaya~1,
198504~St.-Petersburg, Russia.}
\end{footnote}
 }
 \maketitle

\begin{abstract}
\no Consider the operator $ T=-{d^2\/dx^2}+x^2+q(x)$ in $
L^2(\mathbb{R})$, where real functions $q$, $q'$ and
$\int_0^xq(s)\,ds$ are  bounded.
In particular, $q$ is periodic or almost periodic. The spectrum of
$T$ is purely discrete and consists of the simple eigenvalues $\{\mu_n\}_{n=0}^\infty$,
$\mu_n<\mu_{n+1}$.
We determine their asymptotics
$\mu_n=(2n+1)+
(2\pi)^{-1}\int_{-\pi}^{\pi}q(\sqrt{2n+1}\sin\theta)\,d\theta+O(n^{-1/3})$.
\end{abstract}

\section{Introduction and main results}
\setcounter{equation}{0}

Consider the quantum-mechanical harmonic oscillator $T^0=-
{d^2\/dx^2}+x^2$ on $L^2(\mathbb{R})$. It is well known that the spectrum
of $T^0$ is purely discrete and consists of the simple eigenvalues
$\m_n^0\!=\!2n+1$, $n\!\ge\!0$ with   corresponding orthonormed
 eigenfunctions $\p_n^0$.
Define the perturbed operator $ Ty=-y''\!+\!x^2y\!+\!q(x)y$ in
$L^2(\R)$, where  $q$ belongs to the complex Banach space $\cB$
given by
\begin{equation}
\cB=\lt\{ q,q',q_1\in L^\infty(\mathbb{R}): \|q\|_{\cB}= \sup
_{x\in \R}\lt(|q(x)|+|q'(x)|+|q_1(x)|\rt)<\iy \rt \}, \ \ \
q_1(x)\ev \int_0^xq(t)dt.
\end{equation}
In particular, this class includes periodic, almost-periodic
 $q$. If $q\in \cB$ is real,
then $T$ is self-adjoint, its spectrum
 is purely discrete, $\sigma(T)=\{\mu_n\}_{n=0}^\infty$ and
$\m_n=\m_n^0+O(1),  \ n\to \iy$.
Our goal is to determine the asymptotics of $\m_n-\m_n^0$ as
$n\!\to\!\iy$.

 For decaying perturbations
(e.g. $q', xq\in L^2(\R)$) a complete inverse spectral
 theory is obtained in \cite{CKK}, \cite{CKK-CMP}.
At the same time we did not find in the literature any results
concerning periodic and almost-periodic $q$. The existing methods
(e.g. in \cite{CKK}) cannot be used for $q\in \cB$.

Let us number the points of the spectrum so that
$|\mu_n|\le|\mu_{n+1}|$ counting multiplicity.

\begin{theorem} \lb{T1}
For any $q\in  \cB$ the spectrum of operator $ T$ is discrete and
the following asymptotic is fulfilled:
\begin{equation}
\label{MainAsymptotics}
\mu_n=\mu_n^0+\m_n^1+\|q\|_{\cB}O(n^{-{1\/3}}),
\qquad
\m_n^1={1\/2\pi}\int_{-\pi}^{\pi}q(\sqrt{\mu_n^0}\sin\vt)\,d\vt=
\|q\|_{\cB}O(n^{-{1\/4}}).
\end{equation}
\end{theorem}
{\it Remark.} For finite smooth  $q$ the formula (\ref{MainAsymptotics}) gives
$
\mu_n^1=
\frac{1}{\pi\sqrt{\mu_n^0}} \int_{-\infty}^\infty q(s)\,ds,
$
which is the well-known leading term of asymptotics in
this case \cite{CKK}.

\begin{proposition}
\lb{C1}
Let $q\in\cB$ and
$q(x)=\int_{\R}e^{ixt}\,d\nu(t)$ for some Borel
measure $d\nu$ on $\R$ which satisfies the condition
$C_q=\int_{\R}(1+|t|^{-p})\,d\n(t)<\iy$ for some $p>{3\/2}$. Then
\begin{equation}
\lb{asft}
\mu_n^1=
\int_{\R}J_0(t\sqrt \l)\,d\nu(t)=
{\s(\sqrt{\m_n^0})\/(\m_n^0)^{1\/4}}+C_qO(n^{-{3\/4}}),
\end{equation}
where $J_0$ is the Bessel function and
\begin{equation}
 \lb{asft1}
 \s(s)=\sqrt{{2\/\pi }} \int_\R
{\cos(|t|s-{\pi\/4})\/|t|^{\frac{1}{2}}}d\nu(t).
\end{equation}
Moreover, if $q$ has the form
$q(x)=\sum_{k\in \Z}q_ke^{ixt_k}$, 
then
\begin{equation}
\label{ZonEndsAsympt-1}
\s(s)=
\sqrt{\frac{2}{\pi}}
\sum_{k\in \Z}{\frac{q_k}{\sqrt{|t_k|}}}
\cos(s|t_k|+\frac{\pi}{4}).
\end{equation}
\end{proposition}

In Section 2
we introduce the quasiclassical change of variable. In this variable we write
 the integral equation for the fundamental solutions.
In Sections 3 and 4 we prove convergence of the
iteration series for the fundamental solutions in the
 sub-barrier ($x\gtrsim\sqrt{|\l|}$) and over-barrier ($0<x\lesssim\sqrt{|\l|}$)
 regions, respectively. In Section 5 using
 these series we derive  the asymptotics of the Wronskian.
 In Section 6 using this asymptotics   we  prove Theorem~\ref{T1}.
 We prove auxiliary properties of the quasiclassical
 change of variables in the Appendix.

\section{Preliminaries: changes of variables}
\setcounter{equation}{0}

Consider the differential equation
\begin{equation}
 \lb{OurEq}
-y''+(x^2+q(x))y=\l y,\qquad (x,\l)\in\R\ts\C.
\end{equation}
We shall show that there exist fundamental solutions  $\p_\pm$
 which satisfy the asymptotics
\begin{equation}
\label{AnAsDef}
 \p_\pm(x,\l)=(\pm\sqrt{2}x)^{\l-1\/2}e^{-{x^2\/2}}(1+o(1)),\quad
\quad
\p_\pm'(x,\l)=-x(\pm\sqrt{2}x)^{\l-1\/2}e^{-{x^2\/2}}(1+o(1))
\end{equation}
 as $x\to\pm\iy$ and locally uniformly in $\l$. If $q\ev 0$ then these solutions
have the form $\p^0_{\pm}(x,\l)=D_{\l-1\/2}(\pm\sqrt{2}x)$, where
$D_r$ is the Weber (parabolic cylinder) functions (see \cite{AS}).
 We introduce the Wronskian $\{f,g\}=fg'-f'g$.

\begin{theorem}\lb{Tan}
Let $q, q_1\in L^\iy(\R)$. Then

\no i) For any $\l\in\C$ there exist unique  solutions
$\p_\pm(x,\l)$ of (\ref{OurEq}) with the asymptotics (\ref{AnAsDef}).
Moreover, for each $x\in\R$ the functions $\p_\pm(x,\cdot),
\p_\pm'(x,\cdot)$ and $w=\{\p_-,\p_+\}$  are entire.

 \no ii) If $q$ is real, then the operator $T=-{d^2\/dx^2}+x^2+q(x)$
 has only simple eigenvalues.
\end{theorem}
\no {\it Proof.} Consider the function $\p_+$, the proof for
$\psi_-$ is similar. In order to prove that $\psi_+$ is entire
function of $\l$ it is sufficient to show that it is  analytic in
 each disc $\dD(\m)=\{\l\in\C:|\l-\m|\le 1\},\m\in\C$.
For $\l\in \dD(\m)$ we have (see \cite{AS}) the uniform asymptotics
\begin{equation}
\label{AS-psi0} \psi_+^0(x,\l)=g(x)
\left(1+O\left(x^{-2}\right)\right), \quad
{\psi_+^0}'(x,\l)=-xg(x)\left(1+O\left(x^{-2}\right)\right),
\quad x\to+\iy,
\end{equation}
where $g(x)=(\sqrt{2}x)^{{\l-1\/2}}e^{-{x^2\/2}}$. Let $h(x,\l)=
{1\/2\sqrt \pi}\G({1-\l\/2}) (\p^0_-(x,\l)-\sin {\l\/2} \cdot
\p^0_+(x,\l))$; note that
\begin{equation}
\label{AS-phi0}
 h(x,\l)={1\/2xg(x)}(1+O(x^{-1})),
\quad h'(x,\l)={1\/2g(x)}(1+O(x^{-1})),
\quad x\to+\iy,
\end{equation}
(see \cite{AS}) uniformly for $\l\in \dD(\m)$,
 so that $\{\p_+^0,h\}=1$. Define the entire function
$
M(x,y)=h(x,\l)\p_0^+(y,\l)-\p_0^+(x,\l)h(y,\l)
$.
Then a solution of
\begin{equation}
\label{IE-analytic} \p(x,\l)= \p_+^0(x,\l)+ \lim_{t\to\iy}
\int_x^tM(x,y)q(y)\p(y,\l)\,dy
\end{equation}
solves (\ref{OurEq}). We rewrite (\ref{IE-analytic}) in the form
\begin{equation}
\label{IE-analyticPodkruchennoe}
 p(x,\l)=p_0(x,\l)+ \lim_{t\to\iy} \int_x^t
K(x,y)q(y)p(y,\l)\,dy, \qquad x>1,
\end{equation}
where
\begin{equation}
\label{defpp0K}
p={\p(x,\l)\/g(x)}, \ \ \ p_0={\p_+^0(x,\l)\/g(x)}, \ \ \
 \qquad K(x,y)={M(x,y)g(y)\/g(x)}.
\end{equation}
Let $h_0=2xg(x)h(x,\l)$ and
\begin{equation}
\lb{defK01}
K=U-V,\ \ \ \  \ \ \
 U(x,y)={h_0(x)p_0(y)\/2x}{g^2(y)\/g^2(x)},\ \ \ \ \
V(x,y)={p_0(x)h_0(y)\/2y}.
 \end{equation}
In order to study Eq.(\ref{IE-analyticPodkruchennoe}) introduce
the spaces  of functions
$$
\cF_\a=\{ f\in C([1,\iy)): \|f\|_\a\ev \sup_{x\in [1,\iy)} |x|^\a|f(x)|<\iy
\},\qquad \alpha\in\R,
$$
and $\cF_{\alpha,\beta}=\{f\in\cF_\a: f'\in \cF_\beta\}$ with the norm
$\|f\|_{\a,\beta}=\|f\|_\a+\|f'\|_\beta$, $\beta\in\mathbb{R}$.
By (\ref{AS-psi0}), for $\l\in \dD(\m)$ we have the estimate
\begin{equation}
\label{AnEstchi0}
  \|p_0\|_{0,1}\le c<\iy.
\end{equation}
Let $f\in \cF_\a$ for some  $\alpha\in\mathbb{R}$ and
$u=Uqf$. Then we have
\begin{equation}
u(x)=\int_x^\iy{h_0(x)p_0(y)\/2x}{g^2(y)\/g^2(x)}q(y)f(y)dy
={h_0(x)\/2x^\l}\int_x^\iy p_0(y)e^{x^2-y^2}y^{\l-1}q(y)f(y)dy.
\end{equation}
Using  (\ref{AS-psi0}), (\ref{AS-phi0}) and the
estimate $\int_x^\iy e^{-y^2}y^\g dy\le C e^{-x^2}x^{\g-1}$ for
$ x\ge
1$ and $ \g\in \R$ we obtain
\begin{equation}
\label{AnEstKe}
 \|u\|_{\a+2}\le C \|q\|_\iy\|f\|_\a,\qquad
\|u'\|_{\a+1}\le C\|q\|_\iy \|f\|_\a
\end{equation}
uniformly in $\l\in \dD(\m)$.
Here and below $C$ is some absolute constant.

Let $ f\in \cF_{\a,\b}$ for some
$\alpha\in\mathbb{R}$ and $\beta>0$.
Set $v=Vqf$. Then
integration by parts gives
$$
v(x)=\lim_{t\to\iy}\!\!\int_x^t\!\! {p_0(x)h_0(y)\/2y}
q(y)f(y)\,dy=p_0(x)\!\!\int_x^\iy\!\!  [q_1(x)-q_1(y)]
\lt({h_0(y)f(y)\/2y}\rt)' \,dy,
$$
where the last integral converges absolutely. Using
(\ref{AS-psi0}--\ref{AS-phi0}) we
 obtain
\begin{equation}
\label{AnEstKp}
|v(x)|\le CC_q(\|f\|_{\a}x^{-\a-1}+\|f'\|_{\b}x^{-\b}),\ \ \ \
|v'(x)|\le CC_q(\|f\|_{\a}x^{-\a-2}+\|f'\|_{\b}x^{-\b-1})
\end{equation}
for $x\ge 1$, uniformly in $\l\in \dD(\m)$, where
$C_q=\|q\|_\iy+\|q_1\|_\iy$. Thus
\begin{equation}\label{Ns}
K: \cF_{\alpha,\beta}\to\cF_{\alpha',\beta'},\qquad
\quad \a'=\min\{\a+1,\b\}, \quad
\b'=\min\{\a+1,\b+1\}.
\end{equation}
Consider the iterations $p_{n+1}=Kqp_n, n\ge 0$. By
(\ref{AnEstchi0}),  we have $p_0\in\cF_{0,1}$; using
(\ref{defK01}), (\ref{AnEstKe}), (\ref{AnEstKp}) and (\ref{Ns}), we conclude
that
\begin{equation}
\|p_{n+1}\|_{\a_{n+1},\b_{n+1}}\le CC_q
\|p_{n}\|_{\a_n,\b_n},
\end{equation}
where $\alpha_0=0$, $\beta_0=1$,
\begin{equation}
\a_{2n}=\a_0+n, \ \  \b_{2n}=\a_0+1+n, \ \ \a_{2n+1}=\a_0+1+n,\ \
\b_{2n+1}=\a_0+n.
\end{equation}
Using (\ref{AnEstchi0}) we obtain
$$
 |p_{2n}(x)|\le  {(CC_q)^{2n}c x^{-n}},\ \
 |p_{2n}'(x)|\le  {(CC_q)^{2n}c x^{-n-1}},
 $$$$
  \ \ \ \ \ \ \ \ \
  |p_{2n+1}(x)|\le  {(CC_q)^{2n+1}cx^{-n-1}},\ \
 |p_{2n+1}'(x)|\le  {(CC_q)^{2n+1}c x^{-n-1}}.
$$
Hence for $x\ge x_0=(2CC_q)^2$  the series
 $p(x)=\sum_{n\ge 0}p_n(x)$ and $ p'(x)=\sum_{n\ge 0}p_n'(x)$
converge absolutely and uniformly in $\l\in \dD(\m)$; $p(x)$ gives the
solution of Eq.(\ref{IE-analyticPodkruchennoe}). Moreover,
\begin{equation}
\label{AnAsChi}
 p(x)= 1+O(x^{-1}), \qquad p'(x)=O(x^{-1}), \ \ x\to \iy,
\end{equation}
uniformly for $\l\in \dD(\m)$. Therefore  $\p=gp$ is a solution of
(\ref{IE-analytic}). By (\ref{defpp0K}) and (\ref{AnAsChi}), $\p$
satisfies
\er{AnAsDef}.
For each $n\ge 0$ and fixed $x\ge x_0$ the iterations
$p_n(x,\cdot), p_n'(x,\cdot)$ are analytic in $\dD(\m)$. Hence
$p(x,\cdot), p'(x,\cdot)$ are analytic in $\dD(\m)$ for each fixed
$x\ge x_0$.
 By (\ref{defpp0K}), $\p_+(x,\l)$ and
$\p_+'(x,\l)$ are analytic in $\dD(\m)$ for each fixed
$x\ge x_0$. Hence, the solution is also analytic in $\l$
for any fixed $x$ (see this simple fact e.g. in \cite{PT}). Thus
$\p_+(x,\l)$ and $\p_+'(x,\l)$ are entire functions of
$\l$ for any $x\in\R$.

Suppose that $\p_+(x,\l)$ is not unique and denote by $\wt\p_+$
another solution of (\ref{OurEq}) satisfying (\ref{AnAsDef}). Then
$\{\p_+,\wt\p_+\}=0$ and therefore these solutions are linearly
dependent. Since they have the same asymptotics (\ref{AnAsDef}),
$\tilde{\psi}_+=\psi_+$.  Thus $\psi_+$ is unique.

ii)  Let $\l$ be an eigenvalue, which is not simple.
 Then there
exist solutions $\psi_+$ and $\psi$ of (\ref{OurEq}) in
$L^2(\mathbb{R})$. Note that
$\psi$ and $\p'$ are in $L^\infty(\mathbb{R})$. The
asymptotics (\ref{AnAsDef}) yields
$\{\p_+,\p\}=0$. Therefore $\p_+$ and $\p$ are linearly
dependent, which gives contradiction.
 $\BBox$


Theorem \ref{Tan} gives no information on high-energy asymptotics
$\l\to\infty$. To derive those, we introduce some notations and
auxiliary functions.

Throughout the paper we use the following agreements:
\begin{itemize}
    \item {\it The functions $\log z$  and $z^\alpha=e^{\alpha\log z}$
    for $\alpha\in\mathbb{R}$
take their principal values on
$\mathbb{C}\setminus\mathbb{R}_-$.
    \item For $\l\in \ol\C_+\setminus \{0\}$ we set
    $\l=|\l|e^{2i\vt}$,
    $\vt\in[0,{\pi\/2}]$.}

\end{itemize}
 Define the function $\la z\rangle=(1+|z|^2)^{\frac{1}{2}}, z\in \C$.
For any interval $I\ss\R$ we introduced a sector
$S(I)=\{z\in\C:\arg z\in I\}$. Define the function
\begin{equation}
\label{DefXi}
\x(t)=\int_1^t\sqrt{s^2-1}\,ds=
{1\/2}\lt(t\sqrt{t^2\!-1}-\log(t+\sqrt{t^2-1})\lt),\quad  t\in
S(-\frac{\pi}{2},0),
\end{equation}
where $\x(t)>0$ for $t>1$. The function $\xi$ is a conformal
mapping  from $ S(-\frac{\pi}{2},0)$ onto
$\X=\C_-\cup
\{
\Re\x<0, \Im \x \in [0, {\pi\/4})\}$.
The following uniform asymptotics is fulfilled:
\begin{equation}
\label{asympXi}
\x(t)={1\/2}\lt(t^2-{1\/2}-\log 2t +O(|t|^{-2})\rt),\qquad
 |t|\to\infty,\qquad t\in S[-\frac{\pi}{2},0].
\end{equation}
We introduce the function
\begin{equation}
k(t)=\lt({3\/2}\xi(t)\rt)^{\frac{2}{3}},\ \ \ \ \
t\in S(-\frac{\pi}{2},0), \quad
k(0)=-\left(\frac{3\pi}{8}\right)^{\frac{2}{3}}<0.
 \label{ZofT}
\end{equation}
Note that  $k(t)$ is a conformal mapping  from $
S(-\frac{\pi}{2},0)$ onto the domain $\cK$ given by
\begin{equation}
\cK=S[-\frac{2\pi}{3},0) \bigcup \left\{ k\in
S(-\pi,-\frac{2\pi}{3}]: |k|\sin^{{2\/3}}{3\arg
k\/2}<|k(0)|\right\}.
\end{equation}
 By (\ref{asympXi}), the following
asymptotics and estimates are fulfilled:
\begin{equation}
\label{k-from-t}
k(t)=\lt({3\/4}\rt)^{2\/3}t^{4\/3}\lt(1+O(t^{-1})\rt),\ \ t\in S[-\frac{\pi}{2},0],
\qquad
t(k)=\lt({4\/3}\rt)^{1\/2}k^{3\/4}\lt(1+O(k^{-\frac{3}{4}})\rt),\ k\in
\ol\cK,
\end{equation}
\begin{equation}
 |t'(k)| \le C{\la k \ra }^{-{1\/4}},\qquad
 |t''(k)| \le C \la k \ra ^{-{5\/4}},\ \ \ k\in \ol\cK,
\label{k-from-t-est}
\end{equation}
where $t(k)$ is the  inverse  function for $k(t)$. Here and below
$C$ is an absolute constant.

\begin{figure}[h!]
\includegraphics[width=10cm,height=7.5cm ]{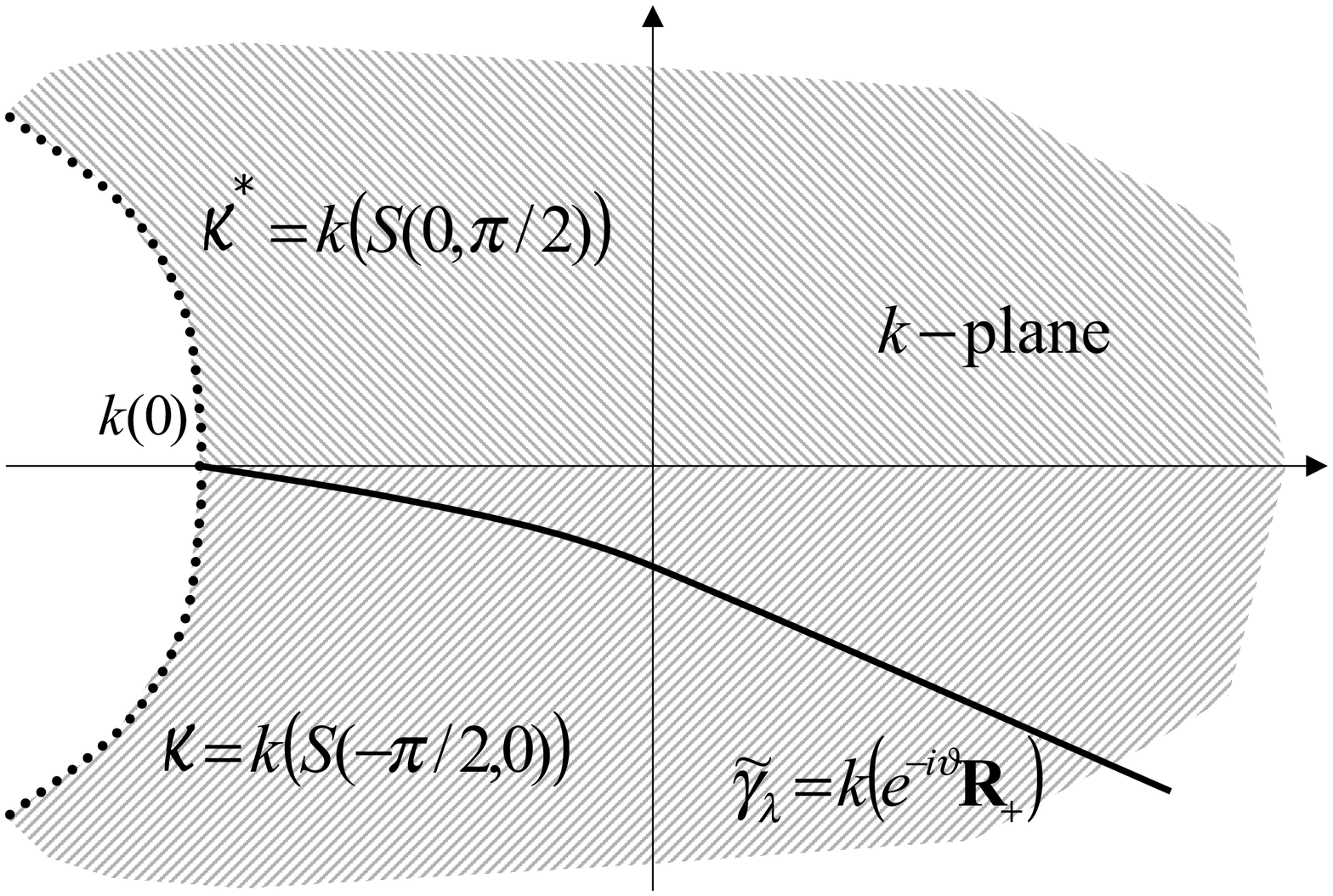}\\
\caption[short caption here]{ The domain $\cK=k(S(-\frac{\pi}{2},0))$
and the curve $\tilde{\gamma}_\l=k(e^{i\vt}\mathbb{R}_+)$.   }
\label{Fig_zeta_plane}  
\end{figure}

\begin{figure}[h!]
\centering
\includegraphics[width=12cm,height=8cm]{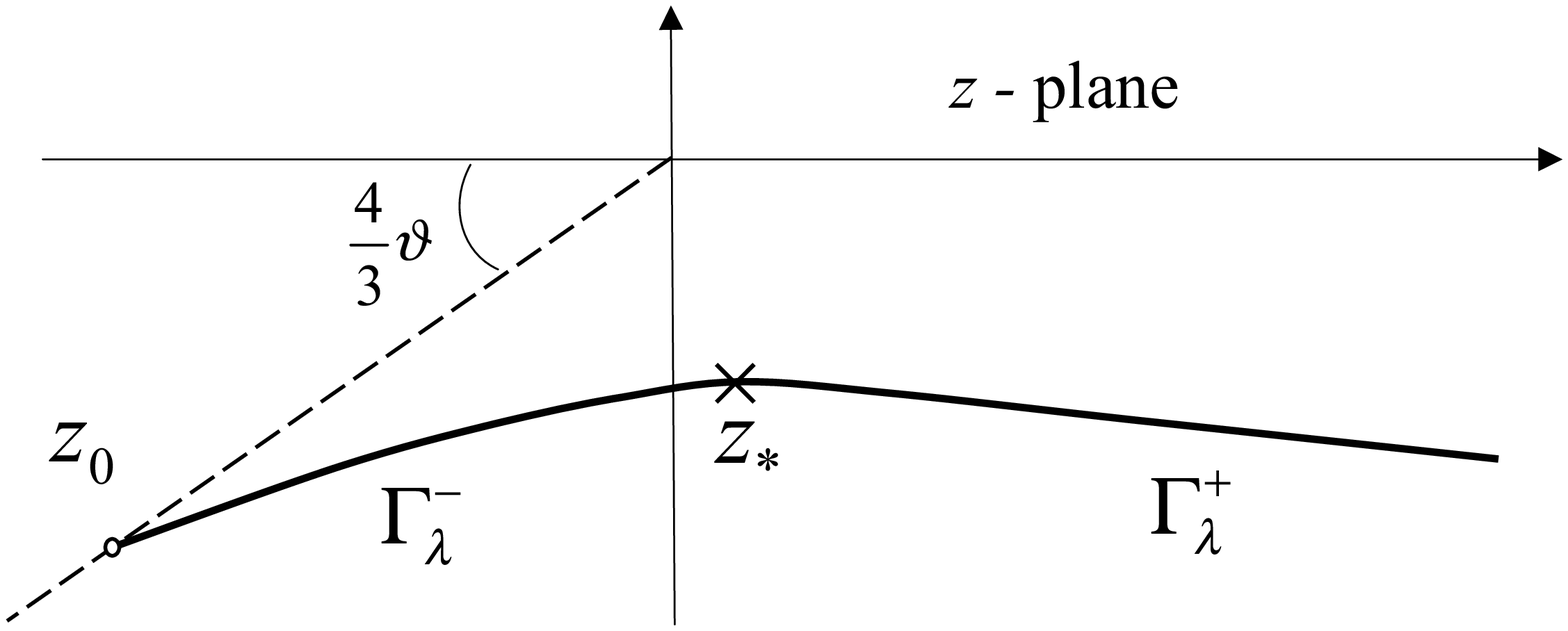}\\
\caption[short caption here]{ The point $z_*$ divides the curve
$\Gamma_\l=\Gamma_\l^-\bigcup\Gamma_\l^+$
(for $\l=|\l|e^{2i\vt}\in\ol\C_+\sm\{0\}$).}
\label{Fig_z_plane_image}  
\end{figure}

Consider the change of variable $x\to k= k(\frac{x}{\sqrt{\l}})$;
it maps $\mathbb{R}_+$ onto the curve
$\tilde{\gamma}_\l=k(e^{i\vt}\mathbb{R}_+)$. The domain $\cK$ and
the curve $\tilde{\gamma}_\l$ are presented on
Fig.~\ref{Fig_zeta_plane}.
By (\ref{OurEq}), the function
$
y_1(k,\l)=\frac{y(\sqrt{\l}t(k))}{\sqrt{t'(k)}}$ solves
\begin{equation}
\label{OurEqInK}
 y_1''(k,\l)-\l^2ky_1(k,\l)=v_0(k)y_1(k,\l)
  +v_q(k,\l)y_1(k,\l),\qquad k\in\tilde{\g}_\l,
\end{equation}
  where
 \begin{equation}
  v_0(k)=t'(k)^{3\/2}
  \lt({d^2\/dt^2}{1\/\sqrt{k'(t)}}\rt)\rt|_{t=t(k)},
  \qquad
  v_q(k,\l)=\l \ t'(k)^2q(\l^{1\/2} t(k)).
\label{eqZeta}
\end{equation}
Using (\ref{k-from-t}) and (\ref{k-from-t-est}) we obtain
\begin{equation}
\label{EstimateV0} |v_0(k)| \le  {C(1+|k|)^{-2}},\ \ \ \ \ k\in
\cK.
\end{equation}
For each $\l\in \ol\C_+\sm\{0\}$ we define the {\bf basic
variable} $ z=\l^{\frac{2}{3}}k $.
 We have the function
\begin{equation}
\label{DefZ} z(x,\l)=\l^{{2\/3}}k(x{\l}^{-{1\/2}}),\qquad \l\in
\ol\C_+\setminus\{0\}, \qquad x\ge0.
\end{equation}
Each mapping $z(\cdot,\l):\mathbb{R}_+\to
\Gamma_\l=z(\mathbb{R}_+,\l)$ is a real analytic isomorphism (see
Fig.~\ref{Fig_z_plane_image}) and see Lemma~\ref{SvaGForAll} about
$\G_\l$. If $\l>0$, then
$\G_\l=[-\l^{\frac{2}{3}}({3\pi\/8})^{2\/3},\iy)$ is a half-line.
Moreover,
\begin{equation}
\label{z-0} z_0\equiv z(0,\l)=
-\l^{{2\/3}}\left(3\pi/8\right)^{\frac{2}{3}}, \qquad \l\in
\ol\C_+\setminus\{0\}.
\end{equation}

For any $\l\in\ol\C_+\sm\{0\}$ and $0\le x_1<x_2$ set $ z_n=z(x_n,\l)$, $n=1,2$.
We define the  curves
$$
\G_\l(z_1,z_2)=\{z: z=z(x,\l), x\in[x_1,x_2]\},\ \ \ \
\G_\l(z_1)\equiv \G_\l(z_1,\iy)=\{z: z=z(x,\l), x\ge x_1\}.
$$
By (\ref{OurEqInK}), the function
$u(z,\l)=y_1({z}{\l^{-\frac{2}{3}}},\l)$ solves
\begin{equation}
\pa_z^2 u(z,\l)-zu(z,\l)=V(z,\l)u(z,\l),\ \ \ V=V_0+V_q, \qquad
\qquad z\in\G_\l,
  \label{EqinZ}
\end{equation}
here and below $\pa_z=\frac{\pa }{\pa z}$ and $V_q$ (linear in
$q$) and $V_0$ (not include $q$) are given by
  \begin{equation}
  \label{EffPotentials}
  V_0(z,\l)={v_0\left({z}{\l^{-\frac{2}{3}}}\right)\/\l^{{4\/3}}},
  \quad  V_q(z,\l)={\r^2(z,\l)q\left(\sqrt{\l}t({z}{\l^{-\frac{2}{3}}})\right)\/\l^{{1\/3}}},
\quad
 \r(z,\l)\ev t'\left({z}{\l^{-\frac{2}{3}}}\right).
\end{equation}
Using (\ref{k-from-t-est}), we obtain for $ \l\in \ol\C_+\sm\{0\}$ and
$z\in S[-\pi,{\pi\/3}]$ the  estimates
\begin{equation}
 \label{EstNu}
  |\r (z,\l)|\le
C\langle
z \l^{-\frac{2}{3}}
\rangle^{-\frac{1}{4}},\quad
|\pa_z\r(z,\l)|\le {|\l|^{-\frac{2}{3}}}
\langle
z \l^{-\frac{2}{3}}
\rangle^{-\frac{5}{4}},
\quad |V_0(z,\l)|\le {C\/|\l|^{4\/3}+|z|^2},
\end{equation}
where $C$ does not depend on $\l$ and $z$.

To analyze (\ref{EqinZ}) we need  well-known properties \cite{AS}
of the Airy functions Ai and Bi:
\begin{equation}
\label{AsAiry} \Ai(z)=\frac{e^{-{2\/3}z^{\frac{3}{2}}}}{2\sqrt[4]{z}}
  \(1+O(z^{-\frac{3}{2}})\),\qquad
 |z|\to\iy,\qquad
 |\arg z|<\pi-\ve,
 \quad \forall\ve>0,
\end{equation}
 \begin{equation}
\lb{BfromA}
\{\Ai(z),\Bi(z)\}=1, \quad
 \Bi(z)=i\lt(2e^{-i\frac{\pi}{3}}\Ai(\o z)-\Ai(z)\rt),\ \ \o=e^{{2\pi
 i\/3}},
 \quad
\end{equation}
\begin{equation}
\lb{ab1}
\Ai(z)=e^{-i\frac{\pi}{3}}\Ai(z\omega)+
e^{i\frac{\pi}{3}}\Ai(z\ol\omega),\quad
\Bi(z)=ie^{-i\frac{\pi}{3}}\Ai(z\omega)-
ie^{i\frac{\pi}{3}}\Ai(z\ol\omega).
\end{equation}
Let $\Gamma\subset\mathbb{C}$ be a smooth curve. For any continuous function
$f$  on $\Gamma$
we denote by $\int_\Gamma f(s)\,ds$  the usual complex line integral.
 We denote by $\int_\Gamma f(s) \,|ds|$  the line integral of $f$
along $\Gamma$ with respect to the arc length
$|ds|=\sqrt{(dx)^2+(dy)^2}$. For integration along the infinite curve
$\Gamma_\l$, defined above, we use the standard notation
 p.v.$\int_{\G_\l(z)} f(s)\,ds=\lim \int_{\G_\l(z,w)} f(s)\,ds$
 as ${w\to\iy, w\in \G_\l^+}$ whenever it exist.

We will study the formal integral equation
 \begin{equation}
\label{IntegralEqinZ}
  u_+(z,\l)=u_0(z)+\text{p.v.}
\int_{\G_\l(z)} J_0(z,s)V(s,\l) u_+(s,\l)\,ds,\quad z\in\G_\l,
\end{equation}
 \begin{equation}
 \lb{defJ}
 u_0(z)=\Ai(z),\ \ \ \ J_0(z,s)=\Ai(s)\Bi(z)-\Ai(z)\Bi(s),\ \ \ z,s\in \C.
 \end{equation}
 We rewrite (\ref{IntegralEqinZ}) in the form
 \begin{equation}
 \lb{IEpodkrychennoe}
  v_+(z)=\ai(z)+
  \text{p.v.}\int_{\G_\l(z)}
   J(z,s)V(s,\l)v_+(s)\,ds,\quad  \ai(z)\ev \Ai(z)e^{\frac{2}{3}z^{\frac{3}{2}}},
   \quad z\in\G_\l,
 \end{equation}
\begin{equation}
\label{PodkrutkaU}
u_+(z)=v_+(z)e^{-\frac{2}{3}z^{\frac{3}{2}}},\qquad
  J(z,s)=J_0(z,s)
e^{{2\/3}(z^{\frac{3}{2}}-s^{\frac{3}{2}})}.
\end{equation}
If $z<0$ and $\l\in\ol\C_+$, then  $z^{\frac{3}{2}}$ takes its values on the
lower side of the cut. This agreement provides continuity as $\arg\l\downarrow0$,
 since for
$\l\in\C_+$   the curve $\Gamma_\l$ lies in the lower
half-plane. By (\ref{AsAiry}) and (\ref{ab1}), the following
estimates are fulfilled:
 \begin{equation}\label{aiest}
|\ai(z)|\le {\co}{\langle z\rangle^{-\frac{1}{4}}}, \qquad \forall z\in \C,\qquad
 \end{equation}
\begin{equation}
|\ai'(z)|\le {\co}{\langle z\rangle^{-\frac{5}{4}}}, \quad |\arg z|\le \pi-\ve,\quad
\forall\ve>0. \label{aiprimest}
\end{equation}
We write (\ref{IEpodkrychennoe}) in the form
\begin{equation}
\label{IEpodkrJ}
     v_+=\ai+\mathbf{J} V v_+,
\end{equation}
where the integral operator $\bf J$ is given by
\begin{equation}\label{J-operator}
 [\mathbf{J}f](z)=\text{p.v.}\int_{\Gamma_\l(z)}
   J(z,s) f(s) ds.
\end{equation}

The next  lemma (proved in the Appendix) gives a splitting of
$\G_\l$. Here and below we fix
\begin{equation}\label{DefDelta}
\delta\in(0,\frac{4}{3}\arccos{2^{-\frac{1}{3}}}).
\end{equation}
\begin{lemma}
\label{Z-star}
For any $\l\in\ol\C_+\sm\{0\}$ there exists a  unique point
$z_*\equiv z_*(\l)\in\Gamma_\l$ such that
\begin{enumerate}
    \item \label{LZ*-1}
    if $0\le\arg\l <\delta$, then
$|z_*|=\min\limits_{z\in\Gamma_\l}\
|z|$,\it
    \item \label{LZ*-2}
     if $\delta\le\arg\l \le\pi$, then
     $z_*=\Gamma_\l\cap\{z:\arg z=-\frac{\pi}{3}\}$.
\end{enumerate}
\end{lemma}

For any $\l\in\ol\C_+\sm\{0\}$ here and below we use $z_*$,
defined by Lemma~\ref{Z-star}. We define
 the point $x_*$ by $z_*=z(x_*,\l)$ and let
  $t_*=\frac{x_*}{\sqrt{\l}}$ and set
\begin{equation}
\label{DefGPM}
\G_\l^-=\G_\l(z(0,\l),z_*),\qquad \G_\l^+=\G_\l(z_*,\iy),\qquad
\G_\l=\G_\l^-\cup\Gamma_\l^+.
\end{equation}
\begin{lemma}
\lb{27}
Let $\l\in\ol\C_+\sm\{0\}$.
 Let $h(x,\l)=|\exp(\frac{2}{3}z(x,\l)^{\frac{3}{2}})|$ for $x\ge0$. Then
\begin{enumerate}

\item \label{2.3-1} if $\l>0$, then $h(\cdot,\l)$ is strictly increasing on
$[x_*,\infty)$ and $h(\cdot,\l)\equiv 1$ on $[0,x_*]$,

\item \label{2.3-2}  if $0<\arg\l\le\pi$, then
$h(\cdot,\l)$ is strictly increasing on $[0,\infty)$.

\end{enumerate}
\end{lemma}

\begin{lemma}
\label{EstExpInt}
Let $\l\in\ol\C_+\sm\{0\}$ and $|\l|\ge 1$. Assume a) $z\in\G_\l$,
$ \d\le\arg\l\le\pi$ or b) $z\in\G_\l^+$. Then  the following
estimates are fulfilled:
\begin{equation}
\label{Exp-2} \int_{\G_\l(z)}
{|e^{-\frac{4}{3}{s}^{\frac{3}{2}}}|}{\langle s\rangle^{-\a}}\,|ds| \le C
{|e^{-\frac{4}{3}z^{\frac{3}{2}}}|}{\langle z\rangle^{-\a-\frac{1}{2}}},
\qquad
\alpha\in\mathbb{R},
\end{equation}
\begin{equation}
\label{pow-2}
\int_{\Gamma_\l(z)}
{\langle s\rangle^{-\a}}\,{|ds|} \le C {\langle z\rangle^{-\a+1}},
\qquad
\a\ge1,
\end{equation}
where  $C$ is independent of $\l$ and $z$.
\end{lemma}

\begin{lemma}
\label{PowLem} Let $\l\in\ol\C_+$ and $|\l|\ge 1$.  Then
 the following
estimates are fulfilled:
\begin{equation}
\label{pow-3Le1}
\int_{\Gamma_\l^-}
{\langle s\rangle^{-\a}}\,{|ds|} \le
\left\{
\begin{array}{ll}
  {C}(1-\alpha)^{-1}|\l|^{\frac{2}{3}(1-\alpha)}\quad &
\hbox{\rm for}\quad 0\le\a<1,\\[2ex]
  \co \log (|\l|+1) \quad &
  \hbox{\rm for}\quad \alpha=1,\\[1ex]
{C}{(\alpha-1)^{-1}} \quad &
 \hbox{\rm for}\quad \alpha>1,\\
\end{array}
\right.
\end{equation}
\begin{equation}\label{EIV0}
\int_{\Gamma_\l} \frac{|ds|}{|\l|^{\frac{4}{3}}+|s|^2}
\le\frac{C}{|\l|^{\frac{2}{3}}},
\end{equation}
where  $C$ is independent of $\l$ and $z$.
\end{lemma}

\section{Analysis of the integral equation}
\setcounter{equation}{0}

In this section we consider the integral equation $v_+=\ai+\mathbf{J} V v_+$
for large $|\l|$ in the following cases:
i) $ 0\le\arg\l\le\pi$, $z\in\G_\l^+$
and ii) $ \d\le\arg \l\le\pi$, $z\in\G_\l^-$.
Moreover, we give the complete analysis for these cases.
The case $ 0\le\arg\l\le\pi$, $z\in\G_\l^-$ is treated in the next section.

For $\l\in \ol\C_+\sm\{0\}$ and $\a,\beta>0$  define the Banach
spaces of functions on $\Gamma_\l$:
\begin{equation}
\label{Def-f-G+}
   \cF_\a^\l=
   \lt\{
f\in C(\G_\l^+): \|f\|_\a\equiv
\sup_{z\in\Gamma_\l^+}\langle z\rangle^\a|f(z)|<\infty
   \rt\}\qquad \hbox{for}\quad |\arg\l|<\delta,
\end{equation}
\begin{equation}
\label{Def-f-allG}
   \cF_\a^\l=
   \left\{
f\in C(\G_\l): \|f\|_\a\equiv
\sup_{z\in\Gamma_\l}\langle z\rangle^\a|f(z)|<\infty
   \right\}\qquad \hbox{for}\quad\delta\le|\arg\l|\le \pi,
\end{equation}
 \begin{equation}
 \lb{DefF}
 \cF_{\a,\b}^\l=\lt\{f\in F_\a^\l: f'\in F_\beta^\l
 \lt\},
 \qquad
 \|f\|_{\a,\beta}=\|f\|_\a+\|f'\|_\beta.
 \end{equation}
 Evidently $\cF_\a^\l\subset \cF_{\a'}^\l$ and
 $\cF_{\a,\b}^\l\subset \cF_{\a',\b'}^\l$ for $\a<\a'$ and
 $\beta<\beta'$.
Now we formulate the main result of this section; its proof is
given in the end of the section.

\begin{theorem}
\label{ConvG+THM}
Let $q,q_1\in L^\iy(\R)$, and  $\l\in
\ol\C_+$.
  Then the equation $v_+=v_0+\mathbf{J} V v_+$, $v_0(z)\equiv\ai(z)$,
has a unique solution $v_+\in\cF_{\frac{1}{4},1}^\l$ for
$
|\l|^{\frac{1}{6}}\ge 2c_0
\left(\|q\|_\infty+\|q_1\|_\infty+1
\right)
$,
where
$c_0>1$ is an absolute constant. Moreover, the solution satisfies
\begin{equation}
\label{G+EstV0}
|v_+(z)|\le
 {\co}{\langle z\rangle^{-\frac{1}{4}}}, \qquad \quad
  |v_+'(z)|\le
 {\co}{\langle z\rangle^{-1}},
\end{equation}
\begin{equation}
\label{G+EstV1}
 |v_+(z)-v_0(z)|\le
\ve
  {\co}{\langle z\rangle^{-\frac{3}{4}}}, \qquad \quad
  |v_+'(z)-v_0'(z)|\le
\ve
 {\co}{\langle z\rangle^{-1}},
\end{equation}
where $ \ve=c_0|\l|^{-\frac{1}{6}}
\left(\|q\|_\infty+\|q_1\|_\infty+1 \right)$.
\end{theorem}
We have the identity
\begin{equation}
\Ai(z\omega)=
e^{\frac{2}{3}z^{\frac{3}{2}}}\ai(z\omega),\qquad
\omega=e^{\frac{2\pi i}{3}},
\qquad
-\pi\le \arg z<{\pi\/3}.
\lb{Aitoai+}
\end{equation}
 Using
(\ref{BfromA}),(\ref{ab1})  and (\ref{Aitoai+}) we write the kernel
 $J(z,s)$, given by (\ref{PodkrutkaU}), in terms of
$a(z)$ and $a(\omega z)$.
As a result we obtain
\begin{equation}
\lb{Jform}
  J(z,s)= -2 ie^{-i\frac{\pi}{3}}\left(
\ai(z)\ai(s\omega)-
e^{\frac{4}{3}(z^{\frac{3}{2}}-s^{\frac{3}{2}})} \ai(z\omega)\ai(s)\right),
\qquad
z,s\in S(-\pi,\frac{\pi}{3}).
\end{equation}
Note that by   \refLem{A7.5-1}{SvaGForAll},
 in both cases i) and ii) we have
$\Gamma_\l(z)\subset S[-\pi+\frac{2}{3}\delta,0]$, so (\ref{Jform}) holds
on $\Gamma_\l(z)$.

Following  (\ref{Jform}),
 we represent $\mathbf{J}V_q$ as the sum of two operators.
 In the  next two Lemmas (\ref{AsJ1} and \ref{AsJE}) we estimate
 these two operators in suitable functional spaces.
 In Lemma~\ref{IsoV0} we estimate
$\mathbf{J}V_0$ (which is asymptotically small in comparison with
$\mathbf{J}V_q$). These
estimates, combined in Lemma~\ref{TL-GL},
give an {\it a priori} estimate for
 $\mathbf{J}V$. In Theorem~\ref{ConvG+THM} we prove
convergence of the iterations series for the equation
$v_+=v_0+\mathbf{J}V v_+$. This gives
 the estimates for
$v_+$ necessary for further  analysis.

For $v_0(z)\equiv\ai(z)$ given by (\ref{IEpodkrychennoe}), due to
(\ref{aiest}) and \refLem{A7.5-1}{SvaGForAll} we have
\begin{equation}
\label{Space-ai-G+}
\|v_0\|_{\frac{1}{4},\frac{5}{4}}\le C.
\end{equation}
uniformly in $\l\in\ol\C\setminus\{0\}$. For any fixed $\l\in
\ol\C_+\sm\{0\}$ and $z_1, z_2\in\Gamma_\l$ such that
$z_j=z(x_j,\l), j=1,2$ and $0\le x_1\le x_2 $ we
 define the function
\begin{equation}
\lb{DQ}
Q(z_2,z_1)\ev {\l^{-\frac{1}{6}}}\int_{z_1}^{z_2}\hat{q}(z,\l)\r
(z,\l)\,dz ,
\end{equation}
where $\rho$ is given by (\ref{EffPotentials}).
 We have the identity
$
 Q(z_2,z_1)=
\int_{x_2}^{x_1} q(x)\, dx$.
Since $q_1(x)=\int_0^xq(t)dt\in L^\iy(\R)$, we have
\begin{equation}
\label{Estq-1}
|Q(z_2,z_1)|\le 2\|q_1\|_\iy\ \ \
{\rm  for\ any}
\ \
z_1,z_2\in
\G_\l.
\end{equation}
 By \refLem{A7.5-2}{SvaGForAll} and \refLem{A7.5-3}{SvaGForAll},  we have
\begin{equation}\label{|z|OnG-}
C_1|\l|^{\frac{2}{3}}\le |z|\le  C_2|\l|^{\frac{2}{3}}\quad
\hbox{for}\quad
z\in\G_\l^-, \quad \delta\le\arg\l\le\pi,
\end{equation}
where $C_1$ and $C_2$ are independent of $z$ and $\l$.

Next we estimate the term of $\mathbf{J}V_q$, corresponding
to the first term in decomposition (\ref{Jform}).

\begin{lemma}
\lb{AsJ1}
 Let  $q, q_1\in L^\iy(\R)$. Assume $\l\in\ol\C_+\setminus\{0\}$, $|\l|\ge1$  and
$f\in\cF_{\a,\beta}^\l$ for $\quad\a>0, \b>{3\/4}$
(that is,
  $f$ is defined on $\Gamma_\l$ for $ \d\le\arg \l\le\pi$ and on
  $\Gamma_\l^+$ for $ 0\le\arg\l\le\delta$).
Then $
g(z,\l)=\text{\rm  p.v.}\int_{\Gamma_\l(z)}
\ai(s\omega)V_q(s)f(s)\, ds\in\cF_{\a,\beta}^\l
$
and satisfies
 \begin{equation}
  |\ai(z)g(z,\l)|\le  C |\l|^{-\frac{1}{6}}{\|q_1\|_\iy}
  \lt\{
{\|f\|_\a}{\langle z\rangle^{-\a-\frac{1}{2}}}+ {\|f'\|_\beta }{\langle z\rangle^{-\b+\frac{1}{2}}}
\rt\}, \label{AB-GL}
 \end{equation}
 \begin{equation}
 \label{ABPrime-GL}
\lt|\ai'(z) g(z,\l) \rt|\le C|\l|^{-\frac{1}{6}}
{\|q_1\|_\iy} \lt\{ {\|f\|_\a}{\langle
z\rangle^{-\a-\frac{3}{2}}}+ {\|f'\|_\beta }{\langle
z\rangle^{-\b-\frac{1}{2}}}
\rt\}.
 \end{equation}
\end{lemma}

\no {\it Proof.}
Consider the case $0\le\arg\l\le\pi$, $z\in\Gamma_\l^+$.
By \refLem{A7.5-1}{SvaGForAll},
  we have
$\Gamma_\l(z)\subset S[-\pi+\frac{2}{3}\delta,0]$,
 so the uniform estimates
(\ref{aiest}) and (\ref{aiprimest}) hold on $\Gamma_\l(z)$
for both $a(z)$ and $a(z\omega)$.
 Writing
 $F(z)=\l^{-\frac{1}{6}}\ai(z\omega)\r(z,\l)$
 ( $\rho$ is given by (\ref{EffPotentials})), integration by parts yields
\begin{equation}
\label{ByParts-1} g(z,\l)=\text{p.v.}\int_{\Gamma_\l(z)}
\left(\pa_s Q(s,z)\right)F(s)f(s)ds =
-\text{p.v.}\int_{\Gamma_\l(z)}Q(s,z)(F(s)f(s))'ds,
\end{equation}
where we used $Q(z,z)=0$ and
$
\lim_{\Gamma_\l^+\ni
w\to\infty} Q(w,z)F(w)f(w)
=0$ (this holds by (\ref{EstNu}),
 (\ref{aiest}) and (\ref{Estq-1})). Thus using
(\ref{EstNu}), (\ref{aiest}),  (\ref{aiprimest})
and (\ref{Estq-1}) we have
$$
|g(z,\l)|\le C{\|q_1\|_\iy\/|\l|^{1\/6}} \lt\{
\int_{\Gamma_\l(z)}
{\|f\|_\a }{\langle s\rangle^{-\a-{5\/4}}}|ds|
$$
$$
+\int_{\Gamma_\l(z)}{\|f\|_\a }
{|\l|^{-\frac{2}{3}}}{\langle\frac{s}{\l^{\frac{2}{3}}}\rangle^{-\frac{5}{4}}}
{\langle s\rangle^{-\a-{1\/4}}}|ds|+
\int_{\Gamma_\l(z)}
\|f'\|_\beta{\langle s\rangle^{-\b-{1\/4}}}{ |ds| }\rt\}.
$$
 Due to  \refLem{A7.5-4}{SvaGForAll} we have
 $|z|=\inf\limits_{s\in\Gamma_\l(z)}|s|$.
Using also  (\ref{pow-2})
 we obtain
$$
|g(z,\l)|\le C\frac{\|q_1\|_\iy}{|\l|^{\frac{1}{6}}}
\lt\{
{\|f\|_\a}{\langle z\rangle^{-\a}} \int_{\Gamma_\l(z)}
{\langle s\rangle^{-{5\/4}}}{|ds|}
+ {\|f\|_\a}{\langle z\rangle^{-\a-{1\/4}}} \int_{\Gamma_\l(z)}
{|\l|^{-{2\/3}}}
{\langle\frac{s}{\l^{\frac{2}{3}}}\rangle^{-\frac{5}{4}}}|ds|
$$
\begin{equation}
\label{G+EstPow}
+ \int_{\Gamma_\l(z)} {\|f'\|_\beta}{\langle s\rangle^{-\b-{1\/4}}} |ds|
\rt\} \le
{C|\l|^{-\frac{1}{6}}\|q_1\|_\iy} \left\{ {\|f\|_\a}{\langle
z\rangle^{-\a-\frac{1}{4}}}+ {\|f'\|_\beta }{\langle
z\rangle^{-\b+\frac{3}{4}}} \right\},
\end{equation}
which together with (\ref{aiest}) and (\ref{aiprimest}) proves
(\ref{AB-GL}) and (\ref{ABPrime-GL}), respectively.

Consider the case $0\le\arg\l\le\delta$, $z\in\Gamma_\l^-$.
By \refLem{A7.5-1}{SvaGForAll},
  we have
$\Gamma_\l(z)\subset S[-\pi+\frac{2}{3}\delta,0]$,
 so the uniform estimates
(\ref{aiest}) and (\ref{aiprimest}) hold on $\Gamma_\l(z)$
for both $a(z)$ and $a(z\omega)$.
We have
$g(z,\l)=g_-(z,\l)+g_+(\l)$,  where
$$
g_+(\l)=\text{p.v.}\int_{\Gamma_\l(z)}\ai(s\omega)V_q(s)f(s)\, ds,\quad
g_-(z,\l)=\int_{\Gamma_\l(z,z_*)}
\ai(s\omega)V_q(s)f(s)\, ds.\quad
$$
Using (\ref{|z|OnG-}) and (\ref{G+EstPow}) for $z=z_*$ we have
\begin{equation}
\label{Eg+Gl-} |g_+(\l)|\le C\frac{\|q_1\|_\infty}{|\l|^{\frac{1}{6}}}
\left\{ \frac{\|f\|_\a}{|z_*|^{\a +\frac{1}{4}}}+ \frac{\|f'\|_\beta
}{|z_*|^{\b-\frac{3}{4}}} \right\} \le
C\frac{\|q_1\|_\infty}{|\l|^{\frac{1}{6}}} \left\{
\frac{\|f\|_\a}{|z|^{\a +\frac{1}{4}}}+ \frac{\|f'\|_\beta
}{|z|^{\b-\frac{3}{4}}} \right\}.
\end{equation}
In order to estimate $g_-$ we integrate by parts
\begin{equation}
\label{ByParts-2} g_-(z,\l)= Q(z_*,z)F(z)f(z)
-\!\!\int_{\G_\l(z,z_*)}\!\!\!\!\!\!\!\!\!\!\!\!
Q(z_*s)\left(F(s)f(s)\right)'ds,\ \ \
F(z)=\l^{-\frac{1}{6}}\ai(z\omega)\r(z,\l),
\end{equation}
since $Q(z_*,z_*)=0$ and $\hat{q}(s)\r(s)=-\pa_s Q(z_*,s)$. Using
(\ref{EstNu}), (\ref{aiest}), (\ref{aiprimest}) and (\ref{Estq-1})
we obtain
$$
|g_-(z,\l)|\le \co\frac{\|q_1\|_\iy}{|\l|^{\frac{1}{6}}} \left\{
\frac{\|f\|_\a}{|z|^{\a +\frac{1}{4}}}+ \int_{\Gamma_\l^-} \left(
 \frac{\|f\|_\a}{|s|^{\a+\frac{5}{4} }} +\frac{\|f\|_\a}{|s|^{\a
+\frac{1}{4}}|\l|^{\frac{2}{3}}} \right)\,|ds| +
\int_{\Gamma_\l^-}
\frac{\|f'\|_\beta }{|s|^{\b+\frac{1}{4}}}\,|ds| \right\}.
$$
By (\ref{pow-3Le1}) and (\ref{|z|OnG-}),
we have
\begin{equation}
\label{Eg-Gl-} |g_-(z,\l)|\le C{\|q_1\|_\infty}{|\l|^{-\frac{1}{6}}}
\left\{ {\|f\|_\a}{|z|^{-\a -\frac{1}{4}}}+ {\|f'\|_\beta
}{|z|^{-\b+\frac{3}{4}}} \right\}.
\end{equation}
Combining (\ref{Eg+Gl-}) and (\ref{Eg-Gl-}) with (\ref{aiest}) gives
 (\ref{AB-GL}). The estimate (\ref{ABPrime-GL}) follows from
(\ref{Eg+Gl-}), (\ref{Eg-Gl-}) and (\ref{aiprimest}).
$\BBox$

For the analysis of the part of
$\mathbf{J}V_q$ corresponding to the second term in (\ref{Jform})
we also  integrate by parts. Let us introduce an analogue of
$Q(z_1,z_2)$:
\begin{equation}
P(z)= \frac{1}{\l^{\frac{1}{6}}} \int_{\Gamma_\l(z,\infty)}
e^{-\frac{4}{3}{s}^{\frac{3}{2}}}
\hat{q}(s,\l)\rho(s)ds =
\int_{x}^{\infty}
e^{-2\l \xi(s/\sqrt{\l})} q(s)\, ds,\quad z\in\Gamma_\l^+,
\end{equation}
where $x=\sqrt{\l}\ t\left({z\/\l^{\frac{2}{3}}}\right)$.
Using (\ref{EstNu}) and (\ref{Exp-2})
for $q\in L^\iy(\R)$ gives
\begin{equation}
|P(z)|\le  \|q\|_\iy \co
\int_{\Gamma_\l(z)}
\frac{
|e^{-\frac{4}{3}{s}^{\frac{3}{2}}} |}{(|\l|^{\frac{1}{6}}+|s|^{\frac{1}{4}})} \, |ds|
\le  \co\|q\|_\iy
{|e^{-\frac{4}{3}{z}^{\frac{3}{2}}}|}{\langle z\rangle^{-\frac{3}{4}}},
\qquad z\in\Gamma_\l^+.
\label{EstP-2}
\end{equation}

Using $|\rho(z)|\le\co$  and
 (\ref{Exp-2}),  we
obtain another estimate
\begin{equation} |P(z)|\le  \|q\|_\iy {C
\/|\l|^{\frac{1}{6}}}
 \int_{\Gamma_\l(z)}
|e^{-\frac{4}{3}{s}^{\frac{3}{2}}} |\, |ds| \le
C\frac{\|q\|_\iy}{|\l|^{\frac{1}{6}}}
{|e^{-\frac{4}{3}{z}^{\frac{3}{2}}}|}{\langle z\rangle^{-\frac{1}{2}}},
\qquad z\in\Gamma_\l^+.
\label{EstP-1}
\end{equation}

\no We estimate the part of $\mathbf{J}V_q$, corresponding to the
second term in the decomposition (\ref{Jform}).


\begin{lemma}
\label{AsJE}
 Let $ q\in L^\iy(\R)$. Assume $\l\in \ol\C_+$, $|\l|\ge1$,  and
 $f\in\cF_{\a,\beta}^\l$ for $\quad\a>0, \b>0$
 (that is,
  $f$ is defined on $\Gamma_\l$ for $ \d\le\arg \l\le\pi$ and on
  $\Gamma_\l^+$ for $ 0\le\arg\l\le\delta$).
Then
$
g(z,\l)= \int_{\Gamma_\l(z)} \ai(s)
e^{-\frac{4}{3}{s}^{\frac{3}{2}}}V_q(s)f(s)\,ds\in\cF_{\a,\beta}^\l
$
and satisfies
 \begin{equation}
  |e^{\frac{4}{3}{z}^{\frac{3}{2}}}\ai(z)g(z,\l)|\le  C
{\|q\|_\iy}{|\l|^{-\frac{1}{6}}} \left\{
{\|f\|_\a}{\langle z\rangle^{-\a-\frac{5}{4}}}+ {\|f'\|_\beta
}{\langle z\rangle^{-\b-\frac{7}{4}}}\right\}, \label{BA-GL}
 \end{equation}
 \begin{equation}
\lt|g(z,\l){d\/dz}\lt(\ai(z\omega)e^{{4\/3}{z}^{\frac{3}{2}}}
\rt) \rt| \le
C{\|q\|_\iy \/|\l|^{\frac{1}{6}}} \lt\{ {\|f\|_\a\/\langle z\rangle^{\a+\frac{3}{4} }}
+{\|f'\|_\beta \/\langle z\rangle^{\b+\frac{5}{4} }}
\rt\}. \label{BAPrime-GL}
\end{equation}
\end{lemma}

{\it Proof.} Assume $ 0\le\arg\l\le\pi$ and $z\in\G_\l^+$.
By \refLem{A7.5-1}{SvaGForAll},
  we have
$\Gamma_\l(z)\subset S[-\pi+\frac{2}{3}\delta,0]$,
 so the uniform estimates
(\ref{aiest}) and (\ref{aiprimest}) hold on $\Gamma_\l(z)$
for both $a(z)$ and $a(z\omega)$.
 Let
$F(z)=\l^{-\frac{1}{6}}\ai (z)\rho(z)$, where $\rho$ is given by (\ref{EffPotentials}).
$$
g-PFf=I_1+I_2,\ \  I_1=\int_{\Gamma_\l(z)} P(s)F'(s)f(s)ds, \qquad
I_2=\int_{\Gamma_\l(z)} P(s)F(s)f'(s)ds.
$$
Using (\ref{EstNu}),
(\ref{aiest}) and (\ref{EstP-2})
we have
\begin{equation}
\label{NITerm1} \left|
P(z)F(z)f(z)
\right| \le \co|\l|^{-\frac{1}{6}}
{\|q\|_\iy}|e^{-\frac{4}{3}{z}^{\frac{3}{2}}}| {\|f\|_\a }{\langle
z\rangle^{-\a-1}}.
\end{equation}
In order to estimate $I_1$ and $I_2$ we use
 (\ref{EstNu}),
(\ref{aiest}), (\ref{aiprimest}), (\ref{Exp-2}) and
 (\ref{EstP-1}).
This gives
$$
|I_1|\le \co \frac{\|q\|_\iy}{|\l|^{\frac{1}{3}}} \int_{\Gamma_\l(z)}
\left(
\left\langle
\frac{s}{\l^{\frac{2}{3}}}
\right\rangle^{-\frac{1}{4}}
\frac{\|f\|_\a |e^{-\frac{4}{3}{s}^{\frac{3}{2}}}|}{\langle s\rangle^{\a+{7\/4}}}+
\frac{|\l|^{-\frac{2}{3}}}{
\left\langle\frac{s}{\l^{\frac{2}{3}}}\right\rangle^{\frac{5}{4}}}
{\|f\|_\a
|e^{-\frac{4}{3}{s}^{\frac{3}{2}}}|\/\langle s\rangle^{\a+{3\/4}}}
\right)|ds|
$$$$
\le  \co{\|q\|_\iy}{|\l|^{-\frac{1}{6}}} |e^{-\frac{4}{3}{z}^{\frac{3}{2}}}|
{\|f\|_\a}{\langle z\rangle^{-\a-2}},
$$
$$
|I_2|\le C {\|q\|_\iy\/|\l|^{\frac{1}{3}}}
\int_{\Gamma_\l(z)}
{\|f'\|_\beta |e^{-\frac{4}{3}{s}^{\frac{3}{2}}}|}{
\left\langle\frac{s}{\l^{\frac{2}{3}}}\right\rangle^{-\frac{1}{4}}}
{|ds|\/\langle s\rangle^{\b+{3\/4}}}\le
\co{\|q\|_\iy\/|\l|^{\frac{1}{6}}} |e^{-\frac{4}{3}{z}^{\frac{3}{2}}}|
{\|f'\|_\beta \/\langle z\rangle^{\b+{3\/2}}}.
$$
The above estimates for $I_1$, $I_2$ and (\ref{NITerm1}) give
\begin{equation}
\label{BA-GL-Cutted}
|g(z,\l)|\le  C |\l|^{-\frac{1}{6}}{\|q\|_\iy}
|e^{-\frac{4}{3}{z}^{\frac{3}{2}}}| \left\{ {\|f\|_\a}{\langle
z\rangle^{-\a-1}}+ {\|f'\|_\beta }{\langle
z\rangle^{-\b-\frac{3}{2}}} \right\}.
 \end{equation}
The last estimate together
with (\ref{aiest}) and (\ref{aiprimest}) implies (\ref{BA-GL}) and (\ref{BAPrime-GL}).

Assume $\d\le\arg \l\le \pi$ and $z\in\G_\l^-$.  Using
$\G_\l=\G_\l^-\cup\G_\l^+$ we  have $g=g_-+g_+$, where
$$ g_-(z,\l)=
  \int_{\Gamma_\l(z,z_*)}
\ai(s) e^{-\frac{4}{3}{s}^{\frac{3}{2}}}V_q(s)f(s)\,ds ,\quad
 g_+(\l)=
  \int_{\Gamma_\l^+}
\ai(s) e^{-\frac{4}{3}{s}^{\frac{3}{2}}}V_q(s)f(s)\,ds .\quad
$$
Using (\ref{|z|OnG-}) and  (\ref{BA-GL-Cutted}) for $z=z_*$  we
obtain
\begin{equation}
\label{Eg+Gl-ExpTerm}
 |g_+(\l)|\le  C
\frac{\|q\|_\iy}{|\l|^{\frac{1}{6}}}|e^{-\frac{4}{3}{z_*}^{\frac{3}{2}}}|
 \left\{ \frac{\|f\|_\a}{|z_*|^{\a+1}}+
\frac{\|f'\|_\beta }{|z_*|^{\b+\frac{3}{2}}}, \right\}
\le  C
\frac{\|q\|_\iy}{|\l|^{\frac{1}{6}}}
|e^{-\frac{4}{3}{z_*}^{\frac{3}{2}}}|
 \left\{ \frac{\|f\|_\a}{|z|^{\a+1}}+
\frac{\|f'\|_\beta }{|z|^{\b+\frac{3}{2}}}, \right\}.
\end{equation}
 Using (\ref{aiest}), (\ref{EffPotentials}),
(\ref{EstNu}), (\ref{Exp-2}) and   (\ref{|z|OnG-})
results in
\begin{equation}
\label{DirEstG-Exp}
|g_-(z,\l)|\le  \co\frac{\|q\|_\iy}{|\l|^{\frac{1}{3}}}
  \int_{\Gamma_\l(z,z_*)}
|e^{-\frac{4}{3}{s}^{\frac{3}{2}}}| \frac{\|f\|_\a}{|s|^{\a +\frac{1}{4}}}\,|ds|
\le \co\frac{\|q\|_\iy}{|\l|^{\frac{1}{6}}} |e^{-\frac{4}{3}{z}^{\frac{3}{2}}}|
\frac{\|f\|_\a}{|z|^{\a +1}}.
\end{equation}

Now (\ref{BA-GL}) and (\ref{BAPrime-GL}) follow from
(\ref{Eg+Gl-ExpTerm}) and (\ref{DirEstG-Exp}) taking into account
(\ref{aiest}), (\ref{aiprimest}) and the fact that,  by Lemma~\ref{27},
$|\exp\{\frac{2}{3}{z(x,\l)}^{\frac{3}{2}}\}|$ is strictly increasing.
$\BBox$

In the following Lemma we estimate the operator $\mathbf{J}V_0$.
 We show that as $\l\to\infty$
  it
 is asymptotically small in comparison with  $\mathbf{J}V_q$.

\begin{lemma}
\label{IsoV0} Let $ \l\in\ol\C_+$, $|\l|\ge1$ and
$f\in\cF_{\a}^\l$ for some $\a>0$ (that is,
  $f$ is defined on $\Gamma_\l$ for $ \d\le\arg \l\le\pi$ and on
  $\Gamma_\l^+$ for $ 0\le\arg\l\le\delta$). Then
  $\mathbf{J} V_0 f\in\cF_{\a+\frac{1}{2}}^\l$ and
\begin{equation}
 |(\mathbf{J} V_0 f)(z,\l)|\le \frac{\co}{|\l|^{\frac{2}{3}}}
\frac{\|f\|_\a}{\langle z\rangle^{\a+{1\/2}}}, \qquad
\left|\frac{\partial}{\partial z}(\mathbf{J} V_0 f)(z,\l)\right|\le
\frac{\co}{|\l|^{\frac{2}{3}}} \frac{\|f\|_\a}{\langle z\rangle^{\a+{3\/2}}}.
\label{V0G+}
\end{equation}

\end{lemma}

\no {\it Proof.}
By \refLem{A7.5-1}{SvaGForAll},
  we have
$\Gamma_\l(z)\subset S[-\pi+\frac{2}{3}\delta,0]$,
 so the decomposition (\ref{Jform}) holds on $\Gamma_\l(z)$.
We estimate the part of $\mathbf{J}V_0$,  corresponding to
the first term in decomposition (\ref{Jform}).  Using
Lemma~\ref{SvaGForAll},
 (\ref{EstNu}), (\ref{aiest}), (\ref{pow-2}) and  the inequality
$
\langle z\rangle|\l|^{\frac{2}{3}} \le
2(|\l|^{\frac{4}{3}}+|z|^2)
$.
 As a result we have
$$
\left|\int_{\Gamma_\l(z,\infty)} \ai(s\omega)V_0(s)f(s)\,
ds\right|\le \co \int_{\Gamma_\l(z,\infty)}
\frac{\|f\|_\a}{\langle s\rangle^{\a+\frac{1}{4}}}\frac{|ds|}{|\l|^{\frac{4}{3}}+|s|^2}
$$
\begin{equation}
\label{EV0G+Pow} \le \frac{\co}{|\l|^{\frac{2}{3}}}
\int_{\Gamma_\l(z,\infty)} \frac{\|f\|_\a}{\langle s\rangle^{\a+\frac{5}{4}}}|ds|
\le \frac{\co}{|\l|^{\frac{2}{3}}} \frac{\|f\|_\a}{\langle z\rangle^{\a+{1\/4}}}
\end{equation}

In order to estimate the part of $\mathbf{J}V_0$, corresponding to
the second term in decomposition (\ref{Jform}),  we
use  (\ref{EstNu}), (\ref{aiest}),
(\ref{Exp-2}) and  the inequality
$
\langle z\rangle|\l|^{\frac{2}{3}} \le
2(|\l|^{\frac{4}{3}}+|z|^2)
$. This gives
$$
\left|
 \int_{\Gamma_\l(z,\infty)}
\ai(s) e^{-\frac{4}{3}{s}^{\frac{3}{2}}}V_0(s)f(s)\,ds\right|
\le \co
\int_{\Gamma_\l(z,\infty)}
 \frac{\|f\|_\a}{|\l|^{\frac{4}{3}}+|s|^2}
\frac{|e^{-\frac{4}{3}{s}^{\frac{3}{2}}}|}{\langle s\rangle^{\a+\frac{1}{4}}}\,|ds|
$$
\begin{equation}
\label{EV0G+Exp} \le \co\frac{\|f\|_\a}{|\l|^{\frac{2}{3}}}
\int_{\Gamma_\l(z,\infty)}
\frac{|e^{-\frac{4}{3}{s}^{\frac{3}{2}}}|}{\langle s\rangle^{\a+\frac{5}{4}}}\,|ds|
\le\frac{\co}{|\l|^{\frac{2}{3}}}
\frac{|e^{-\frac{4}{3}{z}^{\frac{3}{2}}}|\|f\|_\a}{\langle z\rangle^{\a+\frac{7}{4}}}.
\end{equation}
The first estimate in (\ref{V0G+}) follows from (\ref{EV0G+Pow})
and (\ref{EV0G+Exp}) taking into account (\ref{aiest}) and
(\ref{Jform}).

In order to estimate  ${\partial_z}g(z,\l)$ we note
that $\pa_zg(z,\l)= \int_{\G_\l(z)} \pa_zJ(z,s)V_0(s) f(s)\,ds. $
Therefore the second estimate in (\ref{V0G+}) follows from
(\ref{EV0G+Pow}) and (\ref{EV0G+Exp}) taking into account
(\ref{aiprimest}) and (\ref{Jform}). $\blacksquare$

Now in order to  estimate
 the operator $\mathbf{J}V=\mathbf{J}V_q+\mathbf{J}V_0$
 we combine the results of the three previous Lemmas.
\begin{lemma}
\label{TL-GL}
Let $ q, q_1\in L^\iy(\R)$. Assume $\l\in \ol\C_+$, $|\l|\ge1$,  and
 $f\in\cF_{\a,\beta}^\l$ for $\quad\a>0, \b>\frac{3}{4}$ (that is,
  $f$ is defined on $\Gamma_\l$ for $ \d\le\arg \l\le\pi$ and on
  $\Gamma_\l^+$ for $ 0\le\arg\l\le\delta$).
Then $\mathbf{J}Vf\in\cF_{\a,\beta}^\l$ and
 \begin{equation}
 \label{Aqest-GL}
  |(\mathbf{J}Vf)(z,\l)|\le  \co\frac{\|q\|_\iy+\|q_1\|_\iy}{|\l|^{\frac{1}{6}}} \left\{
\frac{\|f\|_\a}{\langle z\rangle^{\a+\frac{1}{2} }} + \frac{\|f'\|_\beta
}{\langle z\rangle^{\b-\frac{1}{2} }} \right\}+ \frac{\co}{|\l|^{\frac{2}{3}}}
\frac{\|f\|_\a}{\langle z\rangle^{\a+{1\/2}}} ,
 \end{equation}
 \begin{equation}
 \label{AqPrimeest-GL}
 \left|{\partial_z}(\mathbf{J}Vf)(z,\l)\right|
 \le \co\frac{\|q\|_\iy+\|q_1\|_\iy}{|\l|^{\frac{1}{6}}} \left\{
\frac{\|f\|_\a}{\langle z\rangle^{\a+\frac{3}{4} }} + \frac{\|f'\|_\beta
}{\langle z\rangle^{\b+\frac{1}{2} }} \right\}+ \frac{\co}{|\l|^{\frac{2}{3}}}
\frac{\|f\|_\a}{\langle z\rangle^{\a+{3\/2}}} .
 \end{equation}
\end{lemma}
\no {\it Proof.}
Recall that
$\mathbf{J}V=\mathbf{J}V_q+\mathbf{J}V_0$.
By \refLem{A7.5-1}{SvaGForAll},
  we have
$\Gamma_\l(z)\subset S[-\pi+\frac{2}{3}\delta,0]$,
 so the decomposition (\ref{Jform}) holds on $\Gamma_\l(z)$.
 Taking into account this decomposition, we deduce that the
combination of
Lemmas~\ref{AsJ1} and \ref{AsJE} gives the
estimate for  $\mathbf{J}V_q$ (the corresponding terms in (\ref{Aqest-GL})
and (\ref{AqPrimeest-GL}) contain  curved brackets). Together with
the estimate
 (\ref{V0G+}) for $\mathbf{J}V_0$ this proves
 (\ref{Aqest-GL}) and (\ref{AqPrimeest-GL}). $\BBox$

\no {\bf Proof of Theorem~\ref{ConvG+THM}.} We present the proof  for $\l\in\ol\C_+$,
for $\l\in\ol\C_-$ it is analogous.
Let
$v_{n+1}=\bJ Vv_{n}, n\ge 0$. Substituting $v_n$,
 $g=v_{n+1}$ in (\ref{Aqest-GL}), (\ref{AqPrimeest-GL})
  and taking into account (\ref{Space-ai-G+}) for
$v_0$ we  obtain
$$
|v_{n+1}(z)|\le
 {\ve\|v_n\|_{\a_n}\/\langle z\rangle^{\a_n+{1\/2}}}+
{\ve\|v_n'\|_{\b_n }\/\langle z\rangle^{\b_n-{1\/2} }},\ \ \ \
|v_{n+1}'(z)|\le {\ve\|v_n\|_{\a_n}\/\langle z\rangle^{\a_n+{3\/4}}}+
{\ve\|v_n'\|_{\b_n}\/\langle z\rangle^{\b_n+{1\/2} }},
$$
where
\begin{equation}
\label{alfas-betas}
\a_0=\frac{1}{4},
\beta_0=\frac{5}{4},
\quad
\a_{n+1}=\min\{\a_n+\frac{1}{2},
\beta_n-\frac{1}{2}\},\quad \beta_{n+1}=\min\{\a_n+\frac{3}{4}, \beta_n+\frac{1}{2}\}.
\end{equation}
Therefore
\begin{equation}\label{ItEstEven}
    |v_{2n}(z)|\le
  \ve^{2n}
  {\|v_0\|_{\frac{1}{4},\frac{5}{4}}}{\langle z\rangle^{-\frac{n}{4}-\frac{1}{4}}},
  \qquad
|v_{2n+1}(z)|\le
  \ve^{2n+1}
 {\|v_0\|_{\frac{1}{4},\frac{5}{4}}}{\langle z\rangle^{-\frac{n}{4}-\frac{3}{4}}},
\end{equation}
\begin{equation}\label{ItEstOdd}
    |v_{2n}'(z)|\le
 \ve^{2n}
  {\|v_0\|_{\frac{1}{4},\frac{5}{4}}}{\langle z\rangle^{-\frac{n}{4}-\frac{5}{4}}},\quad
|v_{2n+1}'(z)|\le
\ve^{2n+1}
{\|v_0\|_{\frac{1}{4},\frac{5}{4}}}{\langle z\rangle^{-\frac{n}{4}-1}},
\end{equation}
and for $\ve<1$ the series
$
v_+(z)=\sum_{n=0}^\infty v_n(z)
$
converges absolutely and is a solution  of
$v_+=\ai+\mathbf{J} V v_+$. For
$\ve<\frac{1}{2}$
 we obtain from (\ref{Space-ai-G+}),
(\ref{ItEstEven}) and (\ref{ItEstOdd})   the estimates
(\ref{G+EstV0}), (\ref{G+EstV1}).

We prove the uniqueness. Suppose that there exists another solution
$v_+^{(1)}\in\cF_{\frac{1}{4},1}^\l$  and let
$y=v_+-v_+^{(1)}\in\cF_{\frac{1}{4},1}^\l$. We have
$y=\mathbf{J}V y$ and therefore $y=(\mathbf{J}V)^{n} y$ for any integer $n\ge 1$;
applying  Lemma~\ref{TL-GL} we obtain
$    |y(z)|\le \co \ve^{n} \|y\|_{\frac{1}{4},1}$.
Taking the limit $n\to\infty$ for
$\ve<1$ we obtain $y=0$.
 $\blacksquare$

\section{Uniform asymptotics}
\setcounter{equation}{0}

In this section we  consider the equation
 $v_+=a+{\bf J}v_+$ for large $|\l|$,  $|\arg\l|\le\delta$ and $z\in\Gamma_\l^-$.
 The case $z\in\Gamma_\l^+$ was treated in the previous section.

For $|\arg\l|\le\delta$ denote by $\cF_-^\l$  the class of
functions $f$ on $\Gamma_\l^-$ such that $f, f'\in
L^\infty(\Gamma_\l^-)$. We also set
$\cF^\l=\cF_-^\l\oplus\cF_{\frac{1}{4},1}^\l$ (for the definition
of $\cF_{\alpha,\beta}^\l$ see (\ref{DefF})). Our main result is

\begin{theorem}
\label{TThmG-} Let $q\in\cB$ and  $|\arg\l|\le\delta$. Then the equation $v_+=v_0+\mathbf{J} V v_+$, $v_0\equiv a$,
 has a  unique solution $v_+\in\cF^\l$ for
$ {|\l|^{{1\/6}}}\ge 2c_0( \|q\|_{\cB}+1 ) $, where $c_0>1$ is an
absolute constant. If, in addition, $(z,\l)\in\Gamma_\l^-\times
S[-\delta,\delta]$, then the following estimates are fulfilled:
\begin{equation}\label{ESmallLV+}
|v_+(z)|\le C{\la z\ra^{-{1\/4}}}, \qquad |v_+'(z)|\le C \la
z\ra^{{1\/4}},
\end{equation}
\begin{equation}
\label{ESmallLV1} |{v}_{1}(z)|\le {C\ve \la z\ra^{-{1\/4}}},
\qquad |{v}_{1}'(z)|\le \co\langle z\rangle^{\frac{1}{4}}\ve ,\ \
\ \ v_1={\bf J} V v_0, \qquad \ve=c_0{|\l|^{-{1\/6}}}(
\|q\|_{\cB}+1),
\end{equation}
\begin{equation}\label{ESmallV+MimV0mimV1}
|v_+(z)-v_0(z)-v_1(z)|\le {\co \ve^2 \langle z\rangle^{-\frac{1}{4}}}, \qquad
|v_+'(z)-v_0'(z)-v_1'(z)|\le \ve^2\co\langle z\rangle^{\frac{1}{4}}.
\end{equation}
\end{theorem}

\begin{corollary}
\label{TThmG-Corollary}
Let $q\in\cB$.
Then
the equation  $u_+=\Ai+{\bf J_0}Vu_+$
has a unique
 solution $u_+(z,\l)$ such that
 $u_+(z,\l)=\Ai(z)(1+o(1))$,
 $\partial_zu_+(z,\l)+\sqrt{z}u_+(z,\l)=\Ai'(z)O(z^{-\frac{5}{4}})$ as
 $\Gamma_\l\ni z\to\infty$ for
 $
{|\l|^{{1\/6}}}\ge 2c_0(
\|q\|_{\cB}+1
)
$,
where $c_0>1$ is an absolute constant.
Moreover,
$u_+(z,\lambda)=e^{-\frac{2}{3}z^{\frac{3}{2}}}v_+(z,\lambda)$.
 The
following estimates for $u_+$ and
$u_1(z,\lambda)=e^{-\frac{2}{3}z^{\frac{3}{2}}}v_1(z,\lambda)$
are fulfilled uniformly in $z\in\Gamma_\l^-$:
\\
If
$|\arg\l|\le\d$, then
\begin{equation}
\label{Estu+} \left|u_+(z,\l)\right|\le \co {
e^{\frac{2}{3}|\Re z^{\frac{3}{2}}|}}{\langle z\rangle^{-\frac{1}{4}}}, \qquad
\left|\partial_zu_+(z,\l)\right|\le \co \langle z\rangle^{\frac{1}{4}}
 e^{\frac{2}{3}|\Re z^{\frac{3}{2}}|},
\end{equation}
\begin{equation}
\label{Estu1}
\quad
|u_1(z,\l)|
\le \co \ve \frac{e^{\frac{2}{3}|\Re
z^{\frac{3}{2}}|}}{\langle z\rangle^{\frac{1}{4}}}, \quad
\left|\partial_zu_1(z,\l)\right|\le \co
\ve\langle z\rangle^{\frac{1}{4}}e^{\frac{2}{3}|\Re z^{\frac{3}{2}}|},
\quad
\ve=c_0
\frac{\|q\|_{\cB}+1}{|\l|^{{1\/6}}},
\end{equation}
\begin{equation}
\label{hvostu+} |u_+(z,\l)-\Ai(z)-u_1(z,\l)|\le
\co \ve^2
{e^{\frac{2}{3}|\Re z^{\frac{3}{2}}|}}{\langle z\rangle^{-\frac{1}{4}}},
\end{equation}
\begin{equation}
\label{hvostu+prime} \left|
\partial_zu_+(z,\l)- \Ai'(z)-
\partial_zu_1(z,\l)\right|\le \co \ve^2 \langle z\rangle^{\frac{1}{4}}e^{\frac{2}{3}|\Re
z^{\frac{3}{2}}|},
\end{equation}
if $\d\le|\arg\l|\le\pi$, then
\begin{equation}\label{U+BigS}
  |u_+(z,\l)-\Ai(z)|\le   \co \ve
{e^{\frac{2}{3}|\Re  z^{\frac{3}{2}}|}}{\langle z\rangle^{-\frac{3}{4}}},\qquad
  |\partial_zu_+(z,\l)-\Ai'(z)|\le \co \ve
 {e^{\frac{2}{3}|\Re z^{\frac{3}{2}}|}}{\langle z\rangle^{-\frac{1}{4}}}.
\end{equation}
\end{corollary}
{\it Proof. } Set
$u_+(z,\lambda)=e^{-\frac{2}{3}z^{\frac{3}{2}}}v_+(z,\lambda)$, where
$v_+$ is given
by Theorem~\ref{ConvG+THM}
(for $\delta<|\arg\l|\le\pi$)
 and
 Theorem~\ref{TThmG-}
(for $|\arg\l|\le\delta$). By (\ref{IEpodkrychennoe}),
(\ref{PodkrutkaU}) and (\ref{G+EstV0}), $u_+$ is a solution of (\ref{IntegralEqinZ})
with the required asymptotics.

 In order to prove uniqueness, we suppose that there
exists another solution  $u_+^{(1)}$ with the same asymptotics
as  $\Gamma_\l\ni z\to\infty$. Then
$v_+^{(1)}(z,\lambda)=e^{\frac{2}{3}z^{\frac{3}{2}}}u_+^{(1)}(z,\lambda)$
is in $\cF^\l$ (for $|\arg\l|\le\delta$) or in
$\cF_{\frac{1}{4},1}^\l$
(for $\delta<|\arg\l|\le\pi$)
 and solves $v_+=v_0+\mathbf{J} V v_+$. Hence,
 by Theorems~\ref{ConvG+THM}
 and
 \ref{TThmG-},
$v_+^{(1)}=v_+$, implying $u_+^{(1)}=u_+$.

 The estimates (\ref{Estu+}--\ref{hvostu+prime}) follow
 from (\ref{ESmallLV+}--\ref{ESmallV+MimV0mimV1}).
 The estimate (\ref{U+BigS}) follows from (\ref{G+EstV1}).
 $\blacksquare$

Below we consider only the case $0\le\arg\l\le\delta$,
the case $-\delta\le\arg\l\le0$ is analogous.

By \refLem{A7.5-1}{SvaGForAll}, $\G_\l^-$ is arbitrarily close to
$\mathbb{R}_-$ as $\arg\l\to0$. Thus we cannot use
(\ref{aiprimest}) in order to estimate the terms in the decomposition
 (\ref{Jform})
for $J(z,s)$.
So we use another representation.
We have
 \begin{equation}
\label{Aitoai-}
 \Ai(z\ol\omega)=
e^{-\frac{2}{3}z^{\frac{3}{2}}}\ai(z\ol\omega),\qquad
 z\in S[-\pi,-{\pi\/3}),
 \qquad
 \omega=e^{\frac{2\pi i}{3}}.
\end{equation}
Using (\ref{BfromA}), (\ref{ab1}) and  (\ref{Aitoai+}),
we obtain from (\ref{defJ}) and (\ref{PodkrutkaU})
\begin{equation}
\label{DK-Ai}
J(z,s)=-2i\left( \ai(z\ol\omega)\ai(s\omega) -
e^{\frac{4}{3}z^{3\/2}-\frac{2}{3}s^{3\/2}} \ai(z\omega)\Ai(s\ol\omega)
\right)
,\qquad
z\in\Gamma_\l^-,\quad s\in \Gamma_\l,
\end{equation}
where,  by \refLem{A7.3-1}{DivG2},
$\Gamma_\l^-\subset S[-\pi,-\frac{\pi}{2}+\frac{5}{12}\delta]
\subset S[-\pi,-\frac{\pi}{3})
$
and
$\Gamma_\l^+\subset S[-\frac{\pi}{2}-\frac{\delta}{4}, 0]
\subset S(-\pi,\pi]$.

As it follows from the decomposition (\ref{DK-Ai}),
the formal operator $\mathbf{J}V$ can be presented as the sum
$\ai(z\ol\omega)\times$(integral
operator)$+e^{\frac{4}{3}z^{3\/2}}\ai(z\omega)\times$(integral
operator). Thus in order to estimate $\mathbf{J}V$  we
estimate these integral operators, introducing for functions on
$\Gamma_\l^-$ a decomposition in the spirit of (\ref{DK-Ai}).

For $f\in\cF_-^\l$ and a fixed decomposition
\begin{equation}
f(z)=\ai(z\ol\omega) f_p(z) +
e^{\frac{4}{3}z^{3\/2}}\ai(z\omega)f_e(z),
\quad
z\in\Gamma_\l^-,
\quad
\l\in S[0,\delta]\bigcap\{\l\in\mathbb{C}: |\l|\ge1\},
\label{DecFGM}
\end{equation}
we introduce the functionals
\begin{equation}
\label{NormFGM-0}
p_0(f,\l)= \sup_{z\in \G_\l^-}
\lt(|f_p(z)|+|e^{\frac{4}{3}z^{\frac{3}{2}}}f_e(z)|\rt), \quad
 p_1(f,\l)= \,\sup_{z\in
\G_\l^-}\langle z\rangle^{1\/2}\lt(|f_p'(z)|+|e^{\frac{4}{3}z^{\frac{3}{2}}}f_e'(z)|
\rt).
\end{equation}
If $f\in\cF_-^\l$ has a decomposition (\ref{DecFGM}), then we have
\begin{equation}
\label{PWEstG-}
|f(z)|\le\co{p_0(f,\l)}{\langle z\rangle^{-\frac{1}{4}}},
\qquad
{|f'(z)|}{\langle z\rangle^{-\frac{1}{4}}}\le
\co\left(p_0(f,\l)+{p_1(f,\l)}{\langle z\rangle^{-1}}\right),
\quad
z\in\Gamma_\l^-.
\end{equation}
If $f\in\cF_-^\l$ has a decomposition (\ref{DecFGM}), then we have
\begin{equation}\label{EstFpFe0}
 |f_p(z)|\le  p_0(f,\l),\qquad
 |f_e(z)|\le   |e^{-\frac{4}{3}z^{\frac{3}{2}}}| p_0(f,\l),
\end{equation}
\begin{equation}\label{EstFpFe1}
 |f_p'(z)|\le  {p_1(f,\l)}{\langle z\rangle^{-\frac{1}{2}}},\qquad
 |f_e'(z)|\le  {|e^{-\frac{4}{3}z^{\frac{3}{2}}}|}{\langle z\rangle^{-\frac{1}{2}}}
 p_1(f,\l).
\end{equation}

Using (\ref{Aitoai+}), (\ref{Aitoai-}) and  the
identity (\ref{ab1})  we obtain for $0\le\arg\l\le\delta$
\begin{equation}
\label{Spac-ai-G-}
 v_0(z)=\ai(z)=e^{i\frac{\pi}{3}}\ai(z\ol\omega) +
e^{-i\frac{\pi}{3}}e^{\frac{4}{3}z^{\frac{3}{2}}}
\ai(z\omega),
\qquad
z\in\Gamma_\l^-\subset S[-\pi,-\frac{\pi}{3}).\qquad
\end{equation}
For this decomposition using (\ref{aiest})
and (\ref{aiprimest}) we obtain
\begin{equation}
\label{Norm-ai-G-}
p_0(v_0,\l)\le2, \qquad p_1(v_0,\l)=0,\qquad 0\le\arg\l\le\delta.
\end{equation}
Below we  estimate $(\mathbf{J} Vf )\Big|_{\Gamma_\l^-}$ in terms of $p_0$
and $p_1$ assuming that $f\in\cF^\l=\cF_-^\l\oplus\cF_{\frac{1}{4},1}^\l$.
Using (\ref{DK-Ai}) for some function $f$ on $\Gamma_\l^-$
 we obtain the following formal
decomposition:
\begin{equation}
\label{DecGGM}
(\mathbf{J} Vf )(z)=\ai(z\ol\omega) g_p(z) +
e^{\frac{4}{3}z^{3\/2}}\ai(z\omega)g_e(z),
\quad
g_p=\mathbf{J}_pVf,
\quad
g_e=\mathbf{J}_eVf,
\quad
z\in\Gamma_\l^-,
\end{equation}
where
 $\l\in S[0,\delta]\bigcap\{\l\in\mathbb{C}: |\l|\ge1\}$
and  the operators $\mathbf{J}_p$ and $\mathbf{J}_e$
are given by
\begin{equation}\label{ABGGM}
(\mathbf{J}_pf)(z)=-2 i
\int_{\Gamma_\l(z)}
\ai(s\omega) f(s)\,ds,
\quad
(\mathbf{J}_ef)(z)=2 i
\int_{\Gamma_\l(z)}
 e^{-\frac{2}{3}s^{\frac{3}{2}}}\Ai(s\ol\omega)f(s)\,ds .
\end{equation}
In the applications below
the integral along the infinite curve $\Gamma_\l^+$
exists in the principal value sense.
In this section
 we will always define
$p_0(\mathbf{J} Vf ,\l)$ and  $p_1(\mathbf{J} Vf ,\l)$
 using the decomposition (\ref{DecGGM}) and (\ref{ABGGM}).

In order to estimate ${\mathbf J_p}V$ and ${\mathbf J_e}V$ we
decompose the integrals in (\ref{ABGGM}) corresponding to the
splitting
$\Gamma_\l(z)=\Gamma_\l(z,z_*)\cup\Gamma_\l^+$.
Thus we have
\begin{equation}\label{IntroH}
 (\mathbf{J}_pf)(z)= (\mathbf{j}_pf)(z)+h_p(f),\qquad
 z\in\Gamma_\l^-\subset S[-\pi,-\frac{\pi}{3}),
\end{equation}
where, by definition
\begin{equation}
\label{Defjpje}
(\mathbf{j}_pf)(z)=-2 i
\int_{\Gamma_\l(z,z_*)}
\ai(s\omega)f(s)\,ds,\quad
h_p(f)=-2i
 \int_{\G_\l^+}
\ai(s\omega) f(s)\,ds.
\end{equation}
A similar decomposition of  ${\mathbf J_e}$
gives
\begin{equation}
\label{DecJe1}
(\mathbf{J}_ef)(z)= (\mathbf{j}_ef)(z)+h_e(f),\qquad
z\in\Gamma_\l^-\subset S[-\pi,-\frac{\pi}{3}),
\end{equation}
where, by definition
\begin{equation}
\label{DecJe2}
(\mathbf{j}_ef)(z)=2 i
\int_{\Gamma_\l(z,z_*)}
 e^{-\frac{2}{3}s^{\frac{3}{2}}}\Ai(s\ol\omega)f(s)\,ds,
\quad
h_e(f)=2 i
 \int_{\G_\l^+}
 e^{-\frac{2}{3}s^{\frac{3}{2}}}\Ai(s\ol\omega)f(s)\,ds.
\end{equation}
Note that the standard asymptotics (\ref{AsAiry}) for
 $\Ai(s\ol\omega)$ fails in the neighborhood of $\arg s=-\frac{\pi}{3}$.
Taking into account (\ref{Aitoai-})
we obtain
\begin{equation}
\label{AGPGM}
(\mathbf{j}_ef)(z)=2 i
\int_{\Gamma_\l(z,z_*)}
e^{-\frac{4}{3}s^{\frac{3}{2}}}\ai(s\ol\omega) f(s)\,ds.
\end{equation}
By (\ref{ab1}), (\ref{IEpodkrychennoe}), (\ref{Aitoai+}) and (\ref{Aitoai-}),
for
$s\in\Gamma_\l^+\subset S(-\pi,\frac{\pi}{3})$ we have
 $ e^{-\frac{2}{3}s^{\frac{3}{2}}}\Ai(s\ol\omega)=
e^{i\frac{\pi}{3}}\ai(s\omega)-
e^{-\frac{4}{3}s^{\frac{3}{2}}}\omega\ai(s)
$. This gives
\begin{equation}
\label{GPE}
h_e(f)=2 i
 \int_{\G_\l^+}
\left(e^{i\frac{\pi}{3}}\ai(s\omega)-
e^{-\frac{4}{3}s^{\frac{3}{2}}}\omega\ai(s)
\right)
f(s)\,ds.
\end{equation}
 Using (\ref{ABGGM}--\ref{GPE}),
 for $z\in\Gamma_\l^-\subset S[-\pi,-\frac{\pi}{3})$
 we have the decomposition
\begin{equation}
\label{DecJVq-G-}
(\mathbf{J}f)(z)=
\ai(z\ol\omega)\left((\mathbf{j}_pf)(z)+h_p(f) \right)+
e^{\frac{4}{3}z^{3\/2}}
\ai(z\omega)
((\mathbf{j}_ef)(z)+h_e(f)).
\end{equation}
 For
$ z_j=z(x_j,\l)\in\Gamma_\l^-$, $j=1,2$ and $\quad x_1<x_2$ we define
\begin{equation}
\label{DefP_pm}
P_{\pm}(z_1,z_2)=\l^{\frac{1}{6}}
\int_{\Gamma_\l(z_1,z_2)}
e^{\pm\frac{4}{3}s^{\frac{3}{2}}}\rho^{-1}(s)V_q(s)\,ds.
\end{equation}

\begin{lemma}
\label{CalculationsGM} Let
$q\in\cB$.  Assume $0\le\arg\l\le\delta$, $|\l|\ge 1$.
Assume that
$f\big|_{\Gamma_\l^-}\in\cF_-^\l$
has a decomposition (\ref{DecFGM})
and  $f\big|_{\Gamma_\l^+}=0$.
 Then
$(\mathbf{J}V_qf)(z)=\ai(z\ol\omega)(\mathbf{j}_pV_qf)(z)+
e^{\frac{4}{3}z^{3\/2}} \ai(z\omega)(\mathbf{j}_eV_qf)(z)\in\cF_-^\l$
and for this decomposition the following estimates are fulfilled:
\begin{equation}
\label{EGM}
p_0(\mathbf{J}V_qf,\l)
\le\co
{\|q\|_\cB}{|\l|^{-\frac{1}{6}}}
\lt( p_0(f,\l)+p_1(f,\l)\log(|\l|+1) \rt),
\end{equation}
\begin{equation}
\label{D-EGM}
p_1(\mathbf{J}V_qf,\l)\le
\co
{\|q\|_\iy}{|\l|^{-\frac{1}{3}}}
p_0(f,\l).
\end{equation}
\end{lemma}
\no {\it Proof.}
By \refLem{A7.3-1}{DivG2}, $\Gamma_\l^-\subset S[-\pi,-\frac{\pi}{3})$,
so (\ref{DK-Ai}) holds.
Using (\ref{DecFGM}), (\ref{DecGGM}), (\ref{IntroH}), (\ref{Defjpje}),
(\ref{DecJe1}),  (\ref{DecJe2})
 and  (\ref{AGPGM})
we obtain
\begin{equation}
\label{I-s}
(\mathbf{j}_pV_qf)(z)=-2i(I_{pp}(z)+I_{pe}(z)),\qquad
(\mathbf{j}_eV_qf)(z)=2 i(I_{ep}(z)+I_{ee}(z)),
\end{equation}
where
$$
I_{pp}(z)=
\int_{\Gamma_\l(z,z_*)}\!\!\!\!\!\!\!\!\!\!\!\!\!\!\!
 \ai(s\omega)
\ai(s\ol\omega)V_q(s)f_p(s)\,ds, \ \
I_{pe}(z)=
\int_{\Gamma_\l(z,z_*)}\!\!\!\!\!\!\!\!\!\!\!\!\!\!\!
\ai(s\omega)^2e^{\frac{4}{3}s^{\frac{3}{2}}}V_q(s)f_e(s)\,ds ,
$$
$$
I_{ep}(z)=
\int_{\Gamma_\l(z,z_*)}\!\!\!\!\!\!\!\!\!\!\!\!\!\!\!
\ai(s\ol\omega)^2e^{-\frac{4}{3}s^{\frac{3}{2}}}
V_q(s)f_p(s)\,ds,\
\ I_{ee}(z)=
\int_{\Gamma_\l(z,z_*)}\!\!\!\!\!\!\!\!\!\!\!\!\!\!\!
\ai(s\omega)\ai(s\ol\omega)V_q(s)f_e(s)\,ds.
$$

We estimate $I_{pp}$. Define the functions
$F_j(s)={\l^{-\frac{1}{6}}} \ai(s\omega) \ai(s\ol\omega) f_j(s)\r(s)$
for $j=e,p$.  Integration by
parts yields
\begin{equation}
I_{pp}(z)= -Q(z_*,z)F_p(z)+
\int_{\Gamma_\l(z,z_*)}
 Q(z_*,s) F_p'(s)ds,
\end{equation}
where $Q$ is given by (\ref{DQ}).
Using
(\ref{EffPotentials}), (\ref{EstNu}),
(\ref{aiest}),  (\ref{aiprimest})  and (\ref{Estq-1}), we obtain
$$
|I_{pp}(z)|\le C\frac{\|q_1\|_\iy}{|\l|^{\frac{1}{6}}} \left[
\frac{|f_p(z)|}{\langle z\rangle^{\frac{1}{2}}}+ \int_{\Gamma_\l^-} \left\{
\frac{|f_p'(s)|}{\langle s\rangle^{\frac{1}{2}}}+  \left(
\frac{|f_p(s)|}{\langle s\rangle^{\frac{3}{2}}}+\frac{|\l|^{-\frac{2}{3}}|f_p(s)|}{\langle s\rangle^{\frac{1}{2}}}
\right) \right\}\,|ds| \right].
$$
Due to  (\ref{pow-3Le1}), (\ref{EstFpFe0})  and (\ref{EstFpFe1}) we have
\begin{equation}
\label{EstIpp0}
|I_{pp}(z)|\le
\co{\|q_1\|_\iy}{|\l|^{-\frac{1}{6}}}
\lt(p_0(f,\l)+p_1(f,\l)
\log(|\l|+1)\rt),
\quad z\in\Gamma_\l^-.
\end{equation}
We estimate  $I_{pp}'(z)=
- \ai(z\omega)
\ai(z\ol\omega)V_q(z)f_p(z)$.
Using
 (\ref{EffPotentials}), (\ref{EstNu})
and (\ref{aiest}) we have
\begin{equation}
\label{EstIpp1}
\left|I_{pp}'(z)\right|\le \co
{\|q\|_\iy|f_p(z)|}{|\l|^{-\frac{1}{3}}\langle z\rangle^{-\frac{1}{2}}},
\quad z\in\Gamma_\l^-.
\end{equation}
The  estimate of $I_{ee}$ is similar. Using  $F_e$
and $Q$  we integrate by parts.
Using (\ref{EffPotentials}), (\ref{EstNu}),
(\ref{aiest}),  (\ref{aiprimest})  and (\ref{Estq-1}),
 we obtain
$$
|I_{ee}(z)|\le C\frac{\|q_1\|_\iy}{|\l|^{\frac{1}{6}}} \left[
\frac{|f_e(z)|}{\langle z\rangle^{\frac{1}{2}}}+
\int_{\Gamma_\l^-}
 \left\{
\frac{|f_e'(s)|}{\langle s\rangle^{\frac{1}{2}}}+  \left(
\frac{|f_e(s)|}{\langle s\rangle^{\frac{3}{2}}}+\frac{|\l|^{-\frac{2}{3}}|f_e(s)|}{\langle s\rangle^{\frac{1}{2}}} \right)
\right\}\,|ds| \right]
$$
We have
$|e^{-\frac{4}{3}z^{3\/2}}|\ge1$ for
$z\in\Gamma_\l^-\subset S[-\pi,-\frac{\pi}{3})$.
Using (\ref{pow-3Le1}),
(\ref{EstFpFe0})  and (\ref{EstFpFe1}) we have
\begin{equation}
\label{EstIee0}
 |I_{ee}(z)|\le \co\frac{\|q_1\|_\iy}{|\l|^{\frac{1}{6}}}
|e^{-\frac{4}{3}z^{3\/2}}| \lt( p_0(f,\l)+p_1(f,\l) \log(|\l|+1)
\rt), \quad z\in\Gamma_\l^-\subset S[-\pi,-\frac{\pi}{3}).
\end{equation}
We estimate $I_{ee}'(z)=-\ai(z\omega)\ai(z\ol\omega)V_q(z)f_e(z)$.
Using
 (\ref{EffPotentials}), (\ref{EstNu})
 and (\ref{aiest}) we obtain
\begin{equation}
\label{EstIee1}
\left|I_{ee}'(z)\right|\le \co
{\|q\|_\iy|f_e(z)|}{|\l|^{-\frac{1}{3}}\langle z\rangle^{-\frac{1}{2}}},
\quad z\in\Gamma_\l^-\subset S[-\pi,-\frac{\pi}{3})
.
\end{equation}
In order to estimate $I_{pe}$ we use $P_+(s,z)$, given by
(\ref{DefP_pm}). Integrating by parts we have
$$
I_{pe}(z)= -P_+(z_*,z)F(z_*)+
\int_{\Gamma_\l(z,z_*)}
P_+(s,z) F'(s)\,ds,
\qquad
F(z)=\l^{-\frac{1}{6}}\ai(z\omega)^2 f_e(z)\rho(z).
$$
Using (\ref{KeyLem1}), (\ref{aiest}), (\ref{aiprimest}), (\ref{EffPotentials})
 and
(\ref{EstNu}) we obtain
$$
|I_{pe}(z)|\le
\co\frac{\|q\|_\cB}{|\l|^{\frac{1}{6}}} \left[
|f_e(z_*)||e^{\frac{4}{3}z_*^{\frac{3}{2}}}|
+\int_{\Gamma_\l^-}
|e^{\frac{4}{3}s^{\frac{3}{2}}}| \lt\{ \frac{|f_e'(s)|}{\langle s\rangle^{\frac{1}{2}}}+
|f_e(s)| \left(
{\langle s\rangle^{-\frac{3}{2}}}+
\frac{|\l|^{-\frac{2}{3}}}{\langle s\rangle^{\frac{1}{2}}}\right)
\rt\}\,|ds| \right].
$$
We have
$|e^{\frac{4}{3}s^{\frac{3}{2}}}|\le1$
 for $s\in\Gamma_\l^-\subset S[-\pi,-\frac{\pi}{3})$.
 Using (\ref{pow-3Le1}) and
(\ref{EstFpFe0}--\ref{EstFpFe1})  we obtain
\begin{equation}
\label{EstIpe0}
|I_{pe}(z)|\le
\co{\|q\|_\cB}{|\l|^{-\frac{1}{6}}}
\lt(
p_0(f,\l)+p_1(f,\l)\log(|\l|+1)
\rt),
\qquad
z\in\Gamma_\l^-
.
\end{equation}
We estimate  $I_{pe}'(z)=-\ai(z\omega)^2e^{\frac{4}{3}z^{\frac{3}{2}}}V_q(z)f_e(z)$.
Using Lemma~\ref{27},
(\ref{EffPotentials}), (\ref{EstNu})
 and (\ref{aiest}) we get
\begin{equation}
\label{EstIpe1}
\left|I_{pe}'(z)\right|\le \co
{\|q\|_\iy|e^{\frac{4}{3}z_*^{\frac{3}{2}}}f_e(z)|}{|\l|^{-\frac{1}{3}}\langle z\rangle^{-\frac{1}{2}}},
\qquad
z\in\Gamma_\l^-.
\end{equation}
In order to estimate  $I_{ep}$  we use $P_-(s,z)$, given by
(\ref{DefP_pm}). Integrating by parts we have
$$
I_{ep}(z)= -P_-(z_*,z)F(z)+
\int_{\Gamma_\l(z,z_*)}
P_-(z_*,s)F'(s)\,ds,
\qquad
F(z)=\l^{-\frac{1}{6}}\ai(z\ol\omega)^2 f_p(z)\rho(z).
$$
Using (\ref{KeyLem2}), (\ref{aiest}), (\ref{aiprimest}),
(\ref{EffPotentials}) and
(\ref{EstNu}) we obtain
$$
|I_{ep}(z)|\le
\co\frac{\|q\|_\cB}{|\l|^{\frac{1}{6}}} \left[
\frac{|f_p(z)e^{-\frac{4}{3}z^{\frac{3}{2}}}|}{\langle z\rangle^{\frac{1}{2}}}+ \right.
\left.
\int_{\Gamma_\l(z,z_*)}
|e^{-\frac{4}{3}s^{\frac{3}{2}}}|
\left\{\frac{|f_p'(s)|}{\langle s\rangle^{\frac{1}{2}}}+ |f_p(s)| \left(
{\langle s\rangle^{-\frac{3}{2}}}+\frac{|\l|^{-\frac{2}{3}}}{\langle s\rangle^{\frac{1}{2}}}\right)
\right\}\,|ds| \right].
$$
By Lemma~\ref{27}, we have
$\max\limits_{\Gamma_\l(z,z_*)}
|e^{-\frac{4}{3}s^{\frac{3}{2}}}|=|e^{-\frac{4}{3}z^{\frac{3}{2}}}|$.
 Using (\ref{pow-3Le1}),
(\ref{EstFpFe0}) and (\ref{EstFpFe1})  we obtain
\begin{equation}
\label{EstIep0}
|I_{ep}(z)|\le
\co{\|q\|_\cB}{|\l|^{-\frac{1}{6}}}|e^{-\frac{4}{3}z^{\frac{3}{2}}}|
\lt(
p_0(f,\l)+p_1(f,\l)
\log(|\l|+1)
\rt),\qquad
z\in\Gamma_\l^-.
\end{equation}
We estimate $I_{ep}'(z)=-\ai(z\ol\omega)^2e^{-\frac{4}{3}z^{\frac{3}{2}}}
V_q(z)f_p(z)$.
Using Lemma~\ref{27},
 (\ref{EffPotentials}), (\ref{EstNu})
 and (\ref{aiest}) we obtain
\begin{equation}
\label{EstIep1}
\left|I_{ep}'(z)\right|\le \co
{\|q\|_\iy|e^{-\frac{4}{3}z^{\frac{3}{2}}}f_p(z)|}{|\l|^{-\frac{1}{3}}\langle z\rangle^{-\frac{1}{2}}},
\qquad
z\in\Gamma_\l^-.
\end{equation}

Due to (\ref{I-s}), (\ref{EstIpp0}), (\ref{EstIee0}), (\ref{EstIpe0})
and (\ref{EstIep0}) we have
$$
p_0(\mathbf{J}V_qf,\l)=
|(\mathbf{j}_pV_qf)(z)|+|e^{\frac{4}{3}z^{\frac{3}{2}}}(\mathbf{j}_eV_qf)(z)|\le
\co\frac{\|q\|_\cB}{|\l|^{\frac{1}{6}}}
\lt( p_0(f,\l)+p_1(f,\l)\log(|\l|+1) \rt),
$$
where $z\in\Gamma_\l^-$. This proves (\ref{EGM}).
Due to (\ref{I-s}), (\ref{EstIpp1}), (\ref{EstIee1}),
(\ref{EstIpe1}) and (\ref{EstIep1})we have
$$
p_1(\mathbf{J}V_qf,\l)=
\left|\partial_z(\mathbf{j}_pV_qf)(z)\right|+
\left|e^{\frac{4}{3}z^{\frac{3}{2}}}\partial_z(\mathbf{j}_eV_qf)(z)\right|\le
\co
\frac{\|q\|_\iy}{|\l|^{\frac{1}{3}}\langle z\rangle^{\frac{1}{2}}}
(|f_p(z)|+|e^{\frac{4}{3}z^{\frac{3}{2}}}f_e(z)|),
$$
where $z\in\Gamma_\l^-$. This proves (\ref{D-EGM}).
$\blacksquare$

\begin{lemma}
\label{BigALLTOGETHER}
Let $q\in\cB$ and $0\le\arg\l\le\d$, $|\l|\ge1$.
Assume $f=f^-+f^+$, where  $f^+=f\big|_{\Gamma_\l^+}\in \cF_{\a,\beta}^\l$ for
$\a>0$, $\beta>\frac{3}{4}$ and $f^-=f\big|_{\Gamma_\l^-}\in \cF_-^\l$.
Then $g=(\mathbf{J}Vf)\big|_{\Gamma_\l^-}\in \cF_-^\l$ and
for the decomposition
$g(z)= \ai(z\ol\omega) (\mathbf{J}_pVf)(z) +
e^{\frac{4}{3}z^{3\/2}}
\ai(z\omega)
(\mathbf{J}_eVf)(z)$
the following estimates are fulfilled:
\begin{equation}
\label{BEsG-}
p_0(g,\l)\le
\co\frac{\|q\|_\cB}{|\l|^{\frac{1}{6}}} \lt(
p_0(f^-,\l)+p_1(f^-,\l)\log(|\l|+1) +\|f^+\|_{\a,\beta}\rt)
+\frac{C}{|\l|^{\frac{2}{3}}}p_0(f^-,\l)
,
\end{equation}
\begin{equation}
\label{EstDerGM}
p_1(g,\l)\le  {\co}{|\l|^{-\frac{1}{3}}}
\left(
 \|q\|_\iy+1
\right)
 p_0(f^-,\l).
\end{equation}
\end{lemma}
\no {\it Proof.}
By (\ref{DecJVq-G-}), we have $g=g_0+g_-+g_+$, where
$$
g_+(z)=\ai(z\ol\omega)h_p(V_qf^+)+
e^{\frac{4}{3}z^{3\/2}} \ai(z\omega)h_e(V_qf^+) ,
$$$$
g_-(z)=\ai(z\ol\omega)(\mathbf{j}_pV_qf^-)(z)+
e^{\frac{4}{3}z^{3\/2}} \ai(z\omega)(\mathbf{j}_eV_qf^-)(z),
$$
$$
g_0(z)= \ai(z\ol\omega) (\mathbf{J}_pV_0f)(z) +
e^{\frac{4}{3}z^{3\/2}}
\ai(z\omega)
(\mathbf{J}_eV_0f)(z).
$$
Firstly, we estimate $g_+$.
From
(\ref{G+EstPow})
and (\ref{BA-GL-Cutted}) we
have
\begin{equation}
|h_p(V_qf)|\le \co{\|q_1\|_\iy}{|\l|^{-\frac{1}{6}}}\|f^+\|_{\a,\beta},
\quad |h_e(V_qf)|\le
 \co(\|q\|_\iy|e^{-\frac{4}{3}z_*^{\frac{3}{2}}}|+\|q_1\|_\iy){|\l|^{-\frac{1}{6}}}
 \|f^+\|_{\a,\beta}.
\label{ESTAPBPGM}
\end{equation}
By  Lemma~\ref{27},  we have
$|e^{\frac{4}{3}(z^{\frac{3}{2}}-z_*^{\frac{3}{2}})}|\le1$
for $z\in\Gamma_\l^-$.
Thus
\begin{equation}
\label{Esthphe}
p_0(g_+,\l)\le
\co(\|q\|_\iy+\|q_1\|_\iy){|\l|^{-\frac{1}{6}}}\|f^+\|_{\a,\beta},
\qquad
p_1(g_+,\l)=0.
\end{equation}
Secondly, we estimate  $g_0$.
Using (\ref{Defjpje}), (\ref{AGPGM}), (\ref{EffPotentials}),
(\ref{EstNu}), (\ref{aiest}), (\ref{EIV0}) and
 Lemma~\ref{27}   gives
\begin{equation}
|(\mathbf{j}_pV_0f)(z)|\le\co
\int_{\Gamma_\l^-}
\frac{p_0(f^-,\l)}{|\l|^{\frac{4}{3}}+|s|^{2}}
\frac{|ds|}{\langle s\rangle^{\frac{1}{2}}}
\le
\co\frac{p_0(f^-,\l)}{|\l|^{\frac{2}{3}}},
\qquad
z\in\Gamma_\l^-,
\label{GPMV0}
\end{equation}
\begin{equation}
|(\mathbf{j}_eV_0f)(z)|\le\co|e^{-\frac{4}{3}z^{\frac{3}{2}}}|
\int_{\Gamma_\l^-}
\frac{p_0(f^-,\l)}{|\l|^{\frac{4}{3}}+|s|^{2}}
\frac{|ds|}{\langle s\rangle^{\frac{1}{2}}}
\le
\co|e^{-\frac{4}{3}z^{\frac{3}{2}}}|\frac{p_0(f^-,\l)}{|\l|^{\frac{2}{3}}}
\qquad
z\in\Gamma_\l^-.
\label{GEMV0}
\end{equation}
Using Lemma~\ref{27},  (\ref{EV0G+Pow}) and (\ref{EV0G+Exp}) we have
\begin{equation}\label{GEMV0H}
|h_p(V_0f)|\le
{\co}{|\l|^{-\frac{2}{3}}}\|f^+\|_\alpha,
\qquad
|h_e(V_0f)|\le
{\co}{|\l|^{-\frac{2}{3}}}|e^{-\frac{4}{3}z_*^{\frac{3}{2}}}|\|f^+\|_\alpha.
\end{equation}
By  Lemma~\ref{27},  we have
$|e^{\frac{4}{3}(z^{\frac{3}{2}}-z_*^{\frac{3}{2}})}|\le1$
for $z\in\Gamma_\l^-$.
Thus substituting (\ref{GPMV0}), (\ref{GEMV0}) and (\ref{GEMV0H})
 in (\ref{DecJVq-G-})  we obtain
 \begin{equation}
 \label{V0G-}
 p_0(g_0,\l)\le\co(p_0(f^-,\l)+\|f^+\|_\a){|\l|^{-\frac{2}{3}}}.
\end{equation}
 Using (\ref{Defjpje}), (\ref{AGPGM}), (\ref{aiest}),
(\ref{EffPotentials}),
 (\ref{EstNu}) and
(\ref{PWEstG-}) we obtain
$$
\left|\partial_z(\mathbf{j}_pV_0f)(z)\right|
\le
\frac{\co}{|\l|^{\frac{4}{3}}}\frac{p_0(f^-,\l)}{\langle z\rangle^{\frac{1}{2}}},
\qquad
\left|\partial_z(\mathbf{j}_eV_0f)(z)\right|
\le
\frac{\co}{|\l|^{\frac{4}{3}}}|e^{-\frac{4}{3}z^{\frac{3}{2}}}|
\frac{p_0(f^-,\l)}{\langle z\rangle^{\frac{1}{2}}},
$$
which proves
\begin{equation}
 \label{V0G-Prime}
p_1(g_0,\l)\le\co{p_0(f^-,\l)}{|\l|^{-\frac{4}{3}}}.
\end{equation}
Finally, apply (\ref{EGM}) to $g_-$; together with
(\ref{Esthphe}) and  (\ref{V0G-})  this gives
(\ref{BEsG-}). Applying (\ref{D-EGM}) to $g_-$ together with
(\ref{Esthphe})
and (\ref{V0G-Prime}) gives (\ref{EstDerGM}).
$\blacksquare$

\no {\bf Proof of Theorem~\ref{TThmG-}. } We
consider the case
$0\le\arg\l\le\delta$, the proof for
$-\delta\le\arg\l\le0$ is similar.
Let
$v_{n+1}=\mathbf{J} V v_{n},n\ge 0$, where $ v_0\ev a$.
Introduce $v_n^\pm=v_n\big|_{\Gamma_\l^{\pm}}$.
By 
(\ref{Aqest-GL}--\ref{AqPrimeest-GL}), for some absolute constant
 $c_0>0$ we have
$$
\|{v}_{n+1}^+\|_{\a_{n+1},\beta_{n+1}}\le
\ve \|{v}_{n}^+\|_{\a_{n},\beta_{n}},
$$
where $\a_n$ and $\beta_n$ are given by (\ref{alfas-betas}).
We estimate  $v_n^-$ in terms of $p_0$ and $p_1$,
using the decomposition
 (\ref{Spac-ai-G-}) for $v_0^-$
 and
 (\ref{DecGGM}), (\ref{ABGGM}) for $v_n^-$, $n\ge1$.
Substituting $f=v_n$ and $g=v_{n+1}$ in
 (\ref{BEsG-}), (\ref{EstDerGM}) and  choosing  $c_0$ sufficiently large,
we obtain
$$
 p_0({v}_{n+1}^-,\l)\le \ve
 \lt(
p_0({v}_{n}^-,\l)
+p_1({v}_{n}^-,\l)\log (|\l|+1)
+\|{v}_{n}^+\|_{\a_{n},\beta_{n}}
 \rt),\quad
p_1({v}_{n+1}^-)\le \ve
\frac{p_0({v}_{n}^-,\l)}{|\l|^{\frac{1}{6}}}.
$$
Using (\ref{Space-ai-G+}) and  (\ref{Norm-ai-G-})
(in particular, $p_1(v_0,\l)=0$)
 we obtain for $v_1$ and $v_2$
$$
 p_0({v}_{1}^-,\l)\le \ve L,
 \qquad
p_1({v}_{1}^-,\l)\le \ve |\l|^{-\frac{1}{6}}L,
\qquad
L=2+\|v_0\|_{\frac{1}{4},\frac{5}{4}},
$$
$$
p_0({v}_{2}^-,\l)\le \ve
\lt(
\ve L +\ve L |\l|^{-\frac{1}{6}}\log (|\l|+1)
+\|{v}_{1}^+\|_{\a_{1},\beta_{1}}
\rt)\le
\ve^2(2+|\l|^{-\frac{1}{6}}\log (|\l|+1))L,
$$
$$
p_1({v}_{2}^-,\l)\le \ve^2|\l|^{-\frac{1}{6}}L.
$$
Increasing the constant $c_0$ and using the induction principle
we obtain for each integer $n\ge0$
\begin{equation}
p_0(v_{n}^-,\l)\le \ve^n
L,\quad
p_1({v}_{n}^-,\l)\le
\frac{1}{|\l|^{\frac{1}{6}}} \ve^n L,
\quad
\|v_{n}^+\|_{\a_{n},\beta_{n}}\le
\ve^n\|v_0\|_{\frac{1}{4},\frac{5}{4}}.
\label{Est-Totaal}
\end{equation}

By Theorem~\ref{ConvG+THM}, for
$\ve<1$ the series
$
v_+\Big|_{\Gamma_\l^+}=\sum_{n=0}^\infty v_n^+
$
converges in $\cF_{\frac{1}{4},1}$-norm and
gives a solution of $v_+=v_0+\mathbf{J}Vv_+$ on $\Gamma_\l^+$.
By (\ref{Est-Totaal}), for
$\ve<1$ the series
$
\sum_{n=0}^\infty p_0(v_n^-,\l)
$
and
$
\sum_{n=0}^\infty p_1(v_n^-,\l)
$ converge; for $\ve\le\frac{1}{2}$ the following estimates are fulfilled:
\begin{equation}
 \label{Est-v_+}
p_0({v}_{+}\big|_{\Gamma_\l^{-}},\l)\le 2 L,\quad
p_1({v}_{+}\big|_{\Gamma_\l^{-}},\l)\le \frac{2 L}{|\l|^{\frac{1}{6}}},\quad
p_0({v}_{1}^-,\l)\le 2 \ve L ,\quad
p_1({v}_{1}^-,\l)\le
\frac{2\ve L}{|\l|^{\frac{1}{6}}},
\end{equation}
\begin{equation}
\label{v-est-Conv}
p_0\lt({v}_+\big|_{\Gamma_\l^{-}}
-{v}_0^--{v}_1^-,\l\rt)
\le
2\ve^2L,\qquad
p_1\lt({v}_+\big|_{\Gamma_\l^{-}}
 -{v}_0^--{v}_1^-,\l\rt)\le {2}{|\l|^{-\frac{1}{6}}}\ve^2 L.
\end{equation}
By (\ref{PWEstG-}), convergence of
$
\sum_{n=0}^\infty p_0(v_n^-,\l)
$
and
$
\sum_{n=0}^\infty p_1(v_n^-,\l)
$
implies convergence of
$
\sum_{n=0}^\infty v_n^-
$
in $C^1$-norm on $\Gamma_\l^-$.
Thus for $\ve<1$ the series
$
v_+=\sum_{n=0}^\infty v_n
$ converges and solves the equation $v_+=v_0+\mathbf{J}Vv_+$ on  $\Gamma_\l$.
Using
(\ref{PWEstG-}) and (\ref{Est-Totaal}) we obtain
 (\ref{ESmallLV+}) and (\ref{ESmallLV1})
 from (\ref{Est-v_+});
 similarly we obtain (\ref{ESmallV+MimV0mimV1}) from (\ref{v-est-Conv}).

We prove uniqueness. Suppose that there exists another solution
$v_+^{(1)}\in\cF^\l=\cF_-^\l\oplus\cF_{\frac{1}{4},1}^\l$.
By Theorem~\ref{ConvG+THM}, we have
 $v_+^{(1)}|_{\G_\l^+}=v_+|_{\G_\l^+}$.
Consider the difference
$y=(v_+-v_+^{(1)})|_{\G_\l^-}\in\cF_{-}^\l$. We have
$y=(\mathbf{J}V)^n y$ for any $n>1$; applying
Lemma~\ref{BigALLTOGETHER} we obtain $    p_0(y,\l)\le \co
\ve^{n}(p_0(y,\l)+p_1(y,\l))$. Taking the limit $n\to\infty$ for
$\ve<1$ gives $y=0$. $\blacksquare$

\section{Asymptotics of the Wronskian}
\setcounter{equation}{0}
In this section we shall determine the asymptotics of
$w(\l)=\{\psi_-,\psi_+\}$ as $\l\to\infty$. To this end we find
the asymptotics of
$u_1(z,\l)=e^{-\frac{2}{3}z^{\frac{3}{2}}}(\mathbf{J}V v_0)(z)$
(given by Corollary~\ref{TThmG-Corollary}) as $\l\to\infty$ in the
sector $|\arg\l|\le\delta$. Introduce an auxiliary function
\begin{equation}
\label{Def-bq}
  E_+(\l)=2 \int_{\G_\l^-}
   \Ai(s\omega)\Ai(s\ol\o) V_q(s)\,ds,
   \qquad
   \omega=e^{\frac{2\pi i}{3}}.
\end{equation}
By (\ref{z-0}), we have
\begin{equation}
\label{Z_0^3/2+}
\frac{2}{3}z_0^{\frac{3}{2}}=\pm i\frac{\pi}{4}\lambda,
   \qquad
   0\le\pm\arg\lambda\le\pi,\qquad
   z_0=z(0,\l)
   =
-\l^{\frac{2}{3}}\left(\frac{3\pi}{8}\right)^{\frac{2}{3}}.
\end{equation}

\begin{lemma}
\label{Asu+} 1. Let $q\in\cB$ and  $|\arg\l|\le\d, |\l|\to\infty$.
Then
\begin{equation}\label{U+ALSmall}
    u_+(z_0,\l)=z_0^{-\frac{1}{4}}
    \left[
\sin\frac{\pi}{4}(\l+1)\left(1+O(\l^{-\frac{1}{6}})\right)
+E_+(\l)\cos\frac{\pi}{4}(\l+1)
    \right]+
    O\left(\frac{e^{\frac{\pi}{4}|\Im\l|}}{\l^{\frac{1}{2}}}\right),
\end{equation}
\begin{equation}\label{U+ALSmallprime}
    \partial_zu_+(z_0,\l)=z_0^{\frac{1}{4}}
    \left[
-\cos\frac{\pi}{4}(\l+1)\left(1+O(\l^{-\frac{1}{6}})\right)
+E_+(\l)\sin\frac{\pi}{4}(\l+1)
    \right]+
    O\left(\frac{e^{\frac{\pi}{4}|\Im\l|}}{\l^{\frac{1}{6}}}\right),
\end{equation}
2. Let $q\in\cB$ and  $\delta\le\pm\arg\l\le\pi$, $|\l|\to\infty$.
Then
\begin{equation}\label{AsU+Up}
    u_+(z_0,\l)=
    \frac{e^{\mp i\frac{\pi}{4}(\l+1)}}{2 z_0^{1\/4}}
    +O\left(\frac{e^{\frac{\pi}{4}|\Im\l|}}{\l^{\frac{2}{3}}}\right),
    \ \
    \partial_zu_+(z_0,\l)= -\frac{z_0^{1\/4}}{2 }
e^{\mp i\frac{\pi}{4}(\l+1)}+
O\left(\frac{e^{\frac{\pi}{4}|\Im\l|}}{\l^{\frac{1}{3}}}\right).
\end{equation}
\end{lemma}
{\it Proof.} We present the proof only for $\Im\l\ge 0$, for
$\Im
\l\le 0$ it is analogous.

1. Let
$0\le\arg\l\le\delta$. We have
$$
u_1(z,\l)=
e^{-\frac{2}{3}z^{\frac{3}{2}}}v_1(z)=
e^{-\frac{2}{3}z^{\frac{3}{2}}}(\mathbf{J}Vv_0)(z)
=
\Ai(z\ol\omega) (\mathbf{J}_pVv_0)(z) +
\Ai(z\omega)(\mathbf{J}_eVv_0)(z).
$$
By (\ref{V0G-}),  $(\mathbf{J}V_0v_0)(z)=O(\l^{-\frac{2}{3}})$.
 Since $V=V_q+V_0$,  we have as $\l\to\infty$
\begin{equation}\label{Asu1}
u_1(z,\l)=  e^{-\frac{2}{3}z^{\frac{3}{2}}}v_1(z)=
e^{-\frac{2}{3}z^{\frac{3}{2}}} (\mathbf{J}V_qv_0)(z) +O\left(
{|\l|^{-\frac{2}{3}}\langle z\rangle^{-\frac{1}{4}}}
{e^{\frac{2}{3}|\Re z^{\frac{3}{2}}|}}
\right).
\end{equation}
Using (\ref{Spac-ai-G-}), (\ref{BfromA}), (\ref{ab1}),
(\ref{Aitoai+}), (\ref{Aitoai-}) and (\ref{IntroH}--\ref{GPE})
 we obtain
$$
 e^{-\frac{2}{3}z^{\frac{3}{2}}}
(\mathbf{J}V_qv_0)(z)=
\Ai(z\ol\omega) (\mathbf{J}_pV_qv_0)(z) +
\Ai(z\omega)(\mathbf{J}_eV_qv_0)(z)
$$
$$
= 2 i \omega\Ai(z)
 \int_{\Gamma_\l^+}
\ai(s \omega)\ai(s)V_q(s)\, ds
- 2 i \Ai(z \omega)
 \int_{\Gamma_\l^+}
e^{-\frac{4}{3}s^{\frac{3}{2}}}\ai^2(s)V_q(s)\, ds
$$
$$
+ 2\Bi(z)
\int_{\Gamma_\l(z,z_*)}
\ai(s \omega)\ai(s \ol\omega)V_q(s)\, ds -
2 i \ol\omega\Ai(z
\ol\omega)
 \int_{\Gamma_\l(z,z_*)}
e^{\frac{4}{3}s^{\frac{3}{2}}}
\ai^2(s\omega)V_q(s)\, ds
$$
$$
+ 2 i e^{i\frac{\pi}{3}}
\Ai(z\omega)
 \int_{\Gamma_\l(z,z_*)}
e^{-\frac{4}{3}s^{\frac{3}{2}}}\ai^2(s
       \ol\omega) V_q(s)\, ds,
$$
where $\omega=e^{\frac{2\pi i}{3}}$. Now we set $z=z_0$
 and write the asymptotics of
$u_1$ in terms of $E_+$.
By Lemma~\ref{Ostatkiu1},
(\ref{Aitoai+}), (\ref{Aitoai-}) and (\ref{aiest}), we have
\begin{equation}
\label{u1-Prel}
u_1(z_0,\l) = 2i \omega\Ai(z_0)
\int_{\Gamma_\l^+}
\ai(s
\omega)\ai(s)V_q(s)\, ds
      + \Bi(z_0)E_+(\l)
      +O\Bigl(
 {\lambda^{-\frac{1}{3}}z_0^{-\frac{1}{4}}}
 {\abs{e^{-\frac{2}{3}z_0^{\frac{3}{2}}}}}
      \Bigr).
\end{equation}
Using (\ref{aiprimest}) we obtain for the derivative
\begin{equation}
\label{u1-Prelprime}
\partial_zu_1(z,\l) = 2 i \omega\Ai'(z_0)
 \int_{\Gamma_\l^+}
\ai(s \omega)\ai(s)V_q(s)\, ds
+
        \Bi'(z_0)E_+(\l)
       +O\Bigl(
{\lambda^{-\frac{1}{3}}} {z^{\frac{1}{4}}}
      \abs{e^{-\frac{2}{3}z^{\frac{3}{2}}}}
      \Bigr).
\end{equation}
From Lemma~\ref{AsJ1}, (\ref{Aitoai+}), (\ref{aiest}) and (\ref{aiprimest})
we deduce  that\\
$
 \int_{\Gamma_\l^+}
 \ai(s \omega)\ai(s)V_q(s)\,
ds=O(\l^{-\frac{1}{6}})
$.
Therefore using (\ref{hvostu+}), (\ref{hvostu+prime}),
(\ref{Z_0^3/2+}), (\ref{Asu1}),
 (\ref{u1-Prel}) and
(\ref{u1-Prelprime}) we obtain
\begin{equation}
\label{Asu+-Ai} u_+(z_0,\l)=\Ai(z_0)(1+O(\l^{-\frac{1}{6}}))+ \Bi(z_0)
E_+(\l)+
O\left({\l^{-\frac{1}{2}}}{e^{\frac{\pi}{4}|\Im\l|}}\right),
\end{equation}
\begin{equation}
\label{Asu+-Aiprime}
\partial_z u_+(z_0,\l)=\Ai'(z_0)(1+O(\l^{-\frac{1}{6}}))+
\Bi'(z_0)E_+(\l)
+O\left({\l^{-\frac{1}{6}}}{e^{\frac{\pi}{4}|\Im\l|}}\right),
\end{equation}
where $z_0$ is given by (\ref{z-0}).
Recall the standard uniform asymptotics of Airy functions \cite{AS}
in the sector
$\abs{\arg z}<{\pi\/3}-\ve$, $\ve>0$, as $z\to\infty$:
\begin{equation}
 \label{Asaire-1}
 \Ai(-z)=z^{-\frac{1}{4}}(\sin \e+O(F(z)),\ \ \ \
\Ai'(-z)=-z^{\frac{1}{4}}(\cos \e+O(F(z))),
\end{equation}
\begin{equation}\label{Asaire-4}
\Bi(-z)=z^{-\frac{1}{4}}(\cos\e+O(F(z)),\ \ \ \
\Bi'(-z)=z^{\frac{1}{4}}(\sin \e+O(F(z))),
\end{equation}
where
$\e={2\/3}z^{3\/2}+{\pi\/4}$ and $F(z)=z^{-{3\/2}}e^{{2\/3}|\Im
z^{3\/2}|}$.
Note that for $0\le\arg\l\le\delta$ we have $-\pi\le\arg z_0\le -\pi+\frac{2}{3}\delta$.
Thus we substitute (\ref{Asaire-1}) and (\ref{Asaire-4}) into
(\ref{Asu+-Ai}--\ref{Asu+-Aiprime}). This gives
 (\ref{U+ALSmall}),(\ref{U+ALSmallprime}).

2. Let $\delta\le\arg\lambda\le\pi$.  By (\ref{U+BigS}), we have
\begin{equation}
\label{WideSector} u_+(z_0,\l)=\Ai(z_0)+
O\left({\l^{-\frac{2}{3}}{e^{\frac{\pi}{4}|\Im\l|}}}\right),\quad
u_+'(z_0,\l)=\Ai'(z_0)+O\left({\l^{-\frac{1}{3}}{e^{\frac{\pi}{4}|\Im\l|}}}\right).
\end{equation}
Note that for $\delta\le\arg\l\le\pi$ we have $-\pi+\frac{2}{3}\delta\le\arg z_0\le 0$.
Thus we apply  (\ref{AsAiry}) and (\ref{Z_0^3/2+}) to
(\ref{WideSector}), which gives (\ref{AsU+Up}).$\blacksquare$

Introduce the function
\begin{equation}\label{Defphi}
\phi(\l)=
2^{\frac{3}{4}}
\left(\frac{\l}{2e}\right)^{\l/4},
\quad
\l\in\mathbb{C}\setminus\mathbb{R}_-,\quad
\qquad \phi(\l)>0\quad \hbox{for} \quad \l>0.
\end{equation}
In the next Lemma we write $\psi_+$, defined by (\ref{AnAsDef}), in terms of
$u_+$.

\begin{lemma}
\label{SvyazUPhi}
Let $q\in\cB$. Let $\psi_+$ be the solution of Eq.(\ref{OurEq}),
satisfying (\ref{AnAsDef}); let $u_+$ be the solution of
Eq.(\ref{IntegralEqinZ}), given by Corollary~\ref{TThmG-Corollary}.
Then for
$
|\l|^{{1\/6}}\ge c_0\left(
\|q\|_{\cB}+1
\right)
$, where $c_0\ge1$ is some absolute constant,  the
following identity holds:
\begin{equation}
\label{Psi+ViaU+}
\psi_+(x,\l)=\phi_+(x,\l),
\quad
where
\quad
\phi_+(x,\l)=
\phi(\l) {u_+(z(x,\l)),\l)\/
\sqrt{z'(x,\l)}},\qquad x\ge 0.
\end{equation}
\end{lemma}
{\it Proof. }
The function $\phi_+$, given by
(\ref{Psi+ViaU+}),  solves Eq.(\ref{OurEq}) by changing
variables according to (\ref{DefZ}). In order to prove
(\ref{Psi+ViaU+}) it is sufficient to demonstrate that $\phi_+$ has
the asymptotics (\ref{AnAsDef}).
Using (\ref{k-from-t}), (\ref{DefZ}), (\ref{AsAiry}),
Corollary~\ref{TThmG-Corollary}
and \eqref{Z_0^3/2+}, we have for
$x\to\infty$
$$
\phi_+(x,\l)
=e^{-\frac{x^2}{2}}
(\sqrt{2}x)^{\frac{\lambda-1}{2}}
\Bigl(1+O\Bigl(\frac{1}{x^{\frac{1}{3}}}\Bigr)\Bigr),
\quad
\frac{\partial \phi_+(x,\l)}{\partial x}
=
-\frac{e^{-\frac{x^2}{2}}}{\sqrt{2}}(\sqrt{2}x)^{\frac{\lambda+1}{2}}
\Bigl(1+O\Bigl(\frac{1}{x^{\frac{1}{3}}}\Bigr)\Bigr),
$$
which proves (\ref{Psi+ViaU+}).
$\blacksquare$

In order to obtain the asymptotics of the Wronskian
$w(\l)=\{\psi_-,\psi_+\}$ we need also the asymptotics
 of the fundamental solution
$\psi_-$ (see Theorem~\ref{Tan}).
We use our results for $x>0$ and consider the
reflected potential
$
q_-(x)=q(-x)$ for $ x\in\mathbb{R}_+
$. We define
$$
V_{q}^-(z)=\l^{-{1\/3}}\r^2(z,\l) \hat{q}_-(z),\qquad
\hat{q}_-(z,\l)\ev q_-(\sqrt{\l}t({z}{\l^{-\frac{2}{3}}})),
$$
where $\rho$ is given by (\ref{EffPotentials}).
 Let $u_-(z,\l)=u_+(z,\l;V_q^-)$,
where $u_+$ is given by Corollary~\ref{TThmG-Corollary}
(with $V_q$ replaced by $V_q^-$).
 By Lemma~\ref{SvyazUPhi},  the fundamental solution $\psi_-$
 is related to $u_-$ by
\begin{equation}\label{TildeUToPsi-}
    \psi_-(-x,\l)=
    \phi(\l)
    \frac{u_-(z(x,\l),\l)}{ \sqrt{z'(x,\l)}}
    ,\qquad
    x\ge0.
\end{equation}
Introduce $E_-(\l)$  by
\begin{equation}
E_-(\l)=2\int_{\G_\l^-}
   \Ai(s\omega)\Ai(s\ol\o) V_q^-(s)\,ds,
   \qquad
\omega=e^{\frac{2\pi i}{3}}.
\end{equation}
By Lemma~\ref{Asu+}, for
$|\arg\l|\ge\delta$ the solution $u_-(z_0,\l)$
has the asymptotics (\ref{AsU+Up}); for
$|\arg\l|\le\delta$ it has the asymptotics (\ref{U+ALSmall}),
(\ref{U+ALSmallprime}) with $E_+$ replaced by $E_-$:
\begin{equation}\label{U+ALSmalltilde}
    u_-(z_0,\l)={ z_0^{-\frac{1}{4}}}
    \left[
\sin\frac{\pi}{4}(\l+1)\left(1+O(\l^{-\frac{1}{6}})\right)
+E_-(\l)\cos\frac{\pi}{4}(\l+1)
    \right]+
    O\left(\l^{-\frac{1}{2}}{e^{\frac{\pi}{4}|\Im\l|}}{}\right),
\end{equation}
\begin{equation}\label{U+ALSmalltildeprime}
    u_-'(z_0,\l)=z_0^{1\/4}
    \left[
-\cos\frac{\pi}{4}(\l+1)\left(1+O(\l^{-\frac{1}{6}})\right)
+E_-(\l)\sin\frac{\pi}{4}(\l+1)
    \right]+
    O\left(\l^{-\frac{1}{6}}{e^{\frac{\pi}{4}|\Im\l|}}{}\right).
\end{equation}
Below we use the function $E=E_-+E_+$.
\begin{lemma}
\label{LemmaAsympWronskian}
Let $q\in\cB$. Then for $|\l|\to\infty$ the following uniform
asymptotics hold:
\begin{equation}
\label{AsWsmallAL}
 w(\l)=\f^2(\l)
\left[ \cos\frac{\pi}{2}\l-E(\l)\sin\frac{\pi}{2}\l+
O\left(\l^{-\frac{1}{3}}{e^{\frac{\pi}{2}|\Im\l|}}\right) \right],\
\ \ \ \ |\arg\l|\le\d
\end{equation}
\begin{equation}
\label{AsWbigAL}
 w(\l)=\frac{\phi^2(\l)}{2} \left[  e^{\mp
i\frac{\pi}{2}\l} +
O\left(\frac{e^{\frac{\pi}{2}|\Im\l|}}{\l^{\frac{1}{6}}}\right) \right]
={\phi^{-2}(-\l)}
\left(4+O({\lambda^{-\frac{1}{6}}})\right),\
\ \ \ \ \ \d\le\pm\arg\l\le\pi.
\end{equation}
\end{lemma}


{\it Proof. }  Using (\ref{DefZ}), (\ref{Psi+ViaU+}) and (\ref{TildeUToPsi-})
we obtain
$$
w(\l)=-\phi^2(\l)\lt(
 u_+(z_0,\l)u_-'(z_0,\l)+u_+'(z_0,\l)u_-(z_0,\l)
\rt)
-
 \phi^2(\l)\frac{u_+(z_0,\l)u_-(z_0,\l)}{16\l k^2(0)}.
$$
 Substituting (\ref{U+ALSmall}--\ref{U+ALSmallprime})
  and (\ref{U+ALSmalltilde}--\ref{U+ALSmalltildeprime})
  in the last identity,  we obtain (\ref{AsWsmallAL});
 substituting (\ref{AsU+Up}) and
 the same formulae for $u_-$, we obtain (\ref{AsWbigAL}).
$\blacksquare$

\section{The proof of Theorem~\ref{T1}}
\setcounter{equation}{0}

Using (\ref{Defphi}) we write the following uniform
asymptotics of the unperturbed Wronskian
$ w_0(\l)=\{\psi_-^0,\p_+^0\}=-\frac{2\sqrt{\pi}}{\Gamma(\frac{1-\l}{2})}$
(see, for example, \cite{AS}):
\begin{align}
\label{Asw-narr-0}
w_0(\lambda)&=\phi^2(\l)
\cos\frac{\pi\lambda}{2}\cdot\left(1+O\left({\lambda^{-1}}\right)\right),
&  \abs{\arg\lambda}&\le\delta
 \\
 \label{Asw-wide-0}
w_0(\lambda)&={\phi^{-2}(-\l)}
\left(4+O\left({\lambda^{-1}}\right)\right),
&  \abs{\arg\lambda}&\ge\delta.
\end{align}

\begin{lemma}
\label{LabelLem}
Let  $q\in\cB$. Then there is $N_0\in\mathbb{Z}$  such that for
each integer $N>N_0$  the operator $T$
 has exactly $N$ simple eigenvalues
in the disc $\{z: |z|<2N\}$ and for each $n>N$, exactly one simple
eigenvalue in the disc
$\{z: |z-\mu_n^0|<{n^{-\frac{1}{6}}}\}.$  There are no other eigenvalues.
\end{lemma}
\no {\it Proof.}
Consider the contours
$
|\l|=2n$, $ \abs{\lambda-\mu_n^0}=\delta_n$, $n>N$,
$n\in\mathbb{N}$, where
$\delta_n={n^{-\frac{1}{6}}}<\frac{\log 2}{\pi}$.
Then
$|\cos\frac{\pi\lambda}{2}|\ge{4}e^{\frac{\pi}{2}\abs{\Im\lambda}}$
for
$|\l|=2n$ and
$
\left|\frac{1}{\pi}\cos\frac{\pi\lambda}{2}\right|
\ge\frac{\delta}{4}e^{\frac{\pi}{2}\abs{\Im\lambda}}$ for
$\abs{\lambda-\mu_n^0}=\delta$,  $\delta<\frac{\log 2}{\pi}$.
 By the asymptotics  (\ref{AsWsmallAL}), \eqref{AsWbigAL},
(\ref{Asw-wide-0}) and (\ref{Asw-narr-0}), there exist
 integer $N_0>\left(\frac{\pi }{\log
2}\right)^6$ such that for any integer $N>N_0$  on these contours
$
\abs{w(\lambda)-w_0(\lambda)}\le \frac{1}{2}\abs{w_0(\lambda)}
$.
It follows that $w(\l)$ does not vanish on these contours.
Hence, by Rouche's theorem, $w(\l)$ has as many roots,
counted with multiplicities, as $w_0(\l)$ in each of the bounded
regions and in the remaining unbounded region. Since $w_0(\l)$ has
only the simple roots $\{\mu_n^0\}_{n=0}^\infty$, the Lemma is proved.
$\BBox$

\begin{lemma}
\label{Est-E}
Let $q\in\cB$.  Then for $|\l|\to \infty$ the following
asymptotics are fulfilled:
\begin{equation}
\label{AsDecrE}
 E(\l+\ve)-E(\l)=O\left({\l^{-\frac{1}{3}}}\right),
 \qquad
\ve=O\left({\l^{-\frac{1}{6}}}\right),
\qquad
-\frac{\delta}{2}\le\arg\l\le\frac{\delta}{2},
\end{equation}
\begin{equation}
\label{AsFormERealPos}
  E(\l)=-\frac{1}{2}
\int_{-1}^{1}\frac{q(t\sqrt{\l})dt}{\sqrt{1-t^2}}+
O\left({\|q\|_\cB}{\l^{-\frac{1}{3}}}\right),\qquad
E(\l)=O\left({\|q\|_\cB}{\l^{-\frac{1}{4}}}\right),
\qquad \l>0.
\end{equation}
\end{lemma}
{\it Proof.} Consider  the case $0\le\arg\l\le\delta$; for
$-\delta\le\arg\l\le0$ the proof is analogous.
 Consider   $E_+$, given by (\ref{Def-bq}).  Recall that
$t_*=\frac{x_*}{\sqrt{\lambda}}$, where
$x_*$ is defined by
$z_*=z(x_*,\lambda)$ (see Lemma~\ref{Z-star} and below). We have $\l=|\l|e^{2i\vt}$. For some $c\in[0,1]$ we set
$t_1=t_*-\frac{c}{\abs{\lambda}^{\frac{2}{3}}}e^{-i\vt}$,
$x_1=\frac{t_1}{\sqrt{\l}}$ and $z_1=z(x_1,\l)$.
The length of $\Gamma_\l(z_1,z_*)$ is
$|\Gamma_\l(z_1,z_*)|=|\l|^{\frac{2}{3}}\int_{[t_1,t_*]}|k'(t)|\,|dt|$.
By \refLem{A7.3-4}{DivG2}, $|t_*|$ is bounded uniformly in $\l$.
Thus using (\ref{k-from-t}) we conclude that
$|\Gamma_\l(z_1,z_*)|\le C$.
 Therefore using (\ref{EffPotentials}), (\ref{EstNu}),
  (\ref{aiest}), (\ref{Aitoai+}) and
 (\ref{Aitoai-})
 gives
\begin{equation}
\label{|G|}
\left|
\int_{\Gamma_\l(z_1,z_*)}
\Ai(s\ol\omega)\Ai(s\omega)V_q(s)\,ds
\right|\le \co\cdot
\frac{\norm{q}_\infty}{\abs{\lambda}^{\frac{1}{3}}}.
\end{equation}
Next, using the asymptotics (\ref{AsAiry}) and
\refLem{A7.5-1}{SvaGForAll}, we have uniformly in
$|\arg\l|\le\delta$
\begin{equation}
\label{A+A-}
\left\lvert\Ai(z\omega)\Ai(z\ol\omega)+
\frac{i}{4\pi z^{\frac{1}{2}}}\right\rvert\le
\frac{\co}{\abs{z}^{\frac{1}{2}}(1+\abs{z})^{\frac{3}{2}}},
\qquad
z\in\Gamma_\l^-\subset S[\pi,-\frac{\pi}{3}].
\end{equation}
Substituting (\ref{|G|}) and (\ref{A+A-}) in (\ref{Def-bq}), we have
\begin{equation}
E_+(\lambda)=2\pi
\int_{\Gamma_\l(z_0,z_1)}
\!\!\!\!\!\!\!\!\!\!\!\!\!\!\!
\Ai(s\omega)\Ai(s\ol\omega)V_q(s)\,ds+O
\left(\frac{\norm{q}_\infty}{\abs{\lambda}^{\frac{1}{3}}}\right)
=-\frac{i}{2}
\int_{\Gamma_\l(z_0,z_1)}
\!\!\!\!\!\!\!\!\!\!\!\!\!\!\!
{V_q(s)}s^{-\frac{1}{2}}\,ds+
O\left(\frac{\norm{q}_\infty}{\abs{\lambda}^{\frac{1}{3}}}\right).
\end{equation} 
Making the substitution $s=k(t)\cdot\lambda^{\frac{2}{3}}$,
$ds=\lambda^{\frac{2}{3}}k'(t)\,dt$ and using (\ref{EffPotentials}) we obtain
\begin{equation}
E_+(\lambda)= -\frac{i}{2}\int_{e^{-i\vt}[0,t_1]}
\frac{q(\sqrt{\lambda}t)}{\sqrt{k(t)}k'(t)}\,dt +
O\left(\frac{\norm{q}_\infty}{\abs{\lambda}^{\frac{1}{3}}}\right).
\end{equation}
Using  (\ref{DefXi}) and (\ref{ZofT}) we have
$
k'(t)\sqrt{k(t)}=\sqrt{t^2-1}=i\sqrt{1-t^2}
$,
where the  last root is positive for $-1<t^2<1$. Hence
$$
E_+(\l) =-\frac{1}{2}\int_0^d\frac{q(t\abs{\l}^{\frac{1}{2}})}
{\sqrt{e^{2i\vt}-t^2}}\, dt+
O\left(\frac{\norm{q}_\iy}{\abs{\l}^{\frac{1}{3}}}\right),\ \ \ \
d=1-{c\/|\l|^{{2\/3}}} .
$$
Using similar arguments for $E_-$, we obtain
\begin{equation}
\label{AsFormE}
 E(\l)=E_c(\l)
+ O\left(\frac{\|q\|_\cB}{\l^{\frac{1}{3}}}\right), \qquad
E_c(\l)=-\frac{1}{2}
\int_{-d}^d\frac{q(t\sqrt{|\l|})dt}{\sqrt{e^{2i\vt}-t^2}}.
\end{equation}

In order to prove \eqref{AsDecrE} we set $c>0$.  We estimate the
partial derivatives of $E_c(\l)$ with respect to real and complex
parts of
$\lambda=\mu+i\nu=|\l|e^{2i\vt}$.
This gives
$\abs{\frac{\partial}{\partial \mu}E_c(\lambda)}\le
\co\frac{\norm{q}_B}{\abs{\lambda}^{\frac{2}{3}}}$,
$\abs{\frac{\partial}{\partial \nu}E_c(\lambda)}\le
\co\frac{\norm{q}_B}{\abs{\lambda}^{\frac{2}{3}}}$.
Therefore as $\l\to\infty$
$$
\abs{E_c(\lambda+\ve)-E_c(\lambda)}\le\abs{\ve}\cdot
\sup_{|\l-\l'|\le\ve} \abs{{\nabla}E_c(\l')}\le
\co\|q\|_{\cB}|\l|^{-{5\/6}},\qquad
\ve=O\left({\abs{\lambda}^{-\frac{1}{6}}}\right),
$$
which together with
\eqref{AsFormE} proves \eqref{AsDecrE}.

Setting $c=0$ and $\l>0$ in (\ref{AsFormE})  proves the first
asymptotics in (\ref{AsFormERealPos}). In order to prove the
second one we use the decomposition
$E_0(\l)=E_1(\l)+I_2(\l)$, $\l>0$, where
$$
I_2(\l)=
-\frac{1}{2}
\int_{1-\l^{-\frac{1}{2}}}^{1}\frac{q(t\sqrt{\l})+q(-t\sqrt{\l})}{\sqrt{1-t^2}}dt.
$$
Using
$\int_{\l^{\frac{1}{2}}t}^{\l^{\frac{1}{2}}-1}q(x)\,
dx=q_1(\l^{\frac{1}{2}}-1)-q_1(\l^{\frac{1}{2}}t)$
we integrate the expression for
 $E_1$ by parts.
This gives $|E_1(\l)|\le
C\frac{\|q_1\|_\infty}{\l^{\frac{1}{4}}}$.
 Direct estimate
gives $|I_2(\l)|\le\frac{\|q\|_\infty}{2\l^{\frac{1}{4}}}$. Combining these
 estimates
 proves the second relation in (\ref{AsFormERealPos}).
$\blacksquare$

{\bf Proof of Theorem \ref{T1}}
By Lemma~\ref{LabelLem}, in each disc
$D_n=\{\l:\abs{\lambda-\m_n^0}\le\frac{1}{n^{\frac{1}{6}}}\}$
there exists exactly one simple eigenvalue $\mu_n$ for $n$
sufficiently large; now we improve this estimate. Using the
asymptotics (\ref{AsWsmallAL}) and
$\abs{\sin(\frac{\pi}{2}\lambda)}\ge \frac{1}{2}$ for $\l\in D_n$, we obtain
\begin{equation}
\label{W-approx}
\cot(\frac{\pi}{2}\mu_n)+E(\mu_n)+
O\left(n^{-\frac{1}{3}}\right)=0,\qquad n\to\infty.
\end{equation}
Using (\ref{AsFormERealPos}), we  write
\begin{equation}
\label{Mu-approx}
\mu_n-\mu_n^0=-\frac{2}{\pi}E(\mu_n^0)+\ve_n,
\qquad
\ve_n=O({n^{-\frac{1}{6}}}),\qquad
n\to\infty.
\end{equation}
Substituting \eqref{Mu-approx} into \eqref{W-approx} we have
\begin{equation}
\frac{\pi}{2}\ve_n
+E(\mu_n)-E(\mu_n^0)=O\left({n^{-\frac{1}{3}}}\right).
\end{equation}
By \eqref{AsDecrE},
$E(\mu_n)-E(\mu_n^0)=O({n^{-\frac{1}{3}}})$.
Thus in (\ref{Mu-approx}) the error term
$\ve_n$ is $O({n^{-\frac{1}{3}}})$, which  yields
$\mu_n=\mu_n^0-\frac{2}{\pi}E(\mu_n^0)+O(n^{-\frac{1}{3}})$.
The change of variable $t=\sin\theta$ gives
$\mu_n^1=-\frac{2}{\pi}E(\mu_n^0)=
(2\pi)^{-1}\int_{-\pi}^{\pi}q(\sqrt{\mu_n^0}\sin\theta)\,d\theta$.
This proves
(\ref{MainAsymptotics}). $\BBox$

\no {\bf Proof of Proposition \ref{C1}.}
Substituting $q(x)=\int_{\R}e^{ixt}\,d\nu(t)$ into (\ref{AsFormE})
and  using the identity for Bessel function
$J_0(z)={1\/\pi}\int_{-\frac{\pi}{2}}^{\frac{\pi}{2}}e^{iz\sin\phi}\,d\f$
(see \cite{AS}) we have
\begin{equation}
\label{muMainProof}
{1\/2\pi}\int_{-\pi}^{\pi}q(\sqrt{\l}\sin\vt)\,d\vt
=\int_{\R}J_0(t\sqrt \l)\,d\nu(t)=I_1+I_2,
\qquad \l>0,
\end{equation}
 where
\begin{equation}
\label{I1I2MainProof}
I_1=\int_{|t|<\ve}J_0(t\sqrt \l)\,d\nu(t), \qquad
I_2=\int_{|t|>\ve} J_0(t\sqrt \l)\,d\nu(t),\ \ \
\ve=\l^{-{3\/4p}},\ \ {3\/4p}<{1\/2}.
\end{equation}
Using $J_0(-z)=J_0(z)$ and the asymptotics (see \cite{AS})
$$
J_0(z)=\sqrt{\frac{2}{\pi
z}}\cos\lt(z-\frac{\pi}{4}\rt)+O\lt(\frac{e^{\abs{\Im
z}}}{z^{\frac{3}{2}}}\rt),\ \ \ \ \ |\arg z|\le{\pi\/2},
$$
we obtain
\begin{equation}
I_2={1\/\l^{\frac{1}{4}}}\int_{|t|>\ve}\sqrt{\frac{2}{\pi
|t|}}\cos\left(\sqrt{\l}|t|-{\pi\/4}
\right)d\nu(t)+{O(1)\/\l^{\frac{3}{4}}}\int_{|t|>\ve} {d\nu(t)\/t^{3\/2}},
\end{equation}
where the last term is $O(\l^{-\frac{3}{4}})$. Next,
\begin{equation}
|I_1|\le C\int_{|t|<\ve} {d\nu(t)}\le
C\ve^p\int_{|t|<\ve}{d\nu(t)\/|t|^p}=C\g\ve^{p}=O(\l^{-\frac{3}{4}}).
\end{equation}
Similarly we have
\begin{equation}
\label{LastEstMainProof}
\left|
\l^{-{1\/4}} \int_{|t|<\ve}\sqrt{{2\/\pi t}}
\cos\left(|t|\sqrt{\l}-\frac{\pi}{4}\right)\,d\nu(t)
\right|
 \le C\l^{-{1\/4}}
\int_{|t|<\ve}{d\nu(t)\/\sqrt{|t|}}\le  C\l^{-{1\/4}}
\ve^{p-{1\/2}}=O(\l^{-{3\/4}}).
\end{equation}
Using (\ref{muMainProof}--\ref{LastEstMainProof}) and setting
$\l=\mu_n^0$ gives
$
\mu_n^1=\frac{\sigma(\sqrt{\mu_n^0})}{(\mu_n^0)^{\frac{1}{4}}} +
O(n^{-{3\/4}})
$
which implies
\er{asft}.  Moreover, substituting
 $d\nu(t)=\sum_{k\in \Z}\d(t-t_k)q_kdt$ into \er{asft1}
 we obtain \er{ZonEndsAsympt-1}.
 $\BBox$

\section{Appendix}
\renewcommand{\theequation}{A.\arabic{equation}}
\setcounter{equation}{0}

For fixed $\arg\l=2\vt$ we have $t=\frac{x}{\sqrt{\l}}\in
e^{-i\vt}\R_+$.
We rewrite $t\in S[-\frac{\pi}{2},0]$  and $\x$ in the form
\begin{equation}
\label{AngularXi} t=re^{-i\vartheta}=1+\eta e^{-i\varphi}, \qquad
\varphi\in[0,\pi], \quad \eta\ge0,
\end{equation}
\begin{equation}
\label{ChVarXi}
\x(t)
=e^{-i\frac{3\varphi}{2}}\int_0^\eta \sqrt{2+se^{-i\varphi}}\cdot
s^{\frac{1}{2}}\,ds,\ \ \ t\in S[-\frac{\pi}{2},0].
\end{equation}

\begin{lemma} \label{Geom-0}
 Let $t\in S[-\frac{\pi}{2},0]$,
$R(t)=|\xi(t)|$,
$\Phi(t)=\arg \xi(t)$.
Then
\begin{enumerate}
\item\label{A7-1}
$t=re^{-i\vt}=1+\eta e^{-i\varphi}$, where $r,\eta\ge0$,
$\vt\in[0,\frac{\pi}{2}]$, $\varphi\in[0,\pi]$,

\item \label{A7-2}
$
\Phi(t)+\frac{3\varphi}{2}\in
\left[
-\frac{\vartheta}{2},0
\right]$,

\item \label{A7-3} if $\eta\le1$, then
$
\frac{2}{3} |t-1|^{\frac{3}{2}}\le R(t)\le  2
|t-1|^{\frac{3}{2}}
$,\\
if $\eta\ge0$, then $\frac{2}{3}\sin^{\frac{3}{2}}\vt\le R(t)$,

\item \label{A7-4}
if $\vt\in(0,\frac{\pi}{2}]$, then
$\bigcup\limits_{r\in\R_+}
\Phi(t)=[-\frac{3\pi}{2},-2\vt)$,

\item \label{A7-5}
 if
$-\pi-\vt\le\Phi(t)
$,
then
$\arg\left( e^{2i\vt}{\partial_r}
\xi(t)\right)\in[-\frac{\pi}{3}+\frac{\vt}{6},\frac{\vt}{2}]$.
\end{enumerate}
\end{lemma}

\no {\it Proof.} The proof of \ref{A7-1} is simple.
 \ref{A7-2}. Consider the integrand in formula (\ref{ChVarXi}).
 For $s\in[0,\eta]$ we have
$\arg\sqrt{2+se^{-i\varphi}}\in[-\frac{\vartheta}{2}, 0] $, hence
$ \Phi(t)+\frac{3\varphi}{2}= \arg \left(\int_0^\eta
\sqrt{2+se^{-i\vp}}\cdot s^{\frac{1}{2}}\,ds\right)\in
[-\frac{\vt}{2},0] $.

 \ref{A7-3}.  Consider the integrand in (\ref{ChVarXi}). For
$s\in[0,\eta]$ we have $ 1<\re\sqrt{2+se^{-i\varphi}} $ and $
\abs{\sqrt{2+se^{-i\varphi}}}<3 $. Using $R(t)=\left| \int_0^\eta
\sqrt{2+se^{-i\varphi}}\cdot s^{\frac{1}{2}}\,ds \right|$ gives
$$
\frac{2}{3}\eta^{\frac{3}{2}}\le R(t)
<3\cdot\frac{2}{3}\eta^{\frac{3}{2}},
\qquad
\eta=|t-1|.
$$
The relation
$\min\limits_{\arg t=-\vt}\abs{t-1}=\sin\vartheta$ for
$\vartheta\in[0,\frac{\pi}{2}]$ finishes the proof.

\ref{A7-4}. Fix $\vt\in(0,\frac{\pi}{2}]$. By direct calculation,
 $\xi(S[-\frac{\pi}{2},0])\subset S[-\frac{3\pi}{2},0]$,
 so $\bigcup\limits_{r\in\R_+}
\Phi(t)\subset S[-\frac{3\pi}{2},0]$. Using
\eqref{asympXi} we have  $\arg \left(e^{2i\vartheta}\xi(t)\right)\to 0$ as
$r\to\infty$. Also we have $\xi(0)=i\frac{\pi}{4}$. Since $\xi(t)$ is continuous,
 this gives
$\bigcup_{r\in\R_+} \Phi(t)\supset[-\frac{3\pi}{2},-2\vt)$. Thus
we need only prove that $\arg
\left(e^{2i\vartheta}\xi(t)\right)<0$.

Consider $e^{2i\vartheta}\xi(t)$, $t=re^{-i\vartheta}$, as
a function of real parameter $r\in\mathbb{R}$.
Note that  by \refLem{A7.2-2}{23},
  $\Im \left(e^{2i\vartheta}\xi(t)\right)$
    strictly decreases in $r$.
Since  $\xi( S[-\frac{\pi}{2},0])\subset S[-\frac{3}{2}\pi,0]$ and $\xi(0)=i\frac{\pi}{4}$,
as $r$ increases $e^{2i\vt}\xi(t)$
  hits only the negative half of the imaginary axis.
Therefore, as soon as $\arg \left(e^{2i\vt}\xi(t)\right)>-\frac{\pi}{2}$, we have
    $\Im \left(e^{2i\vartheta}\xi(t)\right)<0$. Hence
    $\arg
\left(e^{2i\vartheta}\xi(t)\right)<0$,  which
finishes the proof.

\ref{A7-5}.  By (\ref{ChVarXi}), we have
$e^{2i\vartheta}{\partial_r}
\xi(t)=
e^{-i\frac{\varphi}{2}}e^{i\vartheta}\sqrt{2\eta}
\sqrt{1+\frac{\eta e^{-i\varphi}}{2}}$,
where
$\sqrt{1+\frac{\eta e^{-i\varphi}}{2}}\in S[-\frac{\vartheta}{2},0]$.
Therefore
\begin{equation}
\label{Geom-5}
\arg
\left(
e^{2i\vartheta} {\partial_r}
\xi(t)
\right)
\in[-\frac{\varphi}{2}+\frac{\vartheta}{2},-\frac{\varphi}{2}+\vartheta].
\end{equation}
Next, by hypothesis,  $\Phi\in[-\pi-\theta, -2\vartheta]$.
Thus using \ref{A7-2}  we obtain
$-\varphi\in[-\frac{2}{3}(\pi+\vartheta),
-\vartheta]$.
Substituting this into (\ref{Geom-5})
 proves \ref{A7-5}.
$\blacksquare$

\begin{lemma}
\lb{23}
Let $t=re^{-i\vt}\in S[-\frac{\pi}{2},0]$, $R(t)=|\xi(t)|$,
$\Phi(t)=\arg \xi(t)$. Then
\begin{enumerate}
\item\label{A7.2-1}
$\arg{\partial_r}\xi(t)
\in [-\frac{\pi}{2}-\vt,-2\vt)$,

\item \label{A7.2-2}
 if $\vt\in(0,\frac{\pi}{2}]$ and $r\ge0$
 then
$\Im\left( e^{2i\vt}{\partial_r}\xi(t) \right)<0$
and
$\Re\left( e^{2i\vt}{\partial_r}\xi(t) \right)>0$,

\noindent
if $\vt=0$ and  $r\in[0,1)$, then
$\Im\left( e^{2i\vt}{\partial_r}\xi(t) \right)<0$
and $\Re\xi(t)=0$,

\noindent
if $\vt=0$  and $r\in(1,\infty)$, then $\Im\xi(t)=0$ and
$\Re\left( e^{2i\vt}{\partial_r}\xi(t) \right)>0$,

\item  \label{A7.2-3}
if
$\Phi(t)\in (-\pi-2\vt,-\frac{\pi}{2}-\vt)$, then
$
{\partial_r}\Phi(t)>0$,

\item  \label{A7.2-4}
if
$\Phi(t)\in (-\frac{3\pi}{2}-2\vt,-\pi-\vt)$, then
${\partial_r}R(t)<0$,

\item  \label{A7.2-5}
if
$\Phi(t)\in (-\frac{\pi}{2}-2\vt,-\vt)$, then
${\partial_r}R(t)>0$.

\end{enumerate}
\end{lemma}
%

{\it Proof.}
\ref{A7.2-1}. By direct calculation,
$
{\partial_r}\xi(t)= e^{-i\vt}\xi'(t)=e^{-i\vt}\sqrt{t^2-1}
$.
We have
 $(t^2-1)\in S[-\pi,-2\vt)$, so that
 $\arg \sqrt{|t|^2e^{-2i\vt}-1}\in[-\frac{\pi}{2},-\vt)$.

\ref{A7.2-2}. For $\vt\in(0,\frac{\pi}{2}]$ the result follows from \ref{A7.2-1}.
 For $\vt=0$, the result follows from (\ref{DefXi}) by direct calculation.

\ref{A7.2-3}. We have
$
{\partial_r} \Phi(t)=
\frac{|\xi'(t)|}{|\xi (t)|}
\sin\left\{ \arg{\partial_r}\xi (t)
-\Phi(t) \right\}
$.
By \ref{A7.2-1}, ${\partial_r} \Phi(t)$ is strictly positive for
$\Phi(t)\in(-\pi-2\vt,-\frac{\pi}{2}-\vt)$.

\ref{A7.2-4}. and \ref{A7.2-5}. We have
$
{\partial_r}R(t)^2=2 |\xi '(t)||\xi (t)| \cos\left\{
\arg{\partial_r}\xi (t)
-\Phi(t) \right\}
$.
By \ref{A7.2-1}, ${\partial_r}R(t)$ is positive for
$\Phi(t)\in (-\frac{\pi}{2}-2\vt,-\vt)$
 and negative for
 $\Phi(t)\in (-\frac{3\pi}{2}-2\vt,-\pi-\vt)$.
$\blacksquare$

For $\l=|\l|e^{2i\vt}\in\overline{\mathbb{C}}_+\setminus\{ 0 \}$ we have
\begin{equation}\label{Z3-2PowInXi}
  \frac{2}{3}z(x,\l)^{\frac{3}{2}}=\l\xi (t),
  \qquad
  t=\frac{x}{\sqrt{\l}}=re^{-i\vt}\in S[-\frac{\pi}{2},0].
\end{equation}

\begin{lemma}
\label{DivG2}
 For each
 $\l=|\l|e^{2i\vt}\in
 S[0,\delta)\setminus\{0\}$
   there exist a unique $z_*\in \Gamma_\l$ such that
 $|z_*|=\min_{z\in\Gamma_\l}|z|$.
Moreover, the following relations are valid ($x_*$ defined by
$z_*=z(x_*,\l)$):
\begin{enumerate}
\item \label{A7.3-1}
if $\arg\l=0$, then $z_*=0$ and $x_*=\sqrt{\l}$,\\
if $\arg\l\in(0,\delta)$, then
$z_*\in
S[-\frac{\pi}{2}-\frac{\vartheta}{2},-\frac{\pi}{2}+\frac{5}{6}\vt]$,
  \item \label{A7.3-2}
   $|z(\cdot,\l)|$ is strictly decreasing on $[0,x_*)$ and
 strictly  increasing on $(x_*,\infty)$,
  \item \label{A7.3-3}
   if $t=\frac{x}{\sqrt{\l}}$
  for  $x\in [0,x_*)$, then
  $\arg\left( e^{2i\vt}{\partial_r}\xi(t)\right)
  \in[-\frac{\pi}{2},-\frac{\pi}{4}+\frac{3}{4}\vt]$,

\item \label{A7.3-4}
if $t_*=\frac{x_*}{\sqrt{\l}}$, then $|t_*|\le\frac{1}{\sin(\frac{\pi}{3})}$.

\end{enumerate}
\end{lemma}

{\it Proof.}
Let us prove uniqueness of $z_*$. By (\ref{Z3-2PowInXi}),
it is sufficient to show that for each
$\l$
 there exists
a unique solution $t_*$ of ${\partial_r}\abs{\xi(t)}=0$ for
$t=re^{-i\vt}$, $r\in\mathbb{R}_+$. For $\arg\l=0$ direct
calculation gives $z_*=0$ and $x_*=\sqrt{\l}$. Next we show the
existence of $t_*$ in the case
$\l=|\l|e^{2i\vt}\in
 S(0,\delta)\setminus\{0\}$.

Since $\xi(t)\neq 0$, we have
$2{\partial_r}\abs{\xi(t)}=|\xi(t)|^{-1}
\Re\left(
\xi(t)\overline{\frac{\partial \xi(t)}{\partial r}}
\right)
$.
Let us use the representation (\ref{AngularXi}):
$t=re^{-i\vt}=1+\eta e^{-i\varphi}$.
For any $t\in S[-\frac{\pi}{2},0) $ \refLem{A7-2}{Geom-0} implies
$\arg\xi(t)\in
[-\frac{3\varphi}{2}-\frac{\vartheta}{2}, -\frac{3\varphi}{2}]$;
similarly \refLem{A7.2-1}{23} implies
$\arg\frac{\partial }{\partial r}\xi(t) \in
[-\frac{\varphi}{2}-\frac{3\vartheta}{2}, -\frac{\varphi}{2}-\vt]$.
Thus
$\arg{\partial_r}\abs{\xi(t)}^2 \in
[-\varphi+\frac{\vartheta}{2}, -\varphi+\frac{3\vt}{2}]$
and we have
\begin{align}
{\partial_r}\abs{\xi(t)}&> 0 & &\text{for $-\varphi\in
\left(-\frac{\pi}{2}-\frac{\vartheta}{2},
\frac{\pi}{2}-\frac{3\vt}{2}\right)$}, \label{X1.eq} \\
{\partial_r}\abs{\xi(t)}&< 0 & &\text{for $-\varphi\in
\left(-\frac{3\pi}{2}-\frac{\vartheta}{2},
-\frac{\pi}{2}-\vt\right)$}. \label{X2.eq}
\end{align}
By (\ref{X1.eq}) and (\ref{X2.eq}), there exist at least one
solution of ${\partial_r}\abs{\xi(t)}=0$. Moreover, any solution
satisfies
\begin{equation}
\label{Ab.eq}
-{\varphi}\in \text{$\left[-\frac{\pi}{2}-\frac{3{\vartheta}}{2},
-\frac{\pi}{2}-\frac{{\vartheta}}{2}\right]$}.
\end{equation}
Now we show the uniqueness of $t_*$.
We need only verify
that for any $t$,
satisfying
\eqref{Ab.eq},
we have
$
\frac{1}{2}\frac{\partial^2}{\partial r^2}\abs{\xi(t)}^2
=\left|\frac{\partial\xi(t)}{\partial r}\right|^2+
\Re\left\{\overline{\xi(t)}\cdot\frac{\partial^2\xi(t)}{\partial r^2}\right\}>0
$. This is true if
\begin{equation}
\label{SecDer}
\left|{\partial_r\xi(t)}\right|^2>\abs{\xi(t)}
\left|{\partial_r^2\xi(t)}\right|.
\end{equation}
We have
$
\frac{\partial\xi(t)}{\partial r}=e^{-i\vartheta}\sqrt{t^2-1}$,
$
\frac{\partial^2\xi(t)}{\partial r^2}=e^{-3i\vartheta}\frac{r}{\sqrt{t^2-1}}
$.
Therefore \eqref{SecDer} is equivalent to
\begin{equation}\label{SecDer1}
|(t^2-1)^{{3\/2}}/t|>|\xi(t)|.
\end{equation}
Due to \eqref{Ab.eq}, $\abs{t}\le 1$. Therefore we have for $t=1+\eta e^{-i\varphi}$
\begin{equation}
\frac{\abs{t^2-1}^{\frac{3}{2}}}{\abs{t}}\ge\frac{\abs{t-1}^{\frac{3}{2}}\abs{t+1}^{\frac{3}{2}}}{1}=\eta^{\frac{3}{2}}\abs{2+\eta e^{-i\varphi}}^{\frac{3}{2}},
\qquad \abs{2+\eta e^{-i\varphi}}\ge 2\cos\frac{3\vartheta}{2}.
\end{equation}
Thus using $2\vt\in(0,\delta)
\subset[0,\frac{4}{3}\arccos\frac{1}{2^{\frac{1}{3}}})$ we obtain
\begin{equation}
\label{A7.3-unique}
\frac{\abs{t^2-1}^{\frac{3}{2}}}{\abs{t}}\ge
\eta^{\frac{3}{2}}\left(2\cos\frac{3\vartheta}{2}\right)^{\frac{3}{2}}
>2\eta^{\frac{3}{2}}.
\end{equation}
Since $\vt\in[0,\delta)
\subset[0,\frac{\pi}{3}]$, (\ref{Ab.eq})
gives
$\eta\le1$.
Thus we apply \refLem{A7-3}{Geom-0}, which  gives
  $2\eta^{\frac{3}{2}}\ge|\xi(t)|$. Substituting the last estimate in
   (\ref{A7.3-unique})
yields (\ref{SecDer1}) and (\ref{SecDer}). Therefore there exist
 a unique solution $t_*$ of
${\partial_r}\abs{\xi(t)}=0$. Thus $z_*=z(x_*,\l)$ for
$x_*=t_*\sqrt{\l}$.

\ref{A7.3-1}. and \ref{A7.3-2}. Using (\ref{Ab.eq}) and
\refLem{A7-2}{Geom-0} we obtain
  \begin{equation}\label{Argxi(z*)}
\arg\xi(t_*)\in
[-\frac{3\pi}{4}-\frac{11\vartheta}{4},
-\frac{3\pi}{4}-\frac{3\vartheta}{4}].
\end{equation}
By (\ref{Z3-2PowInXi}), this  proves \ref{A7.3-1}.
Uniqueness of $t_*$ and the relation
(\ref{Z3-2PowInXi}) prove \ref{A7.3-2}.

\ref{A7.3-3}. By (\ref{X1.eq}), for  $r<|t_*|$
 we have
$
   -\pi\le-\varphi\le-\frac{\pi}{2}-\frac{\vartheta}{2}
$.
Using \refLem{A7-2}{Geom-0} we obtain
$$
-\frac{\pi}{2}\le
\arg\left(
e^{2i\vt}{\partial_r}\xi(t)
\right)
=\arg
\left( e^{i\vt}
e^{-i\frac{\varphi}{2}}\sqrt{1-\frac{l}{2}e^{-i\varphi}}
\right)
\le \vt-\frac{\varphi}{2}\le-\frac{\pi}{4}+\frac{3\vt}{4}.
$$

\ref{A7.3-4}. Geometric considerations show that in terms of (\ref{AngularXi})
$
r=\frac{\sin\vp}{\sin(\vp-\vt)}
$.
By (\ref{Ab.eq}) and \refLem{A7-2}{Geom-0},
 for $t_*=1+\eta_* e^{-i\phi_*}$ we have
$
\frac{2\pi}{3}-\frac{\vt}{3}\ge\vp_*-\vt\ge\frac{\pi}{3},
$
so that $r_*=\frac{\sin\vp_*}{\sin(\vp_*-\vt)}\le\frac{1}{\sin\frac{\pi}{3}}$.
$\BBox$

\begin{lemma}
\label{PropZstarBigArgL}
 For each $\l=|\l|e^{2i\vt}\in S[\delta,\pi]\setminus \{0\}$ there exist a
  unique $z_*\in \Gamma_\l$ such that
  $\arg z_*=-\frac{\pi}{3}$. Moreover, the following
relations are valid (here $x_*$ defined by $z_*=z(x_*,\l)$):
\begin{enumerate}
\item \label{A7.4-1}
if $x\in[0,x_*)$, then $\arg z(x,\l)<-\frac{\pi}{3}$,\\
if $x\in(x_*,\infty)$, then $\arg z(x,\l)>\-\frac{\pi}{3}$,

\item \label{A7.4-2}
 $|z(\cdot,\l)|$ is strictly increasing on $(x_*,\infty)$,
\item \label{A7.4-3}
if $t_*=\frac{x_*}{\sqrt{\l}}$, then $|t_*|\le\frac{1}{\sin\frac{\pi}{3}}$
uniformly in $\l$.
\end{enumerate}
\end{lemma}

\no {\it Proof.} By \refLem{A7-4}{Geom-0} and (\ref{Z3-2PowInXi}),
$\Gamma_\l$ intersects the ray  $\{z:\arg z=-\frac{\pi}{3}\}$ at
least once. By (\ref{Z3-2PowInXi}), the sector in $z$-plane
$S[-\frac{\pi}{3}-\ve,-\frac{\pi}{3}+\ve]$ for small $\ve>0$
corresponds to
$S[-\frac{\pi}{2}-2\vt-\frac{3\ve}{2},-\frac{\pi}{2}-2\vt-\frac{3\ve}{2}]$
in $\xi$-plane.   By \refLem{A7.2-3}{23},  in this sector
 ${\partial_r}\Phi(t)>0$. Therefore, by (\ref{Z3-2PowInXi}),
$\frac{\partial}{\partial x}\arg z(x,\l)>0$ for
$z\in S[-\frac{\pi}{3}-\ve,-\frac{\pi}{3}+\ve]$.
 Hence $z_*$ is unique and \ref{A7.4-1} holds.

\ref{A7.4-2}. By (\ref{Z3-2PowInXi}), for $x\in(x_*,\infty)$ and $t=\frac{x}{\sqrt{\l}}$
  the hypothesis of
\refLem{A7.2-5}{23} is fulfilled. Therefore $\frac{\partial }{\partial r}R(t)>0$
and $|z(x,\l)|$ is strictly increasing in $x$.

\ref{A7.4-3}. By (\ref{Z3-2PowInXi}), we have $\arg\xi(t_*)=-\frac{\pi}{2}-2\vt$.
 By \refLem{A7-2}{Geom-0}, for $t_*=1+\eta_*e^{-i\varphi_*}$ we have
$\varphi_*-\vt\in[\frac{2\pi}{3}-\frac{\vt}{3},\frac{2\pi}{3} ]$.
Therefore using the  relation
$
r=\frac{\sin\vp}{\sin(\vp-\vt)}
$ in terms of  (\ref{AngularXi}),
 we obtain
$|t_*|\le\frac{1}{\sin\frac{\pi}{3}}$.
$\BBox$

In the next Lemma we analyze the  curves
 $\Gamma_\l^{\pm}$, defined in (\ref{DefGPM}).

\begin{lemma}
\label{SvaGForAll}
Let
$\l=|\l|e^{2i\vt}\in\ol\C_+\setminus\{0\}$. Then
\begin{enumerate}
 \item \label{A7.5-1}
 if $\vt=0$, then
 $\G_\l=[-\l^{\frac{2}{3}}({3\pi\/8})^{2\/3},\iy)$,\\
 if $0<\vt\le\frac{\pi}{2}$, then $\G_\l\subset S[-\pi+{4\/3}\vt,0)$,
 $\Gamma_\l^-\subset S[-\pi+{4\/3}\vt,-\frac{\pi}{3}]$,
$\Gamma_\l^+\subset S[-\frac{\pi}{2}-\frac{\vartheta}{2},0)$,
 \item \label{A7.5-2}
 $\inf_{z\in\G_\l}|z|\ge |\l|^{\frac{2}{3}}\sin\vt$,
 \item \label{A7.5-3}
 $\G_\l^-\subset\{z: |z|\le\co|\l|^{\frac{2}{3}}\}$,
  the length of $\Gamma_\l^-$ satisfies
  $|\G_\l^-|\le \co|\l|^{\frac{2}{3}}$,
\item \label{A7.5-4}
the  function  $|z(\cdot,\l)|$
 is strictly increasing on $[x_*,\infty)$.

\end{enumerate}

\end{lemma}
\no {\it Proof  } \ref{A7.5-1}. Direct calculation yields the
result for $\vt=0$.
 For $0<\vt\le\frac{\pi}{2}$  the assertion on $\Gamma_\l$ follows from \refLem{A7-4}{Geom-0} and
 $z=\l^{\frac{2}{3}}(\frac{3}{2}\xi)^{\frac{2}{3}}$. The assertion on
 $\Gamma_\l^\pm$  for $0\le\arg\l\le\delta$  follows from \ref{A7.5-1}
of the present Lemma
 and \refLem{A7.3-1}{DivG2}.
 For $\delta\le\arg\lambda\le\pi$ it follows from \refLem{A7.4-1}{PropZstarBigArgL}.

\ref{A7.5-2} follows from \refLem{A7-3}{Geom-0} and
$z=\l^{\frac{2}{3}}(\frac{3}{2}\xi)^{\frac{2}{3}}$.

\ref{A7.5-3}.
By \refLem{A7.3-4}{DivG2} and  \refLem{A7.4-3}{PropZstarBigArgL},
 $|t_*|\le\frac{1}{\sin\frac{\pi}{3}}$ uniformly in
 $\l\in\ol\C_+\setminus\{0\}$.
Using the definition (\ref{ZofT}) we conclude that $k(t)$ and  $k'(t)$ for
$t\in[0,t_*]$
are also uniformly bounded. Now the relations
$|\G_\l^-|=|\l|^{\frac{2}{3}}\int_{[0,t_*]}|k'(t)| |dt|$ and
$|z(x,\l)|=|\l|^{\frac{2}{3}}|k(t)|$, $t=\frac{x}{\sqrt{\l}}$
complete the proof.

\ref{A7.5-4}. We deduce from (\ref{Z3-2PowInXi}) that
$\frac{\partial}{\partial x}|z(x_,\l)|>0$ if
${\partial_r}|\xi(t)|>0$
for $t=re^{-i\vt}=\frac{x}{\sqrt{\l}}$.
Thus for $\delta\le\arg\lambda\le\pi$ the result follows from
\refLem{A7.4-2}{PropZstarBigArgL}. For $0\le\arg\l\le\delta$ it follows from
\refLem{A7.3-2}{DivG2}.
$\blacksquare$

\no {\bf Proof of Lemma~\ref{Z-star}. } Case \ref{LZ*-1} (or
\ref{LZ*-2}) follows from Lemma~\ref{DivG2} (or
Lemma~\ref{PropZstarBigArgL}).  $\blacksquare$

\no {\bf Proof of Lemma~\ref{27}.}
Fix $\l=|\l|e^{2i\vt}$. By (\ref{Z3-2PowInXi}), we have
  $h(x)=e^{\Re\l\xi(t)}$
for  $t=\frac{x}{\sqrt{\l}}=re^{-i\vt}$.  By \refLem{A7.2-2}{23},
if either $0<\arg\l\le\pi$, $r\ge0$ or $\arg\l=0$, $r>1$,
then
${\partial_r}\Re\left(\l\xi(t)\right)>0$.
It remains to consider the case
$\vt=0$ and $x\in[0,x_*)$; by \refLem{A7.2-2}{23}, $\Re\xi(t)=0$
and therefore $h=1$.
  $\blacksquare$

In order to prove Lemma~\ref{EstExpInt},
for a fixed $\l=|\l|e^{2i\vt}\in\ol\C\setminus\{0\}$ define the function
$
\Psi(z)
=\frac{2}{3}\frac{z^{\frac{3}{2}}}{|\l|}
$.
$\Psi$ maps $\Gamma_\l(z_1,z_2)$ onto $\gamma_\l(u_1,u_2)$,
$u_{j}=\Psi(z_{j})$, $j=1,2$.
Similarly, we set $\gamma_\l(u)=\Psi(\Gamma_\l(z))$ for $u=\Psi(z)$
and $\gamma_\l^{\pm}=\Psi(\Gamma_\l^{\pm})$.
We also set $u_0=\Psi(z_0)$ and
$u_*=\Psi(z_*)$.

Consider the integral
$\int_{\gamma_\l(u)}
f(v) |dv| $.
By (\ref{Z3-2PowInXi}), we
have the  parametrization
$
\gamma_\l(u)=\left\{v\in\mathbb{C}: v=v(r)\quad for\quad r\in[r_u,\infty],
\quad
 \hbox{where}
\quad
v(r)=e^{2i\vt}\xi(re^{-i\vt}), u=\Psi(z(r_u\sqrt{|\l|},\l))\right\}
$.

Consider the image of the part of $\Gamma_\l$, defined by the hypothesis a) and b) of
Lemma~\ref{EstExpInt}, under the mapping $\Psi$.  Assumption a) transforms into
a')
$\d\le\arg\l\le\pi$ and $u\in\gamma_\l$. Similarly, b) becomes
b') $0\le\arg\l\le\d$, $u\in\gamma_\l^+$.
 Then by \refLem{A7.2-2}{23}, $\Re v(r)$  is strictly increasing in $r$.
Thus we choose the variable  $\varkappa=\Re v$   and obtain
$$
 \int_{\gamma_\l(u)} |f(v)| |dv|=
\int_{\Re u}^\infty |{f}(\varkappa+i\Im v(\varkappa))|
\left|\frac{dv}{d\varkappa}\right|\, d\varkappa,
\qquad
\left|\frac{dv}{d\varkappa}\right|=\sqrt{1+\left(\frac{d\Im
v(\varkappa)}{d\varkappa}\right)^2}.
$$
Let us show that in both cases a') and b') the following estimate holds
:
\begin{equation}
\label{UToKappaEstimate}
 \int_{\gamma_\l(u)} |f(v)| |dv|
 \le\co
\int_{\Re u}^\infty |f(\varkappa+i\Im v(\varkappa))|
\, d\varkappa.
\end{equation}
Let us estimate $\left|\frac{dv}{d\varkappa}\right|$.
In terms of the parametrization $v(r)=e^{2i\vt}\xi(re^{-i\vt})$ we have
$$
\frac{d\Im v(\varkappa)}{d\varkappa}= \frac{ {\partial_r}
\Im
\left(e^{2i\vt}\xi(t) \right)}{ {\partial_r} \Re
\left(
e^{2i\vt}\xi(t)\right)
 }= \tan \arg {\partial_r}
e^{2i\vt}\xi(t).
$$
Consider the two cases a') and b').

a') Let $\d\le\arg\l\le\pi$ and $u\in\gamma_\l$. For each point
$v(r)\in\gamma_\l$ consider
$t=re^{-i\vt}$ such that $v(r)=e^{2i\vt}\xi(t)$.
  By \refLem{A7.2-1}{23}, $\arg{\partial_r}\xi(t)
\in [-\frac{\pi}{2}-\vt,-2\vt)$.
Hence
$\arg {\partial_r}
e^{2i\vt}\xi(t)\in[-\frac{\pi}{2}+\frac{\delta}{2}, 0)$ and
$
\left|\frac{d\Im v(\varkappa)}{d\varkappa}\right|\le\frac{1}{\sin(\frac{\delta}{2})}
$
uniformly for $\d\le\arg \l\le \pi$. Therefore
$\left|\frac{dv}{d\varkappa}\right|$ is also uniformly bounded.

b') Let
$0\le\arg\l\le\d$ and $u\in\gamma_\l^+=\Psi(\Gamma_\l^+)$.
For each point
$v(r)\in\gamma_\l^+$ consider
$t=re^{-i\vt}$ such that $v(r)=e^{2i\vt}\xi(t)$.
 Since $0\le\vt\le\frac{\pi}{7}$,
 (\ref{Argxi(z*)}) (equivalent to \refLem{A7.3-1}{DivG2})
  implies that $-\pi-\vt\le\arg\xi(t)$.
 Therefore by \refLem{A7-5}{Geom-0},
 for $v\in\gamma_\l^+$ we have
$\arg\left( e^{2i\vt}{\partial_r}
\xi(t)\right)\in[-\frac{\pi}{3}+\frac{\vt}{6},\frac{\vt}{2}]$.
Hence
$
\left|\frac{d\Im v(\varkappa)}{d\varkappa}\right|\le\frac{1}{\sin(\frac{\pi}{3})}
$ and
$
\left|\frac{dv}{d\varkappa}\right|
$
is uniformly bounded.

Thus  $\left|\frac{dv}{d\varkappa}\right|$
is uniformly bounded in both cases a') and b'), implying (\ref{UToKappaEstimate}).

\no {\bf Proof of Lemma~\ref{EstExpInt}.} Firstly we prove
(\ref{Exp-2}) for  the case $\d\le\arg\l\le\pi$. By
\refLem{A7.5-2}{SvaGForAll}, we have dist$(\G_\l,\{0\})>$const
uniformly in $\l$. Thus we  replace $\langle \cdot\rangle$ by
$|\cdot|$. The change of variables $v=\P (s)$ in (\ref{Exp-2})
results in the equivalent relation
\begin{equation}
\label{Int-E} \int_{\gamma_\l(u)} \frac{e^{-2|\l|\Re
v}}{|v|^{\frac{2}{3}(\a+\frac{1}{2})}}\,|dv| \le \frac{\co}{|\l|}
\frac{e^{-2|\l|\Re u}}{|u|^{\frac{2}{3}(\a+\frac{1}{2})}}, \quad
u=\frac{2}{3}\frac{z^{\frac{3}{2}}}{|\l|}\in\gamma_\l .
\end{equation}
By (\ref{UToKappaEstimate}), we have the auxiliary estimate
\begin{equation}
\label{EinU} \int_{\gamma_\l(u)} e^{-2|\l|\Re v}\,|dv| \le
\co \frac{e^{-2|\l|\Re u}}{|\l|},
\qquad
u\in\gamma_\l.
\end{equation}
We show that (\ref{Int-E}) follows from (\ref{EinU}). If
$u\in\g_\l^+$, then \refLem{A7.5-4}{SvaGForAll} yields
$|u|=\min\limits_{v\in\gamma_\l(u)}|v|$. Therefore
  (\ref{Int-E}) follows from (\ref{EinU}).
If $u\in\g_\l^-$, then  dist$(\gamma_\l,\{0\})>$const and we have
\begin{equation}
\label{EinU-2}
\int_{\gamma_\l(u)}
\frac{e^{-2|\l|\Re
v}}{|v|^{\frac{2}{3}(\a+\frac{1}{2})}}\,|dv| \le
C
\int_{\gamma_\l(u)} e^{-2|\l|\Re v}\,|dv|,
\qquad
u\in\gamma_\l^-.
\end{equation}
By \refLem{A7.5-3}{SvaGForAll}, $\gamma_\l^-(u)$ is
 uniformly bounded. Therefore,
$|u|$ is bounded on $\gamma_\l^-$,
so (\ref{EinU-2}) and (\ref{EinU})  imply (\ref{Int-E}).

b) We prove (\ref{Exp-2}) for  the case $z\in \G_\l^+$. It
suffices to show that for any $z\in\G_\l^+$ we have
\begin{equation}
\label{EstSmallTG+}
\int_{\Gamma_\l(z)} |e^{-\frac{4}{3}{s}^{\frac{3}{2}}}|\,|ds| \le C
|e^{-\frac{4}{3}z^{\frac{3}{2}}}|, \quad|z|\leq1;\quad
\int_{\Gamma_\l(z)} \frac{|e^{-\frac{4}{3}{s}^{\frac{3}{2}}}|}{|s|^\a}\,|ds| \le C
\frac{|e^{-\frac{4}{3}z^{\frac{3}{2}}}|}{|z|^{\a+\frac{1}{2}}}, 
|z|>1.
\end{equation}
By the change of variable $u=\Psi(z)$ the first estimate in (\ref{EstSmallTG+})
follows from (\ref{EinU}).
For the proof of the second estimate in (\ref{EstSmallTG+})
we observe that by  \refLem{A7.5-4}{SvaGForAll}, for $z\in\Gamma_\l^+$
we have
 $|z|=\min\limits_{v\in\Gamma_\l(z)}|v|$. Hence, (\ref{EstSmallTG+})
 follows from
 (\ref{EinU}).
 This proves (\ref{Exp-2}).

Next we shall prove (\ref{pow-2}).
By the change of variable $u=\Psi(z)$ we have
\begin{equation}
\label{PowLem-A}
\int_{\Gamma_\l(z)}
\frac{|ds|}{\langle s\rangle^\a}=
\left(
\frac{2}{3}
\right)^{\frac{2}{3}(\alpha+\frac{1}{2})}
\frac{I}{|\l|^{\frac{2}{3}(\alpha-1)}},
\quad
I=
\int_{\gamma_\l(u)}
\frac{|dv|}{|v|^{\frac{1}{3}}(\varepsilon+|v|^{\frac{2}{3}})^\alpha},
\quad
u=\frac{2}{3}\frac{z^{\frac{3}{2}}}{|\l|},
\end{equation}
where
$
\ve=(\frac{3}{2}|\l|)^{-\frac{2}{3}}\le\co
$.

Assume $z\in\Gamma_\l^+$, $0<\arg\l\le\pi$.
Set $a=\Re u$ and $b=\Im u$.
 By \refLem{A7.3-1}{DivG2}, \refLem{A7.3-3}{DivG2},  \refLem{A7-5}{Geom-0}
  and definition of $z_*$,
$b\le0$.
By \refLem{A7.2-2}{23}, $\Im v$ is strictly decreasing on $\gamma_\l^+$, so that
 $|b|=\min\limits_{v\in\gamma_\l(u)}|\Im v|$.
Thus using   (\ref{UToKappaEstimate}) we obtain
 $$
 I\le\co J,\qquad
 J=\int_a^\infty
 \frac{d\varkappa}{|\varkappa|^{\frac{1}{3}}(\ve+|b|^{\frac{2}{3}}+|\varkappa|^{\frac{2}{3}})^\alpha}.
 $$
 Consider two cases: $a\ge0$ and $a<0$. Firstly, let $a\ge0$. Then
we have
 $$
 J={3\/2} \int_{a^{{2\/3}}}^\iy \frac{dx}{(\ve+|b|^{\frac{2}{3}}+x)^\alpha}=
 \frac{\frac{3}{2}}{\alpha-1}
  \frac{1}{(\ve+|b|^{\frac{2}{3}}+a^{\frac{2}{3}})^{\alpha-1}}
  \le
  \frac{C}{(\ve+|u|^{\frac{2}{3}})^{\alpha-1}}
 $$
 Therefore
 $$
 \frac{I}{|\l|^{\frac{2}{3}(\alpha-1)}}
 \le
\frac{C}{|\l|^{\frac{2}{3}(\alpha-1)}(\ve+|u|^{\frac{2}{3}})^{\alpha-1}}
 =
 \frac{C}{\langle z\rangle^\alpha},
 $$
 which together with (\ref{PowLem-A}) proves (\ref{pow-2}).
 Secondly, let $a<0$. Again, by
 $x=\varkappa^{\frac{2}{3}}$, we obtain
 $$
 J\le \co
 \int_0^\infty
 \frac{d\varkappa}{\varkappa^{\frac{1}{3}}(\ve+|b|^{\frac{2}{3}}+\varkappa^{\frac{2}{3}})^\alpha}
 \le
 \frac{C}{(\ve+|b|^{\frac{2}{3}})^{\alpha-1}}.
 $$
 Due to \refLem{A7.3-1}{DivG2} and
 \refLem{A7.4-1}{PropZstarBigArgL}, for $u=a+ib\in\gamma_\l^+$
  and $a<0$ we have $|a|\le\co |b|$ uniformly in $0\le\arg\l\le\pi$. Therefore $|u|\le\co b$ and
  $
  \frac{1}{(\ve+|b|^{\frac{2}{3}})^{\alpha-1}}
  \le
  \frac{C}{(\ve+|u|^{\frac{2}{3}})^{\alpha-1}}
  $. Thus
\begin{equation}
\label{PowLem-B}
 \frac{I}{|\l|^{\frac{2}{3}(\alpha-1)}}
 \le
\frac{C}{|\l|^{\frac{2}{3}(\alpha-1)}(\ve+|u|^{\frac{2}{3}})^{\alpha-1}}
 =
 \frac{C}{\langle z\rangle^\alpha},
 \end{equation}
 which together with (\ref{PowLem-A}) proves (\ref{pow-2}).

 Assume $z\in\Gamma_\l^-$ and $\d\le\arg\l\le\pi$.
Recall that $u_*=\frac{2}{3}\frac{z_*^{\frac{3}{2}}}{|\l|}$.
We have $I=I_-+I_+$, where
$$
I_-=
\int_{\gamma_\l(u,u_*)}
\frac{|dv|}{|v|^{\frac{1}{3}}(\varepsilon+|v|^{\frac{2}{3}})^\alpha},
\quad
I_+=
\int_{\gamma_\l^+}
\frac{|dv|}{|v|^{\frac{1}{3}}(\varepsilon+|v|^{\frac{2}{3}})^\alpha},
\quad
u=\frac{2}{3}\frac{z^{\frac{3}{2}}}{|\l|}.
$$
By \refLem{A7.5-2}{SvaGForAll}, we  have
$|z|\ge\sin\frac{\delta}{2}$ for $z\in\Gamma_\l$, so $|u|\ge
\frac{2}{3}\left(\sin\frac{\delta}{2}\right)^{\frac{3}{2}}$ for
$u\in\gamma_\l$. By \refLem{A7.5-3}{SvaGForAll}, the length of the
curve $|\gamma_\l^-|<\co$. Therefore $I_-\le\co$. By
(\ref{PowLem-B}) for $u=u_*$, we have $I_+\le\co$ so that
$I\le\co$. Next, by \refLem{A7.5-3}{SvaGForAll}, $\gamma_\l^-$ is
uniformly bounded, implying $ \co \le
(\ve+|u|^{\frac{2}{3}})^{-1\a} $ and
\begin{equation}
 \frac{I}{|\l|^{\frac{2}{3}(\alpha-1)}}
 \le
\frac{C}{|\l|^{\frac{2}{3}(\alpha-1)}(\ve+|u|^{\frac{2}{3}})^{\alpha-1}}
 =
 \frac{C}{\langle z\rangle^\alpha},
 \end{equation}
 which together with (\ref{PowLem-A}) proves (\ref{pow-2}).
$\blacksquare$

\no {\bf Proof of Lemma~\ref{PowLem}.} First we prove
(\ref{pow-3Le1}). By the change of variable $u=\Psi(z)$, we have
\begin{equation}
\label{PowLem-c}
\int_{\Gamma_\l^-}
\frac{|ds|}{\langle s\rangle^\a}=
\left(
\frac{2}{3}
\right)^{\frac{2}{3}(\alpha+\frac{1}{2})}
\frac{I}{|\l|^{\frac{2}{3}(\alpha-1)}},
\quad
I=
\int_{\gamma_\l^-}
\frac{|dv|}{|v|^{\frac{1}{3}}(\varepsilon+|v|^{\frac{2}{3}})^\alpha},
\end{equation}
where
$
\ve=(\frac{3}{2}|\l|)^{-\frac{2}{3}}\le\co
$.
By \refLem{A7.2-2}{23}, $\Im v$ is strictly decreasing on
$\gamma_\l^-$. Thus we parameterize the last integral by
$\chi=\Im v$, so that  $v(\chi)=\Re v(\chi)+i\chi$.
By \refLem{A7.5-3}{SvaGForAll},
$\gamma_\l^-\subset
\{
u\in\C:|u|<c
\}
$ for some $c>0$ independent of $\l$.
Therefore
\begin{equation}
\label{PowLem-C1}
I\le
2\int_0^{c}
\left|\frac{dv}{d\chi}\right|
\frac{d\chi}{{\chi}^{\frac{1}{3}}
\{\varepsilon+{\chi}^{\frac{2}{3}}\}^\alpha},
\end{equation}
where
$\left|\frac{dv}{d\chi}\right|=
\sqrt{1+\left(\frac{d\Re
v(\chi)}{d\chi}\right)^2}$. Recall that $\l=|\l|e^{2i\vt}$
and $t=re^{-i\vt}$.
In order to estimate
$\left|
\frac{d\Re v(\chi)}{d\chi}
\right|$ we make use of the parametrization\\
$
\gamma_\l(u)=\left\{v\in\mathbb{C}: v=v(r)\ for\  r\in[r_u,\infty],
\
 \hbox{where}
\
v(r)=e^{2i\vt}\xi(re^{-i\vt}), u=\Psi(z(r_u\sqrt{|\l|},\l))\right\}.
$
Thus we have
$$
\frac{d\Re v(\chi)}{d\chi}= \frac{ {\partial_r}
\Re
\left(e^{2i\vt}\xi(t) \right)}{ {\partial_r} \Im
\left(
e^{2i\vt}\xi(t)\right)
 }= \cot\arg
 \left(
  e^{2i\vt}{\partial_r}
\xi(t)
\right).
$$
By \refLem{A7.3-3}{DivG2}, \refLem{A7.4-1}{PropZstarBigArgL},
(\ref{DefDelta}) and  \refLem{A7-5}{Geom-0},
$e^{2i\vt}{\partial_r}
\xi(t)$ is in a sector isolated from the real axis uniformly in $\l$.
Therefore $\left|\frac{d\Re v(\chi)}{d\chi}\right|\le\co$
and
$\left|\frac{dv}{d\chi}\right|\le\co$.
Substituting this estimate in (\ref{PowLem-c}) and making change of variable
$x=\ve^{-\frac{3}{2}}\chi$ we obtain
\begin{equation}
\label{PowLem-D} I\le\frac{C}{\ve^{\alpha-1}}
\int_0^{C_1\ve^{-\frac{3}{2}}}
\frac{dx}{x^{\frac{1}{3}}(1+x^{\frac{2}{3}})^\alpha},\ \ \ \
\ve=(\frac{3}{2}|\l|)^{-\frac{2}{3}}\le\co,
\end{equation}
Recall that $|\l|\ge R>0$. Then the proof of (\ref{pow-3Le1})
follows from (\ref{PowLem-c}) and (\ref{PowLem-D}).

Now we prove (\ref{EIV0}). By the change of variable $u=\Psi(z)$, we have
\begin{equation}
\label{PowLem-c1}
\int_{\Gamma_\l}
\frac{|ds|}{|\l|^{\frac{4}{3}}+|s|^2}=
\frac{I_-+I_+}{|\l|^{\frac{2}{3}}},
\quad
I_\pm=
\int_{\gamma_\l^\pm}
\frac{|dv|}{(\frac{3}{2}|v|)^{\frac{1}{3}}(1+
(\frac{3}{2}|v|)^{\frac{4}{3}})},
\end{equation}
For $I_+$ we use (\ref{UToKappaEstimate}), which gives
\begin{equation}
\label{PowLem-F}
I_+\le\co
\int_0^\infty
\frac{d\varkappa}{\varkappa^{\frac{1}{3}}(1+\varkappa^{\frac{4}{3}})}\le\co.
\end{equation}
In order to estimate $I_-$ we use the parametrization  of
 $\gamma_\l^-$ by $\chi=\Im v$.
 Repeating the arguments used above for the proof of (\ref{pow-3Le1}),
 we conclude that
$\left|\frac{dv}{d\chi}\right|\le\co$ and
\begin{equation}
\label{PowLem-G}
I_-\le\co
\int_0^\infty
\frac{d\chi}{\chi^{\frac{1}{3}}(1+\chi^{\frac{4}{3}})}\le\co.
\end{equation}
Now (\ref{PowLem-F}) and (\ref{PowLem-G}) give $I\le\co$,
which by (\ref{PowLem-c1}) proves (\ref{EIV0}).
$\blacksquare$

Let $P_{\pm}$ be defined by (\ref{DefP_pm}).
\begin{lemma}
\label{EstP+P-}
Let $\l=|\l|e^{2i\vt}\in S[0,\delta]$ and $q, q'\in L^\iy(\R)$. Then
for some absolute constant $C$
the following
estimates are fulfilled:
\begin{equation}
|P_-(z, z_*)|
\le C|e^{-\frac{4}{3}z^{\frac{3}{2}}}|\cdot
\left(\|q'\|_\iy+{\|q\|_\infty}{|\l|^{-\frac{1}{6}}}\right),\quad
z\in\G_\l^-,
 \label{KeyLem2}
\end{equation}
\begin{equation}
|P_+(z_1,z_2)| \le  C|e^{\frac{4}{3}z_2^{\frac{3}{2}}}|\cdot
\left(\|q'\|_\iy+{\|q\|_\infty}{|\l|^{-\frac{1}{6}}}\right)\label{KeyLem1}.
\end{equation}
\end{lemma}
\no {\it Proof.} Let $x_1\le x_2\le x_*$,
$t_{1,2}=\frac{x_{1,2}}{\sqrt{\l}}$ and
$z_{1,2}=z(x_{1,2},\l)\in\Gamma_\l^-$. Let
$\xi_{1,2}=\xi(t_{1,2})$. Then, \refLem{A7.3-2}{DivG2} yields
$|\xi_1-\xi_2|\le2|\xi_1|$. Lemma~\ref{27} gives
$|e^{2\l(\xi_1-\xi_2)}-1|\le2|\l||\xi_1-\xi_2|$ and
$|e^{2\l(\xi_1-\xi_2)}-1|\le\co$. Therefore
\begin{equation}
| (e^{2\l(\xi_1-\xi_2)}-1)/\l\xi_1|\le \co(1+|\l||\xi_1|)^{-1}.
\end{equation}
Using $z_{1,2}=\l^{\frac{2}{3}}(\frac{3}{2}\xi_{1,2})^{\frac{2}{3}}$
we obtain
  the following estimate uniformly in $0\le\arg\l\le\d$:
\begin{equation}\label{Rpm3/2}
  |(e^{{4\/3}(z_1^{{3\/2}}-z_2^{{3\/2}})}-1)z^{-{3\/2}}|
  \le \co {\la z_1\ra }^{{3\/2}}.
  \end{equation}
Introduce the functions
$R_\pm(s,z)=e^{\pm\frac{4}{3}s^{\frac{3}{2}}}-e^{\pm\frac{4}{3}z^{\frac{3}{2}}}$.
 Evidently
 $
\pa_sR_\pm(s,z)=\pm2\sqrt{s}e^{\pm\frac{4}{3}s^{\frac{3}{2}}} $.
Integrating by parts, we have
\begin{equation}\label{}
P_-(z_*,z)= R_-(z,z_*)f(z)+
\int_{\Gamma_\l(z,z_*)}
R_-(s,z_*)f'(s)\,ds,
\quad
f(s)=\r(s){\hat{q}(s)}/(2\sqrt{s}\l^{\frac{1}{6}}).
\end{equation}
Therefore using (\ref{EffPotentials}), (\ref{EstNu}),
(\ref{Rpm3/2}) and $|\frac{d}{ds}\hat{q}(s)|\le
\co{\|q'\|_\infty}{|\l|^{-\frac{1}{6}}}$ we obtain
\begin{equation}\label{EPG-}
|P(z,z_*)|\le
C\frac{|e^{-\frac{4}{3}z^{\frac{3}{2}}}|}{|\l|^{\frac{1}{6}}}
\left[ \frac{\|q\|_\infty}{\la z\ra^{{1\/2}}}+
\int_{\G_\l(z,z_*)}\!\!\!|ds| \left(
\frac{|\l|^{-\frac{1}{6}}\|q'\|_\infty}{\la s\ra^{\frac{1}{2}}}+
\frac{|\l|^{-{2\/3}}\|q\|_\iy}{\la s\ra^{{1\/2}}}+
\frac{\|q\|_\iy}{\la s\ra^{{3\/2}}} \right) \right],
\end{equation}
which together with (\ref{pow-2}--\ref{pow-3Le1}) gives
(\ref{KeyLem2}). Similarly we have
\begin{equation}\label{}
P_+(z_1,z_2)=- R_+(z_1,z_2)f(z_1)- \int_{\G_\l(z_1,z_2)}
R_+(s,z_2)f'(s)ds.
\end{equation}
Again using (\ref{EffPotentials}), (\ref{EstNu}),
(\ref{Rpm3/2}) and
$|\frac{d}{ds}\hat{q}(s)|\le
\co{\|q'\|_\infty}{|\l|^{-\frac{1}{6}}}$ we have
$$
|P(z_1,z_2)|\le
\co\frac{|e^{\frac{4}{3}z_2^{\frac{3}{2}}}|}{|\l|^{\frac{1}{6}}} \left[
\frac{\|q\|_\infty}{(1+|z_2|)^{\frac{1}{2}}}+
\int\limits_{\Gamma_\l(z_1,z_2)}
\left(
\frac{|\l|^{-\frac{1}{6}}\|q'\|_\infty}{\langle s\rangle^{\frac{1}{2}}}+
\frac{|\l|^{-\frac{2}{3}}\|q\|_\infty}{\langle s\rangle^{\frac{1}{2}}}+
\frac{\|q\|_\infty}{\langle s\rangle^{\frac{3}{2}}} \right)\,|ds| \right],
$$
which together with (\ref{pow-2}), (\ref{pow-3Le1}) gives
(\ref{KeyLem1}).
$\blacksquare$

\begin{lemma}
\label{Ostatkiu1}
Let $q\in\cB$. Define $F(z)=\ai^2(ze^{\pm \frac{2\pi i}{3}})V_q(z)$.
Then uniformly in
$z\in\Gamma_\l^-$ and $\l\in S[-\delta,\delta]$
 the following estimates hold for sufficiently large
$|\l|$:
\begin{equation}
\label{GetAs-pr1} \left| \int_{\Gamma_\l(z)}
e^{\frac{4}{3}(z^{\frac{3}{2}}-s^{\frac{3}{2}})}F(s)\,ds \right|
\le\co\frac{\|q\|_\cB}{|\l|^{\frac{1}{3}}},\qquad \left|
\int_{\Gamma_\l(z,z_*)} e^{\frac{4}{3}s^{\frac{3}{2}}}F(s)\,ds
\right| \le\co\frac{\|q\|_\cB}{|\l|^{\frac{1}{3}}}.
\end{equation}
\end{lemma}
\no {\it Proof.} We show the first estimate in
(\ref{GetAs-pr1}).Using (\ref{EffPotentials}), (\ref{EstNu}),
(\ref{aiest}) and (\ref{Exp-2}) we obtain
\begin{equation}
\label{Take As-0} \biggl| \int_{\G_\l^+}
e^{\frac{4}{3}(z^{\frac{3}{2}}-s^{\frac{3}{2}})}F(s)\, ds
\biggr|\le C\cdot\frac{\lVert
q\rVert_\infty}{|\lambda|^{\frac{1}{3}}}.
\end{equation}
Using
$R_-(s,z_*)=e^{-\frac{4}{3}s^{\frac{3}{2}}}-e^{-\frac{4}{3}z_*^{\frac{3}{2}}}$
 we integrate by parts  the integral over
$\Gamma_\l(z,z_*)$. We have
\begin{equation}
  \label{AsQ}
I = \int_{\G_\l(z,z_*)}\!\!\!\! \!\!\!\! \!\!\!\!
e^{\frac{4}{3}(z^{\frac{3}{2}}-s^{\frac{3}{2}})}F(s) ds =
\frac{e^{\frac{4}{3}z^{\frac{3}{2}}}}{2}\biggl[R_-(z,z_*)\frac{F(z)}{\sqrt{z}}
+\int_{\Gamma_\l(z,z_*)}
 R_-(s,z_*)\left(\frac{F(s)}{\sqrt{s}}\right)'\, ds\biggr].
\end{equation}
Therefore, using  Lemma~\ref{27},
$|\frac{d}{ds}\hat{q}(s)|\le
\co{|\l|^{-\frac{1}{6}}}{\|q'\|_\infty}$,
(\ref{EffPotentials}), (\ref{EstNu}), (\ref{aiest}), (\ref{aiprimest})
and (\ref{Rpm3/2}),
 we obtain from (\ref{AsQ})
\begin{equation}
| I| \le
\frac{C}{|\lambda|^{\frac{1}{3}}}
\biggl[
\frac{\|q\|_\infty}{\langle z\rangle^{\frac{1}{2}}}+
\int_{\Gamma_\l^-}
\biggl(\frac{\lVert
q\rVert_\infty}{\langle s\rangle^{\frac{3}{2}+\frac{1}{2}}} +
\frac{|\lambda|^{-\frac{1}{6}}\lVert
q'\rVert_\infty}{\langle s\rangle}+
\frac{|\lambda|^{-\frac{2}{3}}\lVert
q\rVert_\infty}{\langle s\rangle}
\biggr)\,
| ds |
\biggr].
\end{equation}
By Lemma~\ref{PowLem}, for sufficiently large $|\l|$ this implies
$| I|\le {\co}{|\l|^{-\frac{1}{3}}}\|q\|_{\cB} \left(1 +
\abs{\lambda}^{-\frac{1}{6}}{\log(\abs{\lambda}+1)}{} \right) $.
Together with (\ref{Take As-0}) this proves the first estimate in
(\ref{GetAs-pr1}). In order to prove the second estimate in
(\ref{GetAs-pr1})
 we use similar arguments with
$R_+(s,z_*)=e^{\frac{4}{3}s^{\frac{3}{2}}}-e^{\frac{4}{3}z_*^{\frac{3}{2}}}$.
$\blacksquare$


\begin{thebibliography}{c}
\bibitem{AS} Abramowitz, M. and Stegun, A., eds.:
{\it Handbook of Mathematical Functions. } N.Y.: Dover
Publications Inc.

\bibitem{CKK} Chelkak, D., Kargaev, P.,  Korotyaev, E.:
An Inverse Problem for an Harmonic Oscillator Perturbed by
Potential: Uniqueness.   Lett.  Math. Phys.
 {\bf 64}(1),  7-21 (2003)

\bibitem{CKK-CMP} Chelkak D., Kargaev P., Korotyaev E.:
Inverse problem for harmonic oscillator perturbed by potential,
characterization. Comm. Math. Phys. in press



\bibitem{PT}P\"{o}schel, J., Trubovitz, E.: {\it Inverse Spectral
 Theory.}
Boston: Academic Press,  1987


\end{thebibliography}
\end{document}